\documentclass[12pt]{iopart}
\usepackage{graphicx}
\usepackage{amssymb}
\usepackage{cite}
\begin{document}

\topical{ Fingerprints of spin-orbital entanglement in transition metal oxides }

\author{      Andrzej M. Ole\'s  }

\address{     Marian Smoluchowski Institute of Physics, Jagellonian
              University, \\ Reymonta 4, PL-30059 Krak\'ow, Poland }
\address{     Max-Planck-Institut f\"ur Festk\"orperforschung, \\
              Heisenbergstrasse 1, D-70569 Stuttgart, Germany }
\ead{a.m.oles@fkf.mpg.de}

\date{May 10, 2012}

\begin{abstract}
The concept of spin-orbital entanglement on superexchange bonds in
transition metal oxides
is introduced and explained on several examples. It is shown that
spin-orbital entanglement in superexchange models destabilizes the
long-range (spin and orbital) order and may lead either to a disordered
spin-liquid state or to novel phases at low temperature which arise
from strongly frustrated interactions. Such novel ground states cannot
be described within the conventionally used mean field theory which
separates spin and orbital degrees of freedom. Even in cases where the
ground states are disentangled, spin-orbital entanglement occurs in
excited states and may become crucial
for a correct description of physical properties at finite temperature.
As an important example of this behaviour we present spin-orbital
entanglement in the $R$VO$_3$ perovskites, with
$R$=La,Pr,$\cdots$,Yb,Lu, where such finite temperature properties
of these compounds can be understood only using entangled states:
($i$) thermal evolution of the optical spectral weights,
($ii$) the dependence of transition temperatures for the onset of
orbital and magnetic order on the ionic radius in the phase diagram of
the $R$VO$_3$ perovskites, and
($iii$) dimerization observed in the magnon spectra for the $C$-type
antiferromagnetic phase of YVO$_3$.
Finally, it is shown that joint spin-orbital excitations in an ordered
phase with coexisting antiferromagnetic and alternating orbital order
introduces topological constraints for the hole propagation and will
thus radically modify transport properties in doped Mott insulators
where hole motion implies simultaneous spin and orbital excitations.
\\
\\
{\it Published in: Journal of Physics: Condensed Matter \textbf{24}, 313201 (2012).}
\end{abstract}



\maketitle

\section{Introduction: Entanglement in many-body systems}
\label{sec:intro}

Superexchange models with spin-orbital entanglement on superexchange
bonds, discovered by exact diagonalization of finite chains \cite{Ole06},
are a good recent example of entanglement in many-body systems.
Entanglement is inherent to quantum mechanics and occurs in several
systems. In general it means that quantum states have internal
structure and cannot be represented as products of states which belong
to different subspaces of the full Hilbert space \cite{Nie00,Ben06,Hor09}.
This property of quantum states has gained renewed interest in recent
years as it was found in several many-body quantum systems and it was
realized that it may play a role in quantum information. It is shown below
that it leads to measurable consequences in condensed matter systems
with strongly correlated electrons, when orbital degrees of freedom are
active. This new development concerns both model systems and the
physical properties of Mott (or charge transfer) insulators
--- we summarize it shortly in the present topical review.

Entanglement in quantum many-body systems is a broad field
\cite{Ben06} and will not be discussed here as such. In the last
few decades the interest in quantum entanglement has risen sharply in
various formerly disconnected subfields of physics. At present these
different communities come closer to each other in the search for
universal ways of quantifying entanglement and developing algorithms
to treat many-body quantum systems. An interested reader is encouraged
to consult several review articles published recently on this subject
--- we name here only a few which focus on entanglement in:
($i$)
many-body systems \cite{Ami08},
($ii$)
quantum spin systems \cite{Ami09},
($iii$)
interacting fermionic and bosonic many-particle systems \cite{Pes09},
($iv$)
optical lattices \cite{Blo08}, and finally,
($v$)
quantum cryptography and quantum communication \cite{Hor09}.
Entanglement entropy plays a central role in these systems and is
frequently used as a quantitative measure of entanglement \cite{Aff09}.

Spin-orbital entanglement occurs either due to the relativistic on-site
spin-orbit coupling or due to superexchange interactions on the bonds.
While finite spin-orbit coupling introduces on-site entanglement,
a qualitatively new and challenging situation is encountered when
degenerate $3d$ orbitals of transition metal ions are partly filled and
orbital degrees of freedom have to be treated on equal footing with
electron spins in the effective spin-orbital superexchange model
\cite{Kug82}. Here we mainly on the latter but some recent examples of
entangled states in cases with strong spin-orbit coupling will also be
mentioned for completeness at the end.

When degenerate $3d$ orbitals in a transition metal oxide are partly
filled, realistic superexchange includes both orbital and spin degrees
of freedom that are strongly interrelated \cite{Kug82,Ole09}. The
microscopic models designed to describe realistic systems with strongly
correlated and partly localized electrons include as well orbital
interactions which follow from the orbital-lattice coupling and tune
the orbital correlations. These latter interactions are rather strong
in the $e_g$ orbital systems and stabilize the orbital order at rather
high temperature, as for instance in LaMnO$_3$ \cite{Fei99}. It such
cases it is well justified to treat the spin-orbital superexchange in
the mean field (MF) approximation which separates orbital degrees of
freedom and their dynamics from the spin ones \cite{Ole05}. The spin
and orbital degrees of freedom order then in a complementary way and
their order follows the classical Goodenough-Kanamori rules \cite{Goode}.
They were derived long ago from the microscopic insights concerning
the structure of spin-orbital superexchange and predict that the
antiferromagnetic (AF) order coexists with ferro-orbital (FO) order and
ferromagnetic (FM) order coexists with alternating orbital (AO) order.
Out of many examples which follow these rules, we mention here only
the LaMnO$_3$ perovskite, with active $e_g$ orbitals and coexisting
FM/AO order in $ab$ planes and AF/FO order along the $c$ axis
\cite{Fei99,Dag01}. In this case spin and orbital operators indeed
separate as the orbital order sets in at high temperature
$T_{\rm OO}=780$ K and is already saturated when the $A$-type AF
($A$-AF) order occurs at $T_{\rm N}\simeq 140$ K. Therefore, the optical
spectral weights measured in experiment are well described by the MF
decoupling of spin and orbital degrees of freedom \cite{Kov10}. For this
reason simple treatments of the models of manganites which use the MF
approach \cite{Feh04,Dag06} or Hartree-Fock decoupling \cite{Ros07} are
very successful in modeling the complex phase diagram of monolayer,
bilayer and cubic manganites \cite{Tok06}.

In a number of compounds with active orbital degrees of freedom where
strong on-site Coulomb interactions localize electrons (or holes) and
give rise to spin-orbital superexchange, two different types of
long-range order compete with each other. A prominent example of this
behaviour are the $R$VO$_3$ perovskites, where $R$=Lu,Yb,$\cdots$,La.
In the case of perovskite vanadates the cubic symmetry is broken by
GdFeO$_3$-like distortions and the $xy$ orbitals are singly occupied at
all V$^{3+}$ ions, while the second electron occupies the $\{yz,zx\}$
doublet. One finds here two different AF phases for V$^{3+}$ ions in
$d^2$ electronic configuration with $S=1$ spin:
($i$) the $C$-type AF
($C$-AF) phase with staggered AF order in the $ab$ planes accompanied
by the FM order along the $c$ axis, coexisting with  weak AO order of
active $t_{2g}$ orbitals $\{yz,zx\}$, and
($ii$) the $G$-type AF ($G$-AF) phase with staggered AF order along all
three cubic axes coexisting with robust $C$-type orbital order \cite{Miy06}.
In fact, the coexisting orbital and magnetic order in these phases obey
the Goodenough-Kanamori rules along the $c$ axis. However, the situation
in $ab$ planes of this class of compounds is puzzling as
two alternating orders coexist, both for spins and for orbitals.
The reasons of this coexistence are more subtle --- there also $xy$
orbitals are singly occupied at every ion and they are in a FO state, while
the spin order, driven mainly by them, is AF. So once again, the order
in the $ab$ planes can be understood using the Goodenough-Kanamori rules.
However, the AF/AO order may be seen as entangled in the subspace of
$\{yz,zx\}$ orbitals and provides an interesting situation in doped
systems, as we shall discuss below.

Recent interest and theoretical progress in the understanding of
spin-orbital superexchange models was triggered by the observation that
orbital degeneracy significantly enhances quantum fluctuations which
may even suppress long-range order when different types of symmetry
broken states compete with each other near the quantum critical point
\cite{Fei97}. The simplest and paradigmatic model in this context is
the Kugel-Khomskii model introduced long ago for KCuF$_3$ \cite{Kug82},
a strongly correlated insulator with a single hole within degenerate
$e_g$ orbitals at Cu$^{2+}$ ions in the $d^9$ electronic configuration.
This model has two parameters which favour different types of symmetry
broken states:
($i$) Hund's exchange interaction, and
($ii$) the crystal field splitting of $e_g$ orbitals.
When both these parameters
are small the system is driven by its quantum nature --- either
long-range order disappears \cite{Fei97,Fei98} or, for certain
parameters, coexisting spin and orbital order might be stabilized by
order out of disorder mechanism \cite{Kha97}. We shall discuss below
to what extent the order is classical and show that spin-orbital
entanglement manifests itself in the regime of most frustrated
interactions.

First, in this topical review we shall elucidate certain situations with
the spin-orbital entanglement in the ground states (GSs). Such GSs
are very challenging as there are no good methods in the theory to
investigate them in a systematic way. It will become evident that quite
different GSs arise, characterized by overestimated energy
and incorrect correlation functions, when spin-orbital entanglement is
neglected.

Second, even when the GSs are not entangled, entanglement may be
experimentally observed and has important consequences at finite
temperature when the behaviour of the system is driven by low-energy
excited states with spin-orbital entanglement. The existence of such
states is a generic feature of any spin-orbital superexchange model and
therefore the relevant question is only whether such states are
accessible for thermal excitations. It will be shown that a rather exotic
behaviour of the $R$VO$_3$ perovskites cannot be understood without
including the spin-orbital entangled states. This point of view is
supported by several experimental observations:
($i$) the thermal evolution of the optical spectral weights \cite{Miy02},
($ii$) the phase diagram of the $R$VO$_3$ perovskites \cite{Miy06}, and
($iii$) the observed dimerization in the magnon spectra of YVO$_3$
\cite{Ulr03}.

An interesting situation arises also in doped Mott insulators, where
doped holes introduce charge degrees of freedom which perturb the
orbital order and frequently lead to phases with coexisting spin,
charge and orbital order \cite{Ole10}. When orbital degrees of freedom
are quenched, one finds that hole propagation occurs in the $t$-$J$
model via a quasiparticle state that emerges due to quantum
fluctuations in the spin background \cite{Mar91}. In $t$-$J$-like
superexchange models with orbital degrees of freedom, hole propagation
is either entirely suppressed by incoherent processes \cite{Zaa92}, or
occurs by a rather subtle mechanism: either by off-diagonal orbital
hopping in $e_g$ orbital systems \cite{vdB00}, or by next-nearest
neighbour effective hopping in $t_{2g}$ systems \cite{Dag08,Woh08}. When
both spin and orbital degrees of freedom may contribute, the situation
is less clear as scattering on spin-flip processes introduces additional
incoherence in hole propagation \cite{Bal01}. Surprisingly, it was
realized only recently that spin-orbital entanglement introduces
topological constraints for hole propagation in a Mott insulator with
coexisting AF and AO order \cite{Woh09}, and may thus have serious
measurable consequences in doped $R$VO$_3$ perovskites.

The paper is organized as follows. In section 2 we explain general
concepts of:
($i$) intrinsic frustration of orbital interactions,
($ii$) spin-orbital superexchange, and
($iii$) its consequences for the magnetic exchange constants and the
optical spectral weights.
Next we describe spin-orbital entanglement in the GSs of spin-orbital
models in section 3.
Such models are usually employed to explain the magnetic properties of
transition metal oxides \cite{Ole05} and are also used to derive the
optical spectral weights \cite{Kha04}. Therefore, even when the GSs
are disentangled, entangled states have severe consequences on
the experimentally observed properties of some Mott insulators at
finite temperature, and we describe in section 4 the properties of
the perovskite $R$VO$_3$ systems as an example of such a complex
behaviour driven by quantum entanglement. Spin-orbital entanglement may
also occur in GSs in particular parameter regimes, and we provide two
examples of this behaviour in \sref{sec:T=0}:
($i$) the Kugel-Khomskii model on a bilayer \cite{Brz11}, and
($ii$) the $d^1$ model for $t_{2g}$ electrons on a triangular lattice
\cite{Nor08,Cha11}. Entangled states may have also interesting
consequences for hole propagation in ordered states --- here we present
the coupling of a hole to joint spin-orbital excitations \cite{Woh09},
see section 6. The paper is concluded in section 7, where a general
discussion and main conclusions are presented.

\section{Orbital and spin-orbital superexchange}
\label{sec:sex}

\subsection{Intrinsic frustration of orbital interactions}
\label{sec:orbi}

Before presenting the spin-orbital entanglement, we first introduce the
characteristic features of orbital superexchange interactions as
obtained in case of spin polarized (FM) systems. These interactions are
fundamentally different from spin superexchange which has high SU(2)
symmetry and intrinsically frustrated. Frustration is one of the
simplest concepts in physics with far reaching consequences
\cite{Nor09,Bal10}. The main and unusual feature of orbital interactions
is their intrinsic frustration that follows from the directional nature of
superexchange terms which contribute along different bonds and compete
with one another \cite{Fei97}. This type of frustration does not follow
from geometrical frustration and is best understood by considering a
two-dimensional (2D) square lattice. In case of the Ising model
frustration on a square lattice can be achieved, for instance, by
changing signs of interactions along every second column and leaving the
other interactions unchanged. In this case all plaquettes of the 2D
lattice are frustrated as one of the interactions has the wrong sign but
the spins order --- the model is exactly solvable and has long-range
order following the dominating interaction below a finite transition
temperature \cite{Lon80}, being lower than the one of the 2D Ising model.
We emphasize that this frustrated model is exactly solvable because it is
still classical as the interactions concern only commuting $\{S_i^z\}$
spin components.

In contrast, the orbital interactions on a pseudocubic lattice are
{\it quantum\/}, both for $e_g$ and $t_{2g}$ orbitals, because they
involve at least two pseudospin components \cite{vdB04}. Such models
have different (typically cubic) symmetry from both the Z$_2$ symmetry
of Ising and SU(2) symmetry of Heisenberg model, and are in general not
exactly solvable on a 2D square lattice. We begin with the case of $e_g$
orbitals interacting within a 2D $ab$ plane of K$_2$CuF$_4$ compound;
models for three-dimensional (3D) perovskites,
for instance KCuF$_3$ with $d^9$ electronic configurations of Cu$^{2+}$
ions or LaMnO$_3$ with $d^4$ configurations of Mn$^{3+}$ ions discussed
below can be easily obtained as a straightforward generalization of the
2D model using the cubic symmetry of orbital interactions \cite{vdB99}.
Two $e_g$ orbital states,
\begin{equation}
|z\rangle \equiv \frac{1}{\sqrt{6}}\left(3z^2-r^2\right),\hskip 1cm
|x\rangle \equiv \frac{1}{\sqrt{2}}\left(x^2-y^2\right),
\label{eg}
\end{equation}
are the eigenstates of the $\tau^{(c)}_i=\frac12(n_{iz}-n_{ix})$
orbital operator for pseudospin $\tau=\frac12$, where $\{n_{iz},n_{ix}\}$
are hole number operators at site $i$.

The origin of intrinsic frustration in the $e_g$ orbital superexchange
is best realized by considering its form \cite{vdB99},
\begin{equation}
H_{eg} = J_{\rm orb} \sum_{\langle ij\rangle\parallel\gamma} \left(
  \tau^{(\gamma)}_i\tau^{(\gamma)}_j-{1\over 4}\right)\,,
\label{HJ0}
\end{equation}
where the bond is oriented along one of the cubic axes $\gamma=a,b,c$
\cite{vdB99}. Here the orbital interaction $J_{\rm orb}>0$ follows from
the energy of the high-spin charge excitation \cite{Ole05}.
The pseudospin operators take a different form depending on the bond
direction and are defined as follows,
\begin{equation}
\tau^{(a,b)}_i = {1\over 4}\left( -\sigma^z_i \pm \sqrt{3}\sigma^x_i
                           \right), \hskip 1cm
\tau^{(c)}_i   = {1\over 2}   \sigma^z_i,
\label{tabc}
\end{equation}
where $\sigma^{x(z)}_i$ are Pauli matrices and the sign $\pm$ in
$\tau^{(a,b)}_i$ is selected for a bond $\langle ij\rangle$ along $a$
($b$) axis. Thus for the $ab$ plane one has two linear combinations of
$\{\sigma^x_i,\sigma^z_i\}$ Pauli matrices, and these interactions
favour AOs on each bond, being the eigenstates of the $\sigma^x_i$
Pauli matrix as the interactions $\propto\sigma^x_i\sigma^x_j$ are
here the strongest ones. We emphasize that the
interactions in equation \eref{HJ0} are fundamentally different from
the SU(2)-symmetric spin interactions, as they:
($i$) obey only lower cubic symmetry,
($ii$) are Ising-like and select only one component of the pseudospin
interaction along each bond which favours pairs of orthogonal orbitals,
i.e., oriented along the bond ($z$-like) and the orthogonal to it lying
in the plane perpendicular to the bond ($x$-like). This manifests the
intrinsic frustration of orbital interactions in the $e_g$ orbital case
\cite{Fei97}. In fact, the interactions in equation \eref{HJ0} are
Ising-like and classical only in the one-dimensional (1D) model
\cite{Dag04}, but in general they are not. However, due to the gap
which opens in orbital excitations in the 2D model, the quantum
corrections generated by them are rather small \cite{vdB99}.

The essence of orbital frustration which characterizes the $e_g$
orbital superexchange \eref{HJ0} is captured by the 2D compass model
\cite{Nus04} which arises by increasing frustration from the 2D $e_g$
orbital model to the maximal frustration \cite{Cin10}. One considers then
the 2D model (or an exactly solvable model compass ladder \cite{Brz09})
that interpolates between the classical Ising one and
the compass one passing through the $e_g$ orbital model. The orbital
interactions $J_{\rm orb}$ are equal along both rows and columns but
select two orthogonal pseudospin components \cite{Nus04,Mil05}:
\begin{equation}
H_{2D}=J\sum_{\langle ij\rangle\parallel a} \tau^x_i\tau^x_j
      +J\sum_{\langle ij\rangle\parallel b} \tau^z_i\tau^z_j\,.
\label{Hcm}
\end{equation}
Usually one considers the AF case ($J>0$) but the FM model ($J<0$) is
equivalent and equally frustrated. Intersite interactions in the 2D
compass model are descibed by products $\tau^{\alpha}_i\tau^{\alpha}_j$
of pseudospin components with $\alpha=x,y,z$,
\begin{equation}
\tau^x_i = \frac{1}{2}\sigma^{x}_i , \hskip 1cm
\tau^y_i = \frac{1}{2}\sigma^{y}_i , \hskip 1cm
\tau^z_i = \frac{1}{2}\sigma^{z}_i ,
\label{t2g}
\end{equation}
rather than by pseudospin scalar products
${\vec \tau}_i\cdot{\vec \tau}_j$. As explained below, such scalar
products arise for the superexchange interactions with active $t_{2g}$
orbitals degrees of freedom which allow hopping processes for a pair of
them in each 2D plane in the cubic system. Instead in the compass model
\eref{Hcm} the $\tau^x_i\tau^x_j$ interactions for bonds
$\langle ij\rangle$ along the $a$ axis compete with the
$\tau^z_i\tau^z_j$ ones along the $b$ axis \cite{Mil05}. Also the 1D
compass model with alternating $\tau^z_i{2i}\tau^z_{2i+1}$ and
$\tau^x_{2i-1}\tau^x_{2i}$ interactions \cite{Brz07} is intrinsically
frustrated.

Recently the 2D compass model
was investigated by Monte Carlo simulations and the existence of
a phase transition at finite temperature was established \cite{Wen08}.
The ordered GS is degenerate and its different states correspond to
either eigenstates of $\{\tau^x_i\}$ pseudospin components ordered along
the rows, or eigenstates of pseudospin $\{\tau^z_i\}$ components ordered
along the columns \cite{Brz10}. Although this GS is destabilized by
infinitesimal pseudospin interactions, the structure of the lowest excited
states stays unchanged and corresponds to the flips of spin columns \cite{Tro10}.
It has been suggested that the model could serve as an effective model
for protected qubits and such states realized by Josephson arrays
\cite{Dou05} could play a role in quantum communication. Indeed, first
experimental successes in constructing special networks of Josephson
junctions that are designed following the compass model were reported
recently \cite{Gla09}.

Although frustration still increases by going from the 2D to 3D $e_g$
orbital model, there are indications from recent Monte Carlo simulations
that the GS is ordered \cite{Ryn10,Wen11}. Disorder occurs here by
doping which leads to the 3D orbital liquid state \cite{Fei05} that
plays a prominent role in the FM metallic manganites and provides an
explanation for the observed magnon dispersion \cite{Ole02}. The case of
the 3D compass model is more subtle. It was concluded from the high
temperature expansion that the ordered state is excluded here
\cite{Oit11}, but this result was challenged recently \cite{Ryn10} and
further studies are needed to establish whether this model could indeed
serve as an example of an orbital liquid phase.

The orbital models for $t_{2g}$ orbitals contain more terms and were
less studied up to now. The generic form contains scalar products of
two pseudospins $\tau=1/2$ along each direction in the cubic lattice,
defined as in equations \eref{t2g} \cite{Kha00},
\begin{equation}
H_{t2g} = J_{\rm orb} \sum_{\langle ij\rangle} \left(
  {\vec\tau}^{\,(\gamma)}_i\cdot{\vec\tau}^{\,(\gamma)}_j-{1\over 4}\right)\,,
\label{Ht2g}
\end{equation}
where the operators ${\vec\tau}^{\,(\gamma)}_i\equiv
\{\tau^x_i,\tau^y_i,\tau^z_i\}^{(\gamma)}$ depend on the bond direction,
i.e., the pseudospin components are defined here in different subspaces
of the Hilbert space, depending on the pair of active $t_{2g}$ orbitals.
This form follows from the fact that only two $t_{2g}$ orbitals allow
for electron hopping along a given cubic direction $\gamma$, while the
third orbital is inactive, see also below and section 4. For finite Hund's
exchange additional terms arise and the orbital state is disordered
\cite{Khali}. A stable orbital ordered state was found here in the 3D
orbital model for YTiO$_3$, when the spins decouple from $t_{2g}$
orbitals in the FM phase \cite{Kha02}.

\subsection{Spin-orbital superexchange models}
\label{sec:som}

In transition metal compounds with large on-site Coulomb interactions
charge fluctuations are suppressed and electrons partly localize. This
happens when intraorbital Coulomb interaction $U$ is large compared to
the effective $d-d$ hopping element $t$, where $t$ is either the
$(dd\sigma)$ or $(dd\pi)$ effective $d-d$ hopping element for $e_g$ and
$t_{2g}$ systems, respectively, that arises via hybridization with ligand
orbitals.
In the regime of $t\ll U$, correlated (Mott or charge-transfer) insulators
arise in undoped compounds, while doping leads to interesting phenomena
in strongly correlated electron systems with charge fluctuations only
between two neighbouring electronic configurations \cite{Ima98}. Intersite
charge excitations may be then treated within perturbation theory, while
the hopping processes which do not cost high local Coulomb energy $U$ and
occur in doped systems are treated in the restricted Hilbert space.
A well known example of
this description is the $t$-$J$ model, used widely to describe the
physical properties of high-$T_c$ superconductors, but derived one
decade before their discovery \cite{Cha77}.

Here we concentrate on Mott insulators with transition metal ions in
$d^n$ electronic configuration and active orbital degrees of
freedom, where the effective low-energy Hamiltonians contain
spin-orbital superexchange, described within spin-orbital models
\cite{Kug82,Ole09}. Such models are derived using the realistic
multiplet structure of the excited states of transition metal ions
which arise in $d_i^md_j^m\rightleftharpoons d_i^{(m+1)}d_j^{(m-1)}$
intersite charge excitations. As the multiplet structure depends on
the electron number $m$ \cite{Gri71}, with some examples given in
\cite{Ole05}, these models are specific to a given family of compounds.
As a representative example we consider here in more detail the case of
the $R$VO$_3$ perovskites (see section 4), with $S=1$ spin stabilized
by Hund's exchange and active $t_{2g}$ orbitals at V$^{3+}$ ions in
$d^2$ ($m=2$) electronic configuration. In a cubic perovskite all three
$t_{2g}$ orbitals are
degenerate and the kinetic energy electrons is given by:
\begin{equation}
\label{Ht}
H_{t}=- t \sum_{\gamma} \sum_{\langle
ij\rangle{\parallel}\gamma,\alpha(\gamma),\sigma}
\left(d^{\dagger}_{i\alpha\sigma}d^{}_{j\alpha\sigma}+
      d^{\dagger}_{j\alpha\sigma}d^{}_{i\alpha\sigma}\right),
\end{equation}
where $d^{\dagger}_{i\alpha\sigma}$ is electron creation operator
for an electron at site $i$ in orbital state $\alpha$ with spin
$\sigma=\uparrow,\downarrow$. The summation runs over three cubic
axes, $\gamma=a,b,c$, and involves the bonds $\langle ij\rangle\parallel
\gamma$ along them. The hopping $t$ results from transitions via an
intermediate O($2p_{\pi}$) orbital and
conserves the active $t_{2g}$ orbital flavor $\alpha(\gamma)$. Thus the
hopping $t$ is an effective $(dd\pi)$ element that originates from two
subsequent $d-p$ hopping processes along each V$-$O$-$V bond in the
$R$VO$_3$ perovskite structure. For $e_g$ systems the derivation is
similar but the effective hopping elements follow from the hybridization
with O($2p_\sigma$) orbitals. The effective hopping follows from the
charge-transfer model with $p-d$ hybridization $t_{pd}$ and
charge-transfer energy $\Delta$ \cite{Zaa93}, and one expects in the
present vanadate case $t=t_{pd}^2/\Delta\sim 0.2$ eV \cite{Kha01}. Only
two $t_{2g}$ orbitals, labelled by $\alpha(\gamma)$, are active
along each bond $\langle ij\rangle{\parallel}\gamma$ and
contribute to the kinetic energy \eref{Ht}, while the third
one lies in the plane perpendicular to the $\gamma$ axis and the
hopping via the intermediate oxygen $2p_{\pi}$ (or $2p_\sigma$) oxygen
orbital is forbidden by symmetry. This justifies a simplified notation
used below, with the orbitals defined by the axis direction which is
perpendicular to their plane:
\begin{equation}
|a\rangle\equiv |yz\rangle,  \hskip .7cm
|b\rangle\equiv |zx\rangle,  \hskip .7cm
|c\rangle\equiv |xy\rangle.
\label{t2go}
\end{equation}

In this case only $t_{2g}$
orbitals are partly filled by electrons, and it suffices to consider
local Coulomb interactions between $t_{2g}$ electrons at V$^{3+}$
ions described by the degenerate Hubbard Hamiltonian \cite{Ole83},
\begin{eqnarray}
\label{Hint}
H_{\rm int}&=&
   U\sum_{i\alpha}n_{i\alpha  \uparrow}n_{i\alpha\downarrow}
 +\left(U-\frac{5}{2}J_H\right)\sum_{i,\alpha<\beta}n_{i\alpha}n_{i\beta}
\nonumber \\
&+& J_H\sum_{i,\alpha<\beta}
\left( d^{\dagger}_{i\alpha\uparrow}d^{\dagger}_{i\alpha\downarrow}
      d^{       }_{i\beta\downarrow}d^{       }_{i\beta\uparrow}
     +d^{\dagger}_{i\beta\uparrow}d^{\dagger}_{i\beta\downarrow}
      d^{    }_{i\alpha\downarrow}d^{       }_{i\alpha\uparrow}\right)
\nonumber \\
&-&2J_H\sum_{i,\alpha<\beta}\vec{S}_{i\alpha}\cdot\vec{S}_{i\beta}\,.
\end{eqnarray}
Here $n_{i\alpha}=\sum_{\sigma}n_{i\alpha\sigma}$ is the electron
density operator in orbital $\alpha=a,b,c$ at site $i$, and spin operators
${\vec S}_{i\alpha}=\{S_{i\alpha}^x,S_{i\alpha}^y,S_{i\alpha}^z\}$ for
orbital $\alpha$ at site $i$ are
related to fermion operators in the standard way, i.e.,
\begin{equation}
\label{S+z}
S_{i\alpha}^+\equiv d^{\dagger}_{i\alpha\uparrow}d^{}_{i\alpha\downarrow}\,,
\hskip .7cm
S_{i\alpha}^z\equiv\frac12(n_{i\alpha\uparrow}-n_{i\alpha\downarrow})\,.
\end{equation}
The first term in \eref{Hint} describes the largest intraorbital Coulomb
interaction $U$ for a pair of electrons with antiparallel spins in
orbital $\alpha$. The second term stands for the interorbital Coulomb
(density) interaction, the third one is called frequently the
"pair-hopping" term, and the last one is Hund's exchange $J_H$. The
choice of coefficients in \eref{Hint} guarantees that the interactions
satisfy the rotational invariance in the orbital space \cite{Ole83}.
This Hamiltonian is exact when it describes only one representation of
the cubic symmetry group (here $t_{2g}$ orbitals which are partly
occupied in the cubic vanadates) --- then the on-site interactions are
given by two parameters:
($i$) the intraorbital Coulomb element $U$, and
($ii$) Hund's exchange element $J_H$.
These elements may be expressed by the Racah parameters $\{A,B,C\}$. For
$t_{2g}$ electrons considered in section 4 one finds \cite{Ole05,Gri71}:
\begin{eqnarray}
\label{u}
U&=&A+4B+3C\,, \\
\label{jh}
J_H&=&3B+C\,.
\end{eqnarray}
Hund's exchange (and interorbital Coulomb interaction) is in general
anisotropic and depends on the pair of interacting orbital states. For
instance, the corresponding $e_g$ Hund's exchange element is $J_H=4B+C$.
More details are given in \cite{Ole05}.

In the limit of large $U$ ($t\ll U$), the effective low-energy
spin-orbital superexchange interactions arise by considering all the
contributions which originate from possible virtual excitations
$d_i^md_j^m\rightleftharpoons d_i^{m+1}d_j^{m-1}$.
The general structure of spin-orbital superexchange \cite{Ole05},
\begin{equation}
{\cal H}_J = J \sum_{\langle ij \rangle } \left\{
{\hat {\cal J}}_{ij}^{(\gamma)}
\left( {\vec S}_i \cdot {\vec S}_j +S^2\right) +
{\hat {\cal K}}_{ij}^{(\gamma)} \right\},
\label{som}
\end{equation}
involves interactions between SU(2)-symmetric spin scalar products
${\vec S}_i\cdot{\vec S}_j$ on each bond $\langle ij\rangle$,
connecting two nearest-neighbor transition metal ions,
each one coupled to orbital operators $\{{\vec\tau}_i,{\vec\tau}_j\}$.
The orbital operators are given in \sref{sec:orbi} and
obey only much lower symmetry (at most cubic for a cubic lattice).
These operators contribute to the form of orbital operators
${\hat{\cal J}}_{ij}^{(\gamma)}$ and
${\hat{\cal K}}_{ij}^{(\gamma)}$ which depend on the model. They involve
the active orbitals on each bond $\langle ij \rangle$ along direction
$\gamma$.

In order to derive magnetic excitations for the systems with orbital
degeneracy one usually derives magnetic exchange constants for a bond
$\langle ij \rangle$ by averaging over the orbital operators in equation
\eref{som} using the GS $|\Phi_0\rangle$ with decoupled spin and orbital
operators,
\begin{equation}
J_{ij}=\langle\Phi_0|{\hat{\cal J}}_{ij}^{(\gamma)}|\Phi_0\rangle.
\label{Jij}
\end{equation}
This procedure assumes implicitly that spin and orbital operators can be
decoupled from each other in the MF approach and ignores the possibility
of entanglement and composite spin-orbital excitations introduced in
\cite{Fei97}. {\it Inter alia\/}, such excitations play a prominent role in
destabilizing the classical AF long-range order in the $d^9$ spin-orbital
model \cite{Fei98}.

The energy scale for the superexchange is given by
\begin{equation}
\label{J}
J=\frac{4t^2}{U}.
\end{equation}
where $t$ is the relevant effective hopping element and $U$ is the
intraorbital Coulomb element defined in \eref{Hint}. As several charge
excitations contribute to the superexchange \eref{som}, the balance
between competing terms depends on Hund's exchange, namely on
\begin{equation}
\label{eta}
\eta=\frac{J_H}{U}.
\end{equation}
This is the only parameter which decides about the strength of particular
interactions in the superexchange and finally also about the type of
magnetic correlations or symmetry breaking in the GS obtained at orbital
degeneracy and favoured by the model.

We would like to emphasize here that the same charge excitations which
decide about the spin-orbital superexchange \eref{som} contribute as
well to the optical conductivity. In this case they appear at distinct
energies of individual charge excitations that occur at a given ionic
filling, and depend on the multiplet structure of the excited states
arising from intersite charge transitions \cite{Kha04}. Each of these
excitations involves a different intermediate state in the multiplet
structure of at least one of the ions after the charge excitation, i.e.,
either in the $d^{m+1}$ or in the $d^{m-1}$ configuration or in both,
depending on the actual process and on the value of $m$ \cite{Kha04}.
This feature made it possible to relate the averages of these different
excitations to the spectral weights in the optical spectroscopy
\cite{Kha04}, and this principle serves now as a theoretical tool to
analyse and explain the observed
anisotropy and temperature dependence of the spectral weights
in the optical spectra \cite{Ole05}.

As the charge excitations correspond to particular expressions in the
spin-orbital space, it is important to rewrite ther superexchange
Hamiltonian \eref{som} by decomposing it into individual terms
on each bond $\langle ij\rangle$ that stem from
particular excited states labeled by $n$ \cite{Kha04},
\begin{equation}
\label{HJn}
{\cal H}_J = \sum_n\sum_{\langle ij\rangle\parallel\gamma}
             H_n^{(\gamma)}(ij).
\end{equation}
Here the superexchange constant \eref{J} was included in the operators
$H_n^{(\gamma)}(ij)$. As explained above, the spectral weight in the
optical spectroscopy is given in a correlated insulator by the same
virtual charge excitations that contribute also to the superexchange.
They define the individual kinetic energy terms $K_n^{(\gamma)}$ along
the axis $\gamma$, which can be determined from the superexchange
\eref{som} using the Hellman-Feynman theorem \cite{Bae86},
\begin{equation}
\label{hefa}
K_n^{(\gamma)}=-2\big\langle H_n^{(\gamma)}(ij)\big\rangle.
\end{equation}
For convenience, we define them here as positive quantities,
$K_n^{(\gamma)}>0$.

The spectral weights \eref{hefa} are defined by the bond correlation
functions and their changes with increasing temperature which decide
about the {\it temperature dependence\/} of the optical spectrum. To
describe experimental observations it is therefore important to analyse
the various multiplet excitations separately, as they depend on these
correlations in a different way, and will also contribute to a
{\it quite different temperature dependence\/}, as we show in this paper
on the example of LaVO$_3$ in section 4. Such an analysis is of course
possible in each case, as the respective
spin-orbital superexchange models are derived by considering all
different types of charge excitations individually \cite{Ole05}.

In some cases, however, spin dynamics separates from the orbital one in
the GS and MF factorization of spin and orbital operators is indeed
allowed. This happens when the orbital order is stabilized to a large
extent by strong interactions with the lattice which undergoes Jahn-Teller
distortions. A good example of this behaviour is LaMnO$_3$, where the
superexchange and the Jahn-Teller effect help each other \cite{Fei99}
in stabilizing the orbital order which occurs below the relatively high
transition temperature $T_{\rm OO}\simeq 780$ K. The $A$-AF spin order
is observed below a much lower N\'eel temperature $T_{\rm N}\simeq 140$ K
\cite{Kov10}. In this case a very satisfactory description of the
experimental results for the spectral weight distribution in the optical
spectroscopy is obtained using disentangled spin-orbital superexchange,
both for the high spin \cite{Ole05} and low spin \cite{Kov10} optical
excitations. Below we focus on some examples of a more complex behaviour
driven by the spin-orbital entanglement.

\section{Spin-orbital entanglement}
\label{sec:enta}

Before presenting the essence of spin-orbital entanglement, we would
like to remind the reader severe consequences of the MF approximation
used frequently to investigate spin and/or orbital models. In this
approach quantum fluctuations are neglected and only qualitative
conclusions concerning possible symmetry breaking can be drawn. In
spin-orbital systems the MF approach may be applied in different ways,
and here we focus on the decoupling of spin and orbital degrees of
freedom presented below for a single bond in \sref{sec:bond}. When only
spin-orbital coupling is treated in the MF approach, spin and orbital
operators are disentangled. We show in \sref{sec:su2} and \sref{sec:t2g}
that this approximation is unsatisfactory in many situations.

To detect spin-orbital entanglement in the GS we evaluate intersite spin,
orbital and joint spin-orbital bond correlations in several models,
defined as follows for a nearest-neighbour bond $\langle ij\rangle$ (we
keep the notation general here, for the 1D chain $j\equiv i+1$) \cite{Ole06}:
\begin{eqnarray}
\label{sij}
S_{ij}&\equiv&\frac{1}{2S^2}\,\langle{\vec S}_i\cdot {\vec S}_j\rangle, \\
\label{tij}
T_{ij}&\equiv&\langle{\vec\tau}_i\cdot {\vec\tau}_j\rangle, \\
\label{cij}
C_{ij}&\equiv&\frac{1}{2S^2}\,
\Big\{\Big\langle({\vec S}_i\cdot{\vec S}_j)
                ({\vec\tau}_i\cdot{\vec\tau}_j)\Big\rangle
     -\Big\langle{\vec S}_i\cdot{\vec S}_j\Big\rangle
      \Big\langle{\vec\tau}_i\cdot{\vec\tau}_j\Big\rangle\Big\}.
\end{eqnarray}
The above general expressions imply averaging over the exact GS found from
Lanczos diagonalization of a finite cluster and are valid for $S=\frac12$
and $S=1$ encountered in the models for $t_{2g}$ orbitals investigated
in \sref{sec:t2g}. While $S_{ij}$ and $T_{ij}$ correlations indicate
the tendency towards particular spin and orbital order, $C_{ij}$ quantifies
the spin-orbital entanglement --- if $C_{ij}<0$ spin and orbital operators
are entangled and the MF approximation, i.e., decoupling of spin and
pseudospin operators in \eref{som}, cannot be applied as it generates
uncontrollable errors.

\subsection{Exact versus mean field states for a bond}
\label{sec:bond}

The spin-orbital superexchange \eref{som} takes the simplest form when
Hund's exchange is absent ($J_H=0$) --- then the orbital operators
$\hat {\cal J}_{ij}^{(\gamma)}$ which couple to the scalar spin product
$({\vec S}_i \cdot {\vec S}_j)$ are particularly simple. In case
of $e_g$ orbitals they are just projection operators on the active
directional orbital \cite{Ole00}, while for $t_{2g}$ orbitals they give
a scalar product of $\tau=\frac12$ pseudospins which represent two active
orbital flavours along the bond direction \eref{Ht}. As an example, we
consider first a bond in the 1D SU(2)$\otimes$SU(2) model for $S=1$ spins
of V$^{3+}$ ions coupled to $\tau=\frac12$ pseudospins. Here we show that
quantum fluctuations in both spin and orbital system are crucial to
reproduce faithfully the energy spectrum of a bond with the spin-orbital
superexchange,
\begin{equation}
{\cal H}_{12} = \frac12 J\, \left( {\vec S}_1\cdot{\vec S}_2+1\right)
\left( {\vec\tau}_1\cdot{\vec\tau}_2+\frac14\right).
\label{bond}
\end{equation}
Such superexchange interactions are obtained in a perovskite $R$VO$_3$
vanadate along the $c$ axis in absence of Hund's exchange ($\eta=0$)
\cite{Ole07}. The above form is convenient for further discussion and we
neglected here a constant term (which would play a role for the optical
spectral weights \cite{Kha04}).

\begin{figure}[t!]
\begin{center}
    \includegraphics[width=7cm,clip=true]{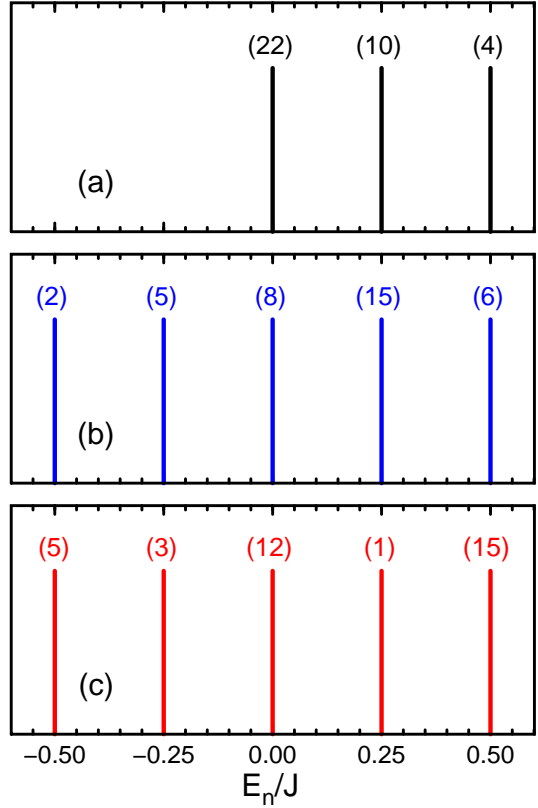}
\end{center}
\caption{
Eigenenergies and their degeneracy (numbers above vertical lines) for a
single spin-orbital superexchange bond in $R$VO$_3$ at $\eta=0$, described by:
(a) classical Ising-like interactions \eref{bondi},
(b) Ising-like interactions for $S=1$ spins but full quantum
interaction for $\tau=\frac12$ pseudospins \eref{bondz}, and
(c) the full quantum model \eref{bond}.
Spin-orbital entanglement is absent here, but orbital fluctuations
change the energy spectrum in a dramatic way and stabilize the
high-spin state.  }
\label{fig:dime}
\end{figure}

While the energy spectrum for a single bond can be easily solved
exactly using the SU(2)$\otimes$SU(2) symmetry and the classification of
quantum states by the total spin
$\vec{\cal S}_t\equiv \vec{S}_1+\vec{S}_2$ and orbital (pseudospin)
$\vec{\tau}_t\equiv\vec{\tau}_1+\vec{\tau}_2$ operators, in a solid or
in an infinite chain this is not the case and one has to employ some
approximation. Frequently the MF theory is used which has
severe limitations as the quantum fluctuations on the bonds are then
neglected. This can be seen by solving the problem of a single bond
\eref{bond}, either by replacing the scalar products of both spin and
orbital operators by their $z$th components (${\cal H}_{12}^{\rm Ising}$),
or by treating in the MF approximation only the (more classical) spin
scalar product for $S=1$ spins (${\cal H}_{12}^z$):
\begin{eqnarray}
\label{bondi}
{\cal H}_{12}^{\rm Ising}&=& \frac12 J\, \Big( S_1^zS_2^z+1\Big)
\left( \tau_1^z\tau_2^z+\frac14\right),  \\
\label{bondz}
{\cal H}_{12}^z&=& \frac12 J\, \Big( S_1^zS_2^z+1\Big)
\left( {\vec\tau}_1\cdot{\vec\tau}_2+\frac14\right).
\end{eqnarray}
Note that the Ising-like Hamiltonian \eref{bondi} is nonnegative by
construction, so the GS energy $E_0=0$ is obtained when at
least one of the following conditions is satisfied:
either $S_1^zS_2^z=-1$ or $\tau_1^z\tau_2^z=-\frac14$.
This property gives a rather high degeneracy $d=22$ of the GS, see
\fref{fig:dime}(a). At the same time the highest energy in the
spectrum with degeneracy $d=4$ is accurately reproduced when
$S_1^zS_2^z=+1$ and $\tau_1^z\tau_2^z=+\frac14$, i.e., for four
possible configurations where quantum fluctuations do not contribute
(2 configurations for spin and 2 for orbital operators). Note that although
the highest energy $E=\frac12 J$ is correct, the degeneracy of this state
at is not correctly reproduced.

The full bond Hamiltonian \eref{bond} has however also negative
eigenvalues and the energy spectrum which includes quantum fluctuations
starts at a much lower energy $E_1=-\frac12 J$, see \fref{fig:dime}(c),
obtained for a pseudospin singlet ${\tau}_t=0$ and the
high spin ${\cal S}_t=2$ state. The exact spectrum, shown in
\fref{fig:dime}(c), is obtained by considering all eigenstates with
${\cal S}=0,1,2$ and $\tau_o=0,1$. In fact, it is even sufficient here
to consider only the exact values of the ${\vec\tau}_1\cdot{\vec\tau}_2$
operator for the orbital singlet ($\tau_o=0$) and orbital triplet
($\tau_o=1$) states and to keep the Ising form of the spin interaction
\eref{bondi} to reproduce {\it all\/} exact eigenenergies for a single
bond, see \fref{fig:dime}(b).
However, only when the spin configurations and the eigenstates of total
spin ${\cal S}$ are used, the degeneracies of all the states in the
spectrum are incorrect, cf. figures \ref{fig:dime}(b) and
\ref{fig:dime}(c). Therefore, we note that the thermodynamic properties
determined in the MF theory are not free from systematic errors. We show
below that full spin and orbital dynamics on the bond plays a very
important role in systems where orbital correlations contribute by
quantum effects.

Yet, there is even a more serious problem which concerns the
spin-orbital superexchange \eref{som} --- the possible entanglement of
spin and orbital operators. Entanglement means here that wave functions
cannot be written as products of spin and orbital states, similar to
entangled spin singlet wave function \cite{Nie00,Ben06,Hor09}, where it
is not just one eigenstate of the $S^z_i$ spin operator at site $i$.
As the wave functions in the present example obey
the SU(2) symmetry in spin and orbital subspace, entanglement does not
occur for a single bond, where only individual spin singlet or orbital
singlet states are entangled.  However, spin-orbital entanglement is a
characteristic feature of any spin-orbital model in a larger system,
both on a finite cluster and in the thermodynamic limit. There the same
spin and orbital operators participate in interactions along several
bonds, and we show below that this is also the case for the interaction
in equation \eref{bond}. The essence of this type of entanglement is
explained in the following section.

\subsection{Entanglement in the SU(2)$\otimes$SU(2) spin-orbital model }
\label{sec:su2}

Before demonstrating spin-orbital entanglement in more realistic models
which apply to systems with active $t_{2g}$ orbital degrees of freedom,
we consider first the SU(2)$\otimes$SU(2) spin-orbital model, with
$S=\frac12$ spins and orbital interactions given by a scalar product
$\vec{\tau}_i\cdot\vec{\tau}_j$ for $\tau=\frac12$ pseudospin operators.
Two components of $\tau=\frac12$ pseudospin stand for two active $t_{2g}$
orbitals along the $\gamma$ axis in a cubic (perovskite) lattice. Indeed,
for a given cubic direction only two $t_{2g}$ orbitals are active
\cite{Zaa93} and the orbital superexchange has SU(2) symmetry when
Hund's exchange is neglected.

Here we shall analyse the energy in the 1D SU(2)$\times$SU(2) model,
\begin{equation}
\label{su2xsu2}
{\cal H}_J=J\sum_{i}
    \Big({\vec S}_i\cdot {\vec S}_{i+1}+x\Big)
    \Big({\vec \tau}_i\cdot {\vec \tau}_{i+1}+y\Big)\,,
\end{equation}
for spins $S=\frac12$ and pseudospins $\tau=\frac12$, using exact
diagonalization of finite chains with periodic boundary conditions (PBC).
The model \eref{su2xsu2} has two parameters $x$ and $y$. Its phase
diagram in the $(x,y)$ plane consists of five distinct phases which
result from the competition between effective AF and FM spin, as well as
AO and FO pseudospin exchange interactions on the bonds \cite{Aff00}.
First of all, the spin and pseudospin correlations are FM/FO, if
$x<-{1\over 4}$ and $y<-{1\over 4}$. Then the GS is characterized by the
maximal values of both total quantum numbers, ${\cal S}_t={\cal T}_t=N/2$,
where $N$ is the chain length; its degeneracy is $d=(N+1)^2$. Two other
phases have either FM spin or FO pseudospin order, accompanied by
alternating order in the other channel, i.e., AF order for the FO phase
and AO order for the FM phase.

A unique feature of FM state is that it is an eigenstate of the
Heisenberg exchange operator. The applies to the orbital interactions,
so the quantum fluctuations are entirely suppressed in the FM/FO phase.
They are also partly suppressed in the FM/AO and AF/FO phases. In all
these situations there is no possibility of joint spin-orbital
fluctuations as the wave function of the GS is exactly known and has no
quantum fluctuations in at least one of the two complementary subspaces.
Under these circumstances separation of spin and orbital operators becomes
exact and the GS is disentangled. This does not concern excited states
\cite{Bal99}, but in this section we are interested only in entanglement in
the GS. Note, however, that at the special point $x=y=-\frac14$ one
finds even three degenerate collective excitations: spin, orbital and
spin-orbital wave, and the latter excitation is robust and
does not decay into separate spin and orbital excitation \cite{Her11}.

\begin{figure}[t!]
\begin{center}
    \includegraphics[width=8cm,clip=true]{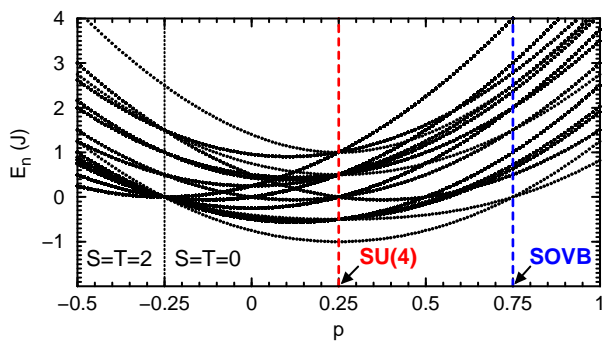}
\end{center}
\caption{
Energy spectrum for the spin-orbital chain \eref{su2xsu2} of $N=4$
sites with periodic boundaru conditions and $x=y=p$, for increasing
$p$. At $p=-\frac14$ the GS changes from high spin-orbital
state (${\cal S}={\cal T}=2$ with degeneracy $d=25$) to spin-orbital
singlet state (${\cal S}={\cal T}=0$ with $d=1$). The point with SU(4)
symmetry ($p=\frac14$ with $d=1$) and the special SOVB point
($p=\frac{3}{4}$ with $d=2$) are marked by vertical dashed lines.
This figure is reproduced from \cite{Pss07}. }
\label{fig:su2xsu2}
\end{figure}

First we consider the variation of the full energy spectrum of a finite
$N=4$ site chain when the FM/FO state changes into the regime dominated
by spin and orbital singlets. Therefere we study the SU(2)$\otimes$SU(2)
model \eref{su2xsu2} along the symmetric line in the parameter space,
$p=x=y$, see \fref{fig:su2xsu2}. Along this line interactions for
$S=\frac12$ spins and $\tau=\frac12$ pseudospins appear on equal footing.
At $p=\frac14$ one finds the high-symmetry SU(4) point and all three
correlation functions are equal: $S_{ij}$ \eref{sij} , $T_{ij}$ \eref{tij},
and
${4\over 3}\langle ({\vec S}_i\cdot{\vec S}_{i+1})
                   ({\vec T}_i\cdot{\vec T}_{i+1})\rangle$ \cite{Fri99}.
On the other hand, at $p=\frac{3}{4}$ the model \eref{su2xsu2} reads as
a product of spin triplet and orbital triplet projection operators at
each bond and its GS is exactly solvable --- one finds two equivalent
states with alternating spin and orbital singlets forming a spin-orbital
valence bond (SOVB) phase \cite{Pss07}. These states may be obtained by
a similar consideration as the Majumdar-Ghosh valence bond (VB) states
in the 1D frustrated $J_1-J_2$ spin model with nearest $J_1$ and
next-nearest neighbour $J_2$ interactions, at $J_2=\frac12 J_1$
\cite{Maj69}. The energy $E_0=0$ is given rigorously by alternating spin
and orbital singlets along the chain.

\begin{figure}[t!]
\begin{center}
    \includegraphics[width=8.4cm,clip=true]{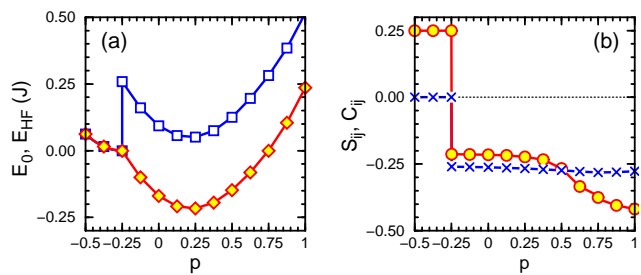}
\end{center}
\caption{ Signatures of entangled spin and orbital operators in the
SU(2)$\otimes$SU(2) model \eref{su2xsu2}, obtained by exact
diagonalization of the 1D chain of $N=8$ sites at $x=y$ with PBC:
(a) the exact GS energy $E_0$ (diamonds) and mean field
energy $E_{\rm MF}$ (squares) per bond;
(b) intersite spin correlations $S_{ij}$ \eref{sij} (circles), and
composite spin-orbital correlations $C_{ij}$ \eref{cij} (crosses);
here orbital correlations \eref{tij} are degenerate with spin ones,
$T_{ij}=S_{ij}$.
Entangled GS is found for $p>-0.25$ \cite{Pss07}. }
\label{fig:entasu2}
\end{figure}

One finds a quantum phase transition (QPT) between the high spin-orbital
FM/FO state (${\cal S}_t={\cal T}_t=2$) and the singlet entangled state
(${\cal S}_t={\cal T}_t=0$) in the 1D model (\ref{su2xsu2}), see
\fref{fig:su2xsu2}. The QPT that occurs at $p=-\frac14$ is first order,
as the energy levels shown in \fref{fig:su2xsu2} cross and all intersite
correlations change abruptly, see \fref{fig:entasu2}(b). At this point
the Hamiltonian \eref{su2xsu2} is a product of spin singlet and orbital
singlet projection operators at each bond, so the FM/FO state has the
lowest possible energy $E_0=0$. But it suffices that either spin or
orbital state is a triplet and thus the degeneracy of the GS
is here much higher. The role of quantum fluctuations in the regime of
$p>\frac14$ is easily recognized by considering further variation of
the GS energy $E_0$ with increasing $p$, see \fref{fig:su2xsu2}.
Classically the energy of the model \eref{su2xsu2} would be minimal at
$p=0$, but the quantum effects shift the energy minimum to the SU(4) point
$p=\frac14$; this state is nondegenerate. The energy $E_0=0$ is obtained
again at $p=\frac{3}{4}$, where the Hamiltonian \eref{su2xsu2} is a
product of spin triplet and orbital triplet projection operators at each
bond --- here one finds the SOVB phase explained above.

While the GS energy $E_0$ per bond of the 1D spin-orbital model
\eref{su2xsu2} is exactly reproduced by the MF energy normalized per one
bond $E_{\rm MF}$  in the entire regime of singlet states
(${\cal S}_t={\cal T}_t=0$) for $p<-\frac14$,
\begin{equation}
\label{emf}
E_{\rm MF}=
    \Big(\Big\langle{\vec S}_i\cdot{\vec S}_{i+1}\Big\rangle+p\Big)
    \Big(\Big\langle{\vec T}_i\cdot{\vec T}_{i+1}\Big\rangle+p\Big),
\end{equation}
one finds large corrections to the MF energy for $p>-\frac14$,
see \fref{fig:entasu2}(a). This confirms that this MF decoupling procedure
is not allowed as joint spin-orbital quantum fluctuations are then ignored
which leads to uncontrollable errors in bond correlations.

To characterize spin-orbital entanglement in the GS we evaluate
intersite spin, orbital and joint spin-orbital bond correlations,
defined in equations \eref{sij}-\eref{cij}, see \fref{fig:entasu2}(b).
Here we use $s=\frac12$ in equations \eref{sij} and \eref{cij}. Note
that spin and orbital correlations are intially weaker than the
classical ones, i.e., $S_{ij}=T_{ij}>-\frac14$, as they are disturbed by
the joint spin-orbital correlations $C_{ij}$ \eref{cij}. The latter are
negative in the entire regime of $p>-\frac14$ and provide the dominating
energy gain in the GS, including the SU(4) symmetric
$p=\frac14$ and the exactly solvable SOVB $p=\frac{3}{4}$ points.
It is of importance that spin-orbital entanglement is related to local
properties of spins and orbitals on a bond. Therefore the entanglement
phase diagram of a finite system is in agrement with the magnetic and
orbital phase diagram of the infinite SU(2)$\otimes$SU(2) model \cite{Chen}.

Another and a more precise quantity to quantify spin-orbital entanglement
is von Neumann entropy defined as follows \cite{Che07},
\begin{equation}
\label{vNS}
{\cal S}_{S\tau}\equiv - \Tr_S\{\rho_S\,\log_2\rho_S\}\,,
\end{equation}
where $\rho_S\equiv \Tr_\tau\{|\Psi\rangle\langle\Psi|\}$ is the reduced
density matrix of the spin part in the state $|\Psi\rangle$ by
integrating out all the orbital degrees of freedom by
$\Tr_\tau\{\cdots\}$. This measure captures well the correlation between
the two types of degrees of freedom, and when spin and orbital operators
factorize one finds ${\cal S}_{S\tau}=0$. It has been shown that von
Neumann entropy makes a jump at the QPTs between the disentangled and
entangled states \cite{Che07} and may therefore be used to investigate
the phase diagram of the SU(2)$\otimes$SU(2) model. We suggest that QPTs
in more complex systems of strongly correlated electrons
may be investigated by calculating von Neumann entropy in the future.

\subsection{Entanglement in $t_{2g}$ spin-orbital models }
\label{sec:t2g}

Entanglement plays an important role in realistic spin-orbital models for
$t_{2g}$ orbital degrees of freedom which may be considered as being more
quantum than the respective models for $e_g$ electrons. This is a
consequence of two orbital flavours being active along each cubic axis,
while in $e_g$ systems there is {\it de facto\/} just one directional
orbital which is active along a given cubic axis while the other orbital
is inactive, so these latter models are more classical.

\begin{figure}[t!]
\begin{center}
    \includegraphics[width=7.5cm,clip=true]{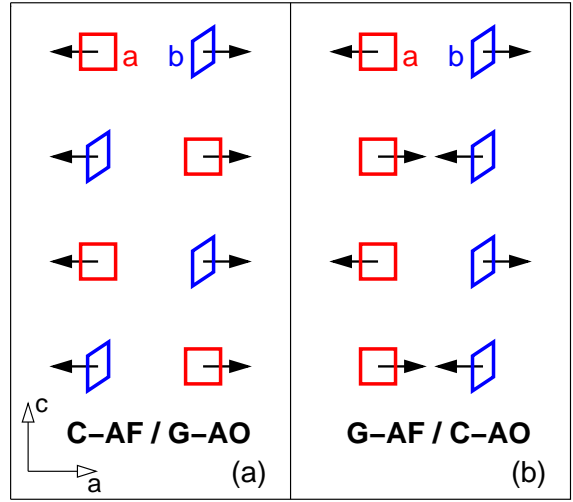}
\end{center}
\caption{ Schematic view of the Goodenough-Kanamori rules on the example
of two AF phases observed in the $R$VO$_3$ perovskites,
with complementary magnetic (arrows) and orbital (squares) order of
active $t_{2g}$ orbitals $\{a,b\}$ in the $ac$ plane:
(a) $C$-AF spin order accompanied by $G$-AF orbital order;
(b) $G$-AF spin order accompanied by $C$-AF orbital order.
Both spins and orbitals alternate along the $b$ axis (not shown).
This figure is reproduced from \cite{Ole09}.
}
\label{fig:good}
\end{figure}

The best known example of transition metal oxides with the physical
properties controlled by spin-orbital entanglement are the $R$VO$_3$
perovskites. Two magnetic phases compete with each other at low
temperature, and one finds the $C$-AF phase accompanied by $G$-AO order
in compounds with a large ionic radius $r_R$ of $R$ ions, i.e., for
Pr,$\cdots$,La, while for $R$ ions with smaller ionic radius the $G$-AF
phase accompanied by $C$-AO order is more stable at $T=0$ and the $C$-AF
phase occurs only in a window of finite temperature. In these phases
both the magnetic moments and the occupied orbitals alternate in the
$ab$ planes along both cubic axes, but the order along the $c$ axis is
different, as shown in \fref{fig:good}. Note that along the $c$ axis
the Goodeneough-Kanamori rules are obeyed. The situation looks different
for the $ab$ planes, where the magnetic moments and occupied orbitals
alternate, see also section 6, but one should keep in mind that inactive
$c$ orbitals are singly occupied at every site and rather strong AF
superexchange arises along both the $a$ and $b$ axis due to their
excitations. Thus, one may classify this case as FO order of $c$ orbitals
accompanied by AF order of $S=1$ spins, and once again the
Goodeneough-Kanamori rules are followed.

Although the above discussion shows that the GS of the $R$VO$_3$
compounds does not include spin-orbital entanglement, it is of
interest to investigate the spin orbital models for the perovskite
systems with active $t_{2g}$ orbitals:
($i$) the titanate model ($d^1$ model) valid for the $R$TiO$_3$
perovskites \cite{Kha00,Khali}, and
($ii$) the vanadate model ($d^2$ model) which describes the $R$VO$_3$
perovskites \cite{Kha01,Ole07}.
Spin-orbital entanglement arises along the $c$ axis, where both active
$t_{2g}$ orbitals $\{a,b\}$ contribute and may lead to entangled states.
To avoid additional complications due to partly occupied orbitals, one
may assume that the $c$ orbitals are empty at every site in the $d^1$
model, while they are occupied in the $d^2$ model --- in both cases they
cannot lead to any entangled states.

The spin-orbital models for the cubic perovskites with active $t_{2g}$
orbital degrees of freedom at either Ti$^{3+}$or V$^{3+}$ ions are of the
general form given in equation \eref{som}. The orbital operators
${\hat{\cal J}}_{ij}^{(\gamma)}$ and
${\hat{\cal K}}_{ij}^{(\gamma)}$ are rather complex and depend on the
multiplet structure of the Ti$^{2+}$ and V$^{2+}$ excited states,
respectively. They include the terms which break the SU(2) symmetry of
the orbital interactions present at $\eta=0$, reducing it to the cubic
symmetry. Their explicit form may be found in the original publications.
Here we give only the simplified SU(2)$\otimes$SU(2) form of the
interactions for the bonds along the $c$ axis,
\begin{equation}
\label{somt2g}
{\cal H}_J=\frac12 J\sum_{i}
    \Big({\vec S}_i\cdot {\vec S}_{i+1}+S^2\Big)
    \Big({\vec\tau}_i\cdot {\vec\tau}_{i+1}+\frac14\Big)\,,
\end{equation}
where spin-orbital entanglement is expected. For $S=\frac12$ and $S=1$
the above general form reproduces the respective limits obtained for
the $d^1$ and $d^2$ models at $\eta=0$; otherwise it is approximate.
We assume below that
\begin{equation}
\label{nabc}
 n_{ia}+n_{ib}=1
\end{equation}
for $t_{2g}$ orbitals \eref{t2go} in case of the $d^1$ model, and $n_{ic}=0$.

In the $R$VO$_3$ perovskites, the crystal-field splitting breaks the
cubic symmetry in distorted VO$_6$ octahedra, as suggested by the
electronic structure calculations \cite{And07} and derived using the
point charge model \cite{Hor08}. One finds again the same filling of
$\{a,b\}$ orbitals as given in equation \eref{nabc}, but $n_{ic}=1$.
This defines the $t_{2g}$ orbital degrees of freedom in both cases as
$\{a,b\}\equiv\{yz,zx\}$ orbitals along every cubic direction, and the
superexchange \eref{som2} contains the orbital operators
${\vec\tau}_i=\{\tau_i^+,\tau_i^-,\tau_i^z\}$ (and their components).

\begin{figure}[t!]
\begin{center}
    \includegraphics[width=8.2cm,clip=true]{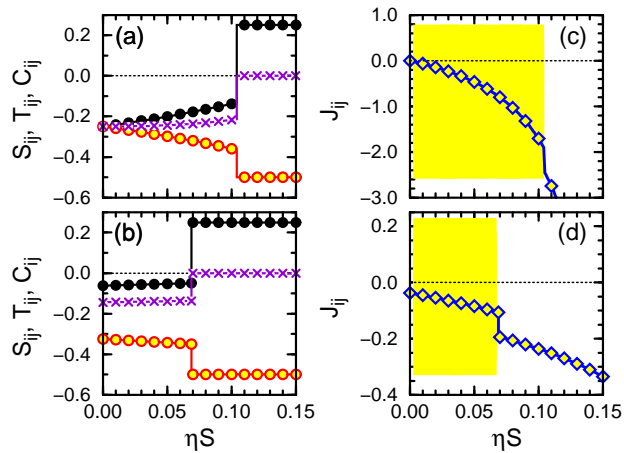}
\end{center}
\caption{ Evolution of: (a-b) intersite spin $S_{ij}$ \eref{sij} (filled
circles), orbital $T_{ij}$ \eref{tij} (empty circles), and spin-orbital
$C_{ij}$ \eref{cij} ($\times$) correlations, and (c-d) exchange
constants $J_{ij}$ (\ref{Jij}) along the $c$ axis, obtained by
exact diagonalizaton of spin-orbital model on a 1D chain of $N=4$
sites with periodic boundary conditions, for increasing Hund's
exchange $\eta$. Panels (a,c) for $S=1/2$; panels (b,d) for $S=1$.
In the shaded areas of (c) and (d) the negative spin correlations
$S_{ij}<0$ do not follow the sign of the exchange constant
$J_{ij}<0$, and the classical Goodenough-Kanamori rules are
violated.
This figure is reproduced from \cite{Ole09}.
 }
\label{fig:enta}
\end{figure}

A method of choice to demonstrate spin-orbital entanglement is here
again exact diagonalization of finite chains, performed for both the
$d^1$ and $d^2$ model \cite{Ole06}. In the $d^1$ model the Hamiltonian
at $\eta=0$ reduces to the SU(4) model, and indeed all three bond
correlation functions are equal for $N=4$ sites \cite{Ole06},
$S_{ij}=T_{ij}=C_{ij}=-0.25$ as shown in \fref{fig:enta}(a). For larger
systems these correlations are also equal but somewhat weaker and one
finds $S_{ij}=T_{ij}=C_{ij}=-0.21502$ in the thermodynamic limit
\cite{Fri99}. By a closer inspection one obtains that the GS wave
function for the four-site cluster is close to a total spin-orbital
singlet, involving a linear combination of
(spin singlet/orbital triplet) and
(spin triplet/orbital singlet)
states for each bond $\langle ij\rangle$.
This result manifestly contradicts the celebrated Goodenough-Kanamori
rules \cite{Goode}, as both spin and orbital correlations have the same
sign. When $\eta$ increases, the charge fluctuations which contribute
to the superexchange concern different states in the multiplet structure
breaks the SU(4) symmetry --- one finds that the bond correlations are
then different and $T_{ij}<C_{ij}<S_{ij}<0$ in the regime of spin
singlet (${\cal S}_t=0$) GS. Here the Goodenough-Kanamori rule which
suggests complementary spin/orbital correlations is still violated.

The vanadate $d^2$ model (for $S=1$ spins) \cite{Ole07} behaves also in
a similar way in a range of small values of $\eta$, with all three
$S_{ij}$, $T_{ij}$ and $C_{ij}$ correlations being negative, see
\fref{fig:enta}(b). Most importantly,
the composite spin-orbital correlations are here finite ($C_{ij}<0$)
which implies that spin and orbital variables are {\it entangled\/}, and
the MF factorization of the GS into spin and orbital part
fails. In this regime the spin and orbital correlations are both
negative and contradict the Goodenough-Kanamori rules \cite{Goode} of
their complementary behaviour. Only for sufficiently large $\eta$ do
the spins reorient at the QPT to the FM GS, and decouple from the
orbitals. In this regime, corresponding to the experimentally observed
$C$-AF phase of LaVO$_3$ (and other cubic vanadates), spin-orbital
entanglement ceases to exist in the GS. However, as we will see below, it
has still remarkable consequences in experiments at finite temperature,
where entangled spin-orbital excited states contribute and decide about
the thermodynamic properties.

A crucial observation concerning the applicability of the
Goodenough-Kanamori rules to the quantum models of $t_{2g}$ electrons
in one dimension can be made by comparing spin exchange constants
$J_{ij}$ calculated from  the exact GS $|\Phi\rangle$,
\begin{equation}
J_{ij}=\langle\Phi|{\hat{\cal J}}_{ij}^{(\gamma)}|\Phi\rangle\,,
\label{Jijex}
\end{equation}
with intersite spin correlations $S_{ij}$ \eref{sij} obtained also
exactly. One finds that exchange interaction
which is formally FM ($J_{ij}<0$) in the orbital-disordered phase at low
values of $\eta$ [see \fref{fig:enta}(c) and \fref{fig:enta}(d)] is in
fact accompanied by AF spin correlations ($S_{ij}<0$), so $J_{ij}S_{ij}>0$
and the GS energy calculated in the MF theory is {\it de facto enhanced\/}
by this term \cite{Ole06}, contrary to what happens in reality.

In contrast, similar analysis (not shown) performed for the $d^9$
spin-orbital model \eref{som} derived for Cu$^{2+}$ ions with $e_g$
orbital degrees of freedom in KCuF$_3$ \cite{Ole00}, gave correctly
$J_{ij}S_{ij}<0$. Hence, in spite of enhanced quantum fluctuations of
the spin-orbital nature \cite{Fei97}, one finds here that spin
correlations follow the sign of the exchange constant \cite{Ole06}.
This remarkable difference between
$t_{2g}$ and $e_g$ systems originates from composite spin-orbital
fluctuations, which are responsible for the {\it `dynamical'\/} nature
of exchange constants in the former case. They exhibit large
fluctuations around the average value, measured by the variance,
\begin{equation}
\label{delJ}
\delta J\equiv \Big\{
\langle\Phi|({\hat{\cal J}}_{ij}^{(\gamma)})^2|\Phi\rangle
-J_{ij}^2\Big\}^{1/2}\,.
\end{equation}
Here again the average is calculated exactly from the Lanczos
diagonalization of a finite chain of length $N=4$ sites.

As an illustrative example, we give here the values found in the
$d^1$ and $d^2$ model at $\eta=0$ \cite{Ole06}.
While the average spin exchange constant is small in both cases
($J_{ij}\simeq 0$ for $d^1$, $J_{ij}\simeq -0.04$ for $d^2$),
$\hat{\cal J}_{ij}^{(\gamma)}$ fluctuates widely over both positive
and negative values. In the $d^1$ model the fluctuations between
($S=0$/$T=1$) and ($S=1$/$T=0$) wave functions on the bond are so large
that $\delta J=1$ ! They survive even quite far from the high-symmetry
SU(4) point (at $\eta\simeq 0.1$), and stabilize spin-orbital singlet
phase in a broad regime of $\eta$. Also in the $d^2$ model the orbital
bond correlations change dynamically from singlet to triplet,
resulting in $\delta J>|J_{ij}|$, with
$\delta J=\frac{1}{4}\{1-(2T_{ij}+\frac{1}{2})^2\}^{1/2}\simeq 0.25$,
while these fluctuations are small for $d^9$ model involving
$e_g$ orbitals, see also \sref{sec:d9}.

We emphasize that spin and orbital correlations on the bonds, as well
as composite spin-orbital correlations which occur in spin-orbital
entangled states for realistic parameters with finite Hund's exchange,
determine the magnetic and optical properties of titanates and vanadates.
These correlations change with increasing temperature as then also
excited states contribute and decide about their thermal evolution.
Therefore, the correct theoretical description of experimental results is
challenging and requires adequate treatment of excited states which
captures their essential features, including their possible
entanglement. This makes simple approaches based on the MF decoupling
of spin and orbital operators not reliable and requires either exact
diagonalization of finite clusters or advanced numerical methods such
as multiscale entanglement renormalization ansatz (MERA) \cite{Vid07}.

In the next section we show that composite spin-orbital fluctuations
play a crucial role in the $R$VO$_3$ perovskites. They are responsible
for the temperature dependence of the optical spectral weights in
LaVO$_3$ \cite{Kha04}, contribute to the remarkable phase diagram of
these systems \cite{Hor08} and trigger spin-orbital dimerization in the
$C$-AF phase of YVO$_3$ in the intermediate temperature regime
\cite{Hor03}. Remarkably, all these properties including the observed
dimerization in the magnetic excitations may be seen as signatures of
spin-orbital entanglement in the excited states which becomes relevant
at finite temperature.

\section{Entangled states in the RVO$_3$ perovskites}
\label{sec:rvo3}

\subsection{Optical spectral weights for LaVO$_3$}
\label{sec:optic}

The coupling between spin and orbital operators in the spin-orbital
superexchange may be so strong in some cases that it leads to a phase
transition modifies the magnetic order and excitations at finite
temperature --- an excellent example of this behaviour are the $R$VO$_3$
perovskites, as explained below. Although the $C$-AF phase, observed in
the entire family of $R$VO$_3$ compounds \cite{Miy06}, where
$R$=Lu,$\cdots$,La stands for a rare earth atom, satisfies to some extent
the Goodenough--Kanamori rules \cite{Goode}, with FM order along the $c$
axis where the active $a$ and $b$ orbitals \eref{t2go} alternate --- the
$G$-AO order of $\{a,b\}$ orbitals is very weak here
and the orbital fluctuations play a very important role \cite{Kha01}.
This situation is opposite to the frozen and classical AO order in
LaMnO$_3$, which can explain both the observed magnetic exchange
constants \cite{Ole05} and the distribution of the optical spectral
weights \cite{Kov10}. In LaVO$_3$ the FM exchange interaction is enhanced
far beyond the usual mechanism following from the splitting between the
high-spin and low-spin states due to finite Hund's exchange $J_H$.
Evidence of orbital fluctuations in the $R$VO$_3$ perovskites was also
found in pressure experiments, which show a distinct competition between
the $C$-AF and $G$-AF spin order, accompanied by the complementary $G$-AO
and $C$-AO order of $\{a,b\}$ orbitals \cite{Miy06}.

The spin and orbital order along the $c$ axis are not entangled in the
GS of the $R$VO$_3$ perovskites (due to either FM or FO order), but
entangled states contribute at
finite temperature. As the first manifestation of spin-orbital
entanglement in the cubic vanadates at finite temperature we discuss
briefly the evaluation of the low-energy optical spectral weight from the
spin-orbital superexchange for LaVO$_3$, following equation \eref{hefa}.
The superexchange operator ${\cal H}_J$ \eref{HJn} is here considered for
a bond $\langle ij\rangle\parallel\gamma$, and arises as a superposition
of individual $d_i^2d_j^2\rightleftharpoons d_i^3d_j^1$ charge
excitations to different spin states in the upper Hubbard subbands
labeled by $n$ \cite{Kha04}.
One finds the superexchange terms $H^{(c)}_{n,ij}$ for a bond
${\langle ij\rangle}$ along the $c$ axis,
\begin{eqnarray}
\label{H1c} H_{n,ij}^{(c)}&=&-\frac{1}{3}Jr_1 \left(2\!+\!\vec
S_i\!\cdot\!\vec S_j\right)
\left(\frac{1}{4}-\vec \tau_i\cdot\vec \tau_j\right),                  \\
\label{H2c} H_{n,ij}^{(c)}&=&-\frac{1}{12}J\left(1-\vec S_i\!\cdot\!\vec
S_j\right) \left(\frac{7}{4}-\!\tau_i^z\tau_j^z\!-\!\tau_i^x\tau_j^x\!
+\!5\tau_i^y \tau_j^y\right),                                           \\
\label{H3c} H_{n,ij}^{(c)}&=&-\frac{1}{4}Jr_3 \left(1-\!\vec
S_i\!\cdot\!\vec S_j\right)
\left(\frac{1}{4}+\tau_i^z\tau_j^z\!+\tau_i^x\tau_j^x -\tau_i^y
\tau_j^y\!\right),
\end{eqnarray}
and $H^{(ab)}_{n,ij}$ for a bond in the $ab$ plane,
\begin{eqnarray}
\label{H1a} H_{n,ij}^{(ab)}&=&-\frac{1}{6}Jr_1\!\left(2+\vec
S_i\!\cdot\!\vec S_j\right)
\left(\frac{1}{4}-\tau_i^z\tau_j^z\right),                 \\
\label{H2a} H_{n,ij}^{(ab)}&=&-\frac{1}{16}J\left(1-\vec
S_i\!\cdot\!\vec S_j\right) \left(\frac{19}{6}\mp
\tau_i^z\mp\tau_j^z-\frac{2}{3}\tau_i^z\tau_j^z\right),           \\
\label{H3a} H_{n,ij}^{(ab)}&=&-\frac{1}{16}Jr_3\left(1-\vec
S_i\!\cdot\!\vec S_j\right) \left(\frac{5}{2}\mp\tau_i^z
\mp\tau_j^z+2\tau_i^z\tau_j^z\right).
\end{eqnarray}
When the spectral weight is evaluated following equation \eref{hefa},
it is reasonable to try first the MF approximation and to separate spin
and orbital correlations from each other. The spectral weights require
then the knowledge of spin correlations $S_{ij}$ \eref{sij}: along the
$c$ axis in \eref{H1c}-\eref{H3c}, and within the $ab$ planes in
\eref{H1a}-\eref{H3a}, as well as the corresponding intersite orbital
correlations $\langle\vec{\tau}_i\cdot\vec{\tau}_j\rangle$ and
$\langle{\tau}_i^\alpha{\tau}_j^\alpha\rangle$, with $\alpha=x,y,z$.
From the form of the above superexchange contributions one sees that
high-spin excitations $H^{(\gamma)}_{n,ij}$ support the FM coupling
while the low spin ones, $H^{(\gamma)}_{2,ij}$ and
$H^{(\gamma)}_{3,ij}$, contribute with AF couplings. The high-spin
spectral weight \eref{hefa} in the MF approximation is given by
\begin{equation}
\label{wc21}
w_{c1}^{\rm MF}=\frac{2}{3}Jr_1
\Big\langle\vec S_i\!\cdot\!\vec S_j+2\Big\rangle
\left\langle\frac{1}{4}-\vec\tau_i\cdot\vec\tau_j\right\rangle\,.
\end{equation}

\begin{figure}[t!]
\begin{center}
\includegraphics[width=7.5cm]{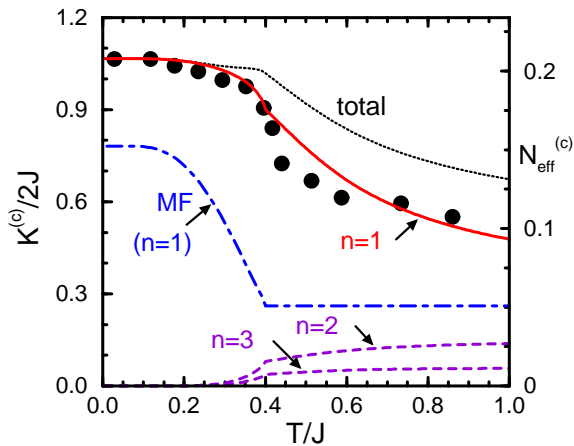}
\end{center}
\caption{Optical spectral weights $K_n^{(c)}$ \eref{hefa} for the
high-spin optical subband ($n=1$; solid line), low-spin optical subbands
($n=2,3$; dashed lines) and total $K^{(c)}$ (dotted line) obtained from
the excitations to different Hubbard subbands along the $c$ axis
\eref{H1c}-\eref{H3c}. Filled circles show the effective carrier number
$N_{\rm eff}^{(c)}$ (in the energy range $\omega<3$ eV which corresponds
to the high-spin excitations) for LaVO$_3$, presented in Fig. 5 of
\cite{Miy02}. Dashed-dotted line shows the spectral weight
$K_1^{(c)}$ obtained from the MF decoupling \eref{wc21}.
Parameters: $\eta=0.12$, $V_c=0.9J$, $V_{ab}=0.2J$. }
\label{fig:optic}
\end{figure}

The low-energy optical spectral weight for the polarization along the
$c$ axis $K_{\rm 1,exp}^{(c)}$ decreases by a factor close to two when
the temperature increases from $T\simeq 0$ to $T=300$ K \cite{Miy02} ---
this change is much larger than the one observed in LaMnO$_3$
\cite{Kov10}. However, the theory based on the MF decoupling of the spin
and orbital degrees of freedom gives only a much smaller reduction of
the weight close to 33\% when a frozen orbital order (similar to
LaMnO$_3$) with $\langle\vec\tau_i\cdot\vec\tau_j\rangle=-\frac14$ is
assumed and has to fail in explaining the experimental data \cite{Ole05}.
In spite of weakening spin and orbital intersite correlations when the
temperature increases in the experimental range $0<T<300$ K, this
variation is clearly not sufficient to describe the experimental data.
Instead, when both spin and orbital correlations are reduced and vanish
above the common transition temperature $T_{\rm N1}=T_{\rm OO}$, the MF
theory predicts that the spectral weight decrease fast and do not change
above $T_{\rm N1}$, see \fref{fig:optic}, contrary to experiment.

On the other hand, when a cluster method is used to determine the optical
spectral weight from the high-spin superexchange term (\ref{H1c}) by
including orbital as well as joint spin-and-orbital fluctuations along
the $c$ axis, the temperature dependence of the spectral weight resulting
from the theory persists above $T_{\rm N1}$ and follows the experimental
data \cite{Kha04}, see \fref{fig:optic}. In this approach a cluster
of $N=4$ sites is solved exactly with the MFs originating from the spin
and orbital order below $T_{\rm N1}$ and $T_{\rm OO}$, respectively, and
a free cluster is solved at high temperature when the long-range order
vanishes. This reflects the realistic temperature dependence with spin
and orbital correlations being finite in this latter regime, in contrast
to the one-site MF approach.

The satisfactory description of the experimental data shown in
\fref{fig:optic} may be considered as a remarkable success of the theory
based on the spin-orbital superexchange model derived for the $R$VO$_3$
perovskites. It proves that spin-orbital entangled states contribute in a
crucial way in the entire regime of finite temperature. In addition, the
theoretical calculation predicts that the high energy spectral weight
($n=2,3$) is low for the polarization along the $c$ axis. The spectral
weight in the $ab$ planes behaves in the opposite way --- it is small
at low energy, and large at high energy (but not as large as the
low-energy one for the $c$ axis). This weight distribution and its
anisotropy between the $c$ and $ab$ directions reflects the nature of
spin correlations on the bonds, which are FM and AF in these two
directions. We expect that future experiments will confirm these theory
predictions for the $ab$ polarization.

\subsection{Phase diagram of the $R$VO$_3$ perovskites}
\label{sec:phd}

The phase diagram of the $R$MnO$_3$ perovskites \cite{Goo06} indicates
that spin and orbital energy scales separate which makes it possible
to describe the experimental data for the magnetic exchange constants
and the optical spectral weights using the disentangled spin-orbital
superexchange \cite{Ole05,Kov10}. In contrast, in the $R$VO$_3$
perovskites the phase diagram suggests the proximity of spin and
orbital energy scales \cite{Miy06}. Experimental studies have shown that
the $C$-AF order is common to the entire family of the $R$VO$_3$
vanadates, and in general the magnetic transition occurs soon below the
orbital transition when the temperature is lowered further, i.e.,
$T_{\rm N1}<T_{\rm OO}$. LaVO$_3$ is an exception here and these transitions
occur almost simultaneously, with $T_{\rm N1}\simeq T_{\rm OO}$ \cite{Miy06}.

When the ionic radius $r_R$ decreases, the N\'eel temperature $T_{\rm N1}$
also decreases, while the orbital transition temperature $T_{\rm OO}$
increases, passes through a maximum close to YVO$_3$, and next decreases
when LuVO$_3$ is approached \cite{Miy06}. One finds that the $C$-AF
order develops in LaVO$_3$ below $T_{N1}\simeq 143$ K, and is almost
immediately followed by a weak structural transition stabilizing the weak
$G$-AO order at $T_{\rm OO}\simeq 141$ K \cite{Miy06}. This provides a
constraint on the theoretical model and is an experimental challenge to
the theory which was addressed using the spin-orbital superexchange model
extended by the orbital-lattice coupling \cite{Hor08}.

In order to unravel the physical mechanism responsible for the surprising
decrease of $T_{\rm OO}$ from YVO$_3$ to LuVO$_3$ one has to analyse in
more detail the evolution of GdFeO$_3$ distortions for decreasing ionic
radius $r_R$ \cite{Hor08}. Such distortions are common for the
perovskites \cite{Pav05}, and one expects that they should increase
when the ionic radius $r_R$ decreases, as observed in the $R$MnO$_3$
perovskites \cite{Goo06}. In the $R$VO$_3$ family the distortions
are described by two subsequent rotations of VO$_6$ octahedra:
($i$) by an angle $\vartheta$ around the $b$ axis, and
($ii$) by an angle $\varphi$ around the $c$ axis.
Increasing angle $\vartheta$ causes a decrease of
V--O--V bond angle along the $c$ direction, being $\pi-2\vartheta$,
and leads to an orthorhombic lattice distortion $u=(b-a)/a$, where
$a$ and $b$ are the lattice parameters of the $Pbnm$ structure of
$R$VO$_3$. By the analysis of the structural data for the $R$VO$_3$
perovskites \cite{Ree06,Sag06} one finds the following empirical
relation between the ionic radius $r_R$ and the angle $\vartheta$:
\begin{equation}
\label{r}
r_R=r_0-\alpha\sin^2\vartheta\,,
\end{equation}
where $r_0=1.5$ \AA{} and $\alpha=0.95$ \AA{} are the empirical
parameters. This allows one to use the angle $\vartheta$ to
parametrize the dependence of the microscopic parameters of the
Hamiltonian and to investigate the transition temperatures
$T_{\rm OO}$ and $T_{\rm N1}$ as functions of the ionic radius $r_R$.

The spin-orbital model introduced in \cite{Hor08} to describe the phase
diagram of $R$VO$_3$ was thus extended with respect to its original
form \cite{Kha01} and reads:
\begin{eqnarray}
\label{som2}
{\cal H}\!&=&J\sum_{\langle ij\rangle\parallel\gamma}
\left\{\Big({\cal J}_{ij}^{(\gamma)}{\vec S}_i\!\cdot\!{\vec S}_j+S^2\Big)
+ {{\cal K}}_{ij}^{(\gamma)}\right\}
+E_z(\vartheta)\!\sum_i\!e^{i{\vec R}_i{\vec Q}}\tau_i^z
\nonumber \\
&-& V_{c}(\vartheta)\sum_{\langle ij\rangle\parallel c}
\tau_i^z\tau_j^z
+V_{ab}(\vartheta)\sum_{\langle ij\rangle\parallel ab}\tau_i^z\tau_j^z
\nonumber \\
&-& gu\sum_i\tau_i^x+\frac12 N K\{u-u_0(\vartheta)\}^2\,,
\end{eqnarray}
where $\gamma=a,b,c$ labels the cubic axes, and the orbital operators
${\cal J}_{ij}^{(\gamma)}$ and ${\cal K}_{ij}^{(\gamma)}$ are
given in \cite{Ole07}. The superexchange
$\propto J$ is supplemented by the crystal field term $\propto E_z$, the
orbital interaction terms $\propto V_c$ and $\propto V_{ab}$ induced by
lattice distortions, and the orbital-lattice term $\propto g$ which is
counteracted by the lattice elastic energy $\propto K$. All these terms
are necessary in a realistic model \cite{Hor08} to reproduce the complex
dependence of the orbital and magnetic transition temperature on the
ionic size in the $R$VO$_3$ perovskites.

The crystal-field splitting $\propto E_z$ between $a$ and $b$ orbital
energies in equation \eref{som2} is given by the pseudospin $\tau_i^z$
operators,
\begin{equation}
\label{tauz}
\tau_i^z=\frac{1}{2}(n_{ia}-n_{ib})\,,
\end{equation}
which refer to two active orbital flavors $\{a,b\}$ in $R$VO$_3$.
It is characterized by the vector ${\vec Q}=(\pi,\pi,0)$ in reciprocal
space and favours the $C$-AO order. Thus, this splitting competes with
the (weak) $G$-AO order supporting the observed $C$-AF phase at
temperature $T<T_{\rm N1}$, effectively weakening this type of
magnetic order.

As for instance in LaMnO$_3$, the orbital order in the $R$VO$_3$
perovskites arises due to joint orbital interactions which are a
superposition of the superexchange and interactions induced by lattice
distortions. These latter terms are twofold:
($i$) intersite orbital interactions $\{V_{ab},V_c\}$ (which
originate from the coupling to the lattice), and
($ii$) orbital-lattice coupling $\propto g$ which induces orbital
polarization $\langle\tau_i^x\rangle\neq 0$ for finite lattice
distortion $u$. The orbital interactions induced by the distortions of
the VO$_6$ octahedra and by the GdFeO$_3$ distortions of the lattice,
$V_{ab}>0$ and $V_c>0$, also favour the $C$-AO order (like the crystal
field term with $E_z>0$). Note that $V_c>0$ counteracts the orbital
interactions included in the superexchange via ${\hat K}_{ij}^{(c)}$
operators.

The last two terms in equation \eref{som2} are particularly important
for ions with small ionic radii $r_R$. They describe the linear coupling
$\propto g>0$ between active $\{a,b\}$ orbitals and the orthorhombic
lattice distortion $u$. The elastic energy which counteracts lattice
distortion $u$ is given by the force constant $K$, and $N$ is the number
of $V^{3+}$ ions. The coupling
\begin{equation}
\label{gef}
g_{\rm eff}\equiv gu
\end{equation}
may be seen as a
transverse field in the pseudospin space which competes with the
Jahn-Teller terms $\{V_{ab},V_c\}$. While the eigenstates of $\tau_i^x$
operator, $\frac{1}{\sqrt{2}}(|a\rangle\pm|b\rangle)$, cannot be realized
due to the competition with all the other terms, increasing lattice
distortion $u$ (increasing angle $\vartheta$) gradually modifies the
orbital order and intersite orbital correlations towards this type order.

Except for the superexchange parameter $J$, all the parameters in the
extended spin-orbital model \eref{som2} depend on the tilting angle
$\vartheta$.
In case of $V_{c}$ one may argue that its dependence on the angle
$\vartheta$ is weak, and a constant $V_{c}(\vartheta)\equiv 0.26J$ was
chosen in \cite{Hor08} in order
to satisfy the experimental constraint that the magnetic and orbital
order appear almost simultaneously in LaVO$_3$ \cite{Miy06}. The
experimental value $T_{\rm N1}^{\rm exp}=143$ K for LaVO$_3$
\cite{Miy06} was fairly well reproduced in the present model taking
$J=200$ K, the value which is also consistent with the magnon energy
scale \cite{Ulr03}. The functional dependence of the remaining two
parameters $\{E_z(\vartheta),V_{ab}(\vartheta)\}$ on the tilting angle
$\vartheta$ was derived from the point charge model \cite{Hor08} using
the structural data for the $R$VO$_3$ series \cite{Ree06,Sag06} ---
one finds:
\begin{eqnarray}
\label{Ez}
E_z(\vartheta)&=&J\,v_z\,\sin^3\vartheta\cos\vartheta\,, \\
\label{vab}
V_{ab}(\vartheta)&=&J\,v_{ab}\,\sin^3\vartheta\cos\vartheta\,.
\end{eqnarray}
Finally, the effective coupling to the lattice distortion
$g_{\rm eff}(\vartheta)$ \eref{gef} has to increase faster with the
ncreasing angle $\vartheta$ as otherwise the nonmonotonous dependence of
$T_{\rm OO}$ on $\vartheta$ (or on the ionic radius $r_R$) cannot be
reproduced by the model, and the following dependence was shown
\cite{Hor08} to give a satisfactory description of the phase diagram of
the $R$VO$_3$ perovskites:
\begin{equation}
\label{geff}
g_{\rm eff}(\vartheta)=J\,v_{g}\,\sin^5\vartheta\cos\vartheta\,.
\end{equation}
Altogether, magnetic and orbital correlations described by the
spin-orbital model \eref{som2}, and the magnetic $T_{\rm N1}$ and orbital
$T_{\rm OO}$ transition temperatures, depend on three parameters:
$\{v_z,v_{ab},v_g\}$.

\begin{figure}[t!]
\begin{center}
\includegraphics[width=8cm]{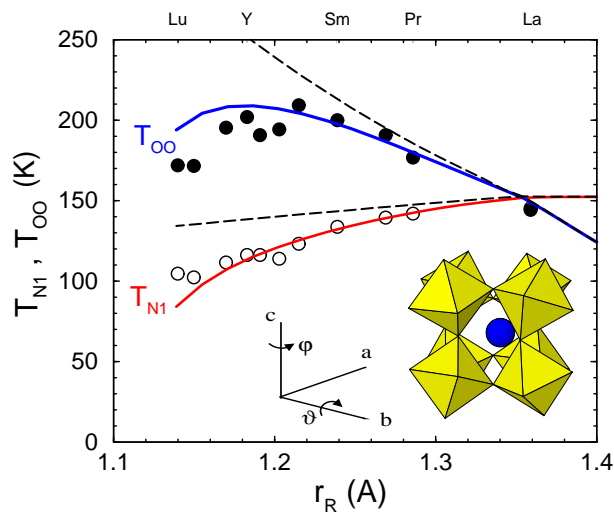}
\end{center}
\caption{The orbital transition temperature $T_{\rm OO}$ and N\'eel
transition temperature $T_{\rm N1}$ for the onset $C$-AF order as
functions of varying ionic size $r_R$ in the $R$VO$_3$ perovskites,
as obtained in experiment (full and empty circles) and from the
theory (solid lines). Dashed lines show $T_{\rm OO}$ and $T_{\rm N1}$
obtained under neglect of orbital-lattice coupling (at $g=0$).
The inset shows the GdFeO$_3$-type distortion, with the rotation angles
$\vartheta$ and $\varphi$ corresponding to the data of YVO$_3$
\cite{Ree06}. This figure is reproduced from \cite{Hor08}. }
\label{fig:phd}
\end{figure}

Due to the spin-orbital entanglement which is activated by finite
temperature in the $R$VO$_3$ perovskites, it is crucial to design
the MF approach in such a way that the spin-orbital coupling is
described {\it beyond\/} the factorization of spin and orbital
operators. As usually, the correct MF treatment of the orbital and
magnetic phase transitions in the $R$VO$_3$ perovskites requires the
coupling between the on-site orbital,
$\langle\tau^z\rangle_G\equiv\frac12|\langle\tau^z_i-\tau^z_j\rangle|$,
and spin order parameters in the $C$-AF phase,
$\langle S_i^z\rangle_C$, as well as including a composite spin-orbital
$\langle S_i^z\tau_i^z\rangle$ order parameter, similar to that
introduced before for the $R$MnO$_3$ perovskites \cite{Fei99}. However,
the on-site MF approach including the above coupling \cite{Sil03} does
not suffice for the $R$VO$_3$ compounds as the orbital singlet correlations
$\langle{\vec\tau}_i\cdot{\vec\tau}_j\rangle$ on the bonds along the $c$
axis play here a crucial role in stabilizing the $C$-AF phase
\cite{Kha01} and the orbital fluctuations are also important
\cite{God07}. Therefore, the minimal physically acceptable approach to
the present problem is a self-consistent calculation od spin and orbital
correlations for an embedded bond $\langle ij\rangle$ along the $c$ axis,
coupled by the MF terms to its neighbours along all
three cubic axes \cite{Hor08}. This procedure, with properly selected
model parameters, led to the successful description of the experimental
phase diagram \cite{Miy06}, see \fref{fig:phd}.
The evolution of the orbital correlations with varying temperature and
with decreasing $r_R$ plays a prominent role in the success of the
theoretical description of the phase diagram of the $R$VO$_3$ perovskites.
One finds that indeed the orbital order occurs first at a higher
temperature and is followed by the magnetic transition in the $R$VO$_3$
perovskites with a smaller ionic radius $r_R$, to the left of LaVO$_3$.

The non-monotonous dependence of the orbital transition temperature
$T_{\rm OO}$ on the ionic radius $r_R$ may be understood as follows.
$T_{\rm OO}$ increases first with decreasing ionic radius $r_R$ as the
Jahn-Teller term in the $ab$ planes, $V_{ab}(\vartheta)$, increases and
induces the orbital correlations which stabilize the $G$-AO order. The
coupling $g_{\rm eff}(\vartheta)$ to the lattice \eref{geff} is then
rather weak, with $\langle\tau^x_i\rangle\simeq 0.03$ in LaVO$_3$, but
$g_{\rm eff}$ increases faster than the interaction $V_{ab}(\vartheta)$
\eref{vab}. Finally, the former term dominates and the $G$-AO order
parameter is almost equal to the competing with it "transverse" moments,
$\langle\tau^x_i\rangle\simeq\langle\tau^z_i\rangle_G$.
Therefore, the $G$-AO order gets weaker and the transition temperature
$T_{\rm OO}$ is reduced. Note that in the entire parameter range the
orbital order parameter
$\langle\tau^x_i\rangle\simeq\langle\tau^z_i\rangle_G$ is substantially
reduced from the classical value $\langle\tau^z\rangle_{G,\rm max}=\frac12$
by singlet orbital fluctuations, being for instance
$\langle\tau^z_i\rangle_G\simeq 0.32$ and 0.36 for LaVO$_3$ and LuVO$_3$.

It is quite remarkable that the magnetic exchange constants
$\{J_{ab},J_c\}$ are modified solely by the changes in the orbital
correlations described above. The superexchange constant $J$ does not
change and the reductions of $T_{\rm N1}$ with decreasing $r_R$ follows
only from the evolution of the orbital state \cite{Hor08}. One finds that
also the width of the magnon band, given by$W_{C-{\rm AF}}=4(J_{ab}+|J_c|)$
at $T=0$, is reduced by a factor close to 1.8
from LaVO$_3$ to YVO$_3$. This also agrees qualitatively with surprisingly
low magnon energies observed in the $C$-AF phase of YVO$_3$ \cite{Ulr03}.

Summarizing, the microscopic model \eref{som2} is remarkably successful
in describing gradual changes of the orbital and magnetic correlations
under increasing Jahn-Teller interactions and the coupling to the
lattice which both suppress the orbital fluctuations along the $c$ axis,
responsible for rather strong FM spin-orbital superexchange \cite{Kha01}.
It describes well the systematic experimental trends for both orbital
and magnetic transitions in the $R$VO$_3$ perovskites \cite{Hor08},
and is able to reproduce the observed non-monotonic variation of the
orbital transition temperature $T_{\rm OO}$ for decreasing ionic radius
$r_R$. Another consequence of the spin-orbital entanglement in the
perovskite vanadates, the spin-orbital dimerization along the $c$ axis
in YVO$_3$, is shortly discussed in the next subsection.

\subsection{Peierls dimerization in YVO$_3$}
\label{sec:dim}

The third and final example of the spin-orbital entanglement at finite
temperature in the family of vanadate perovskites is the existence of a
remarkable first order magnetic transition at $T_{\rm N2}=77$ K from the
$G$-AF to the $C$-AF spin order with rather exotic magnetic properties,
found in YVO$_3$ \cite{Ren00}. This magnetic transition is surprising
and rather unusual as the staggered moments are approximately parallel
to the $c$ axis in the $G$-AF phase, and reorient above $T_{\rm N2}$ to
the $ab$ planes in the $C$-AF phase, with some small alternating $G$-AF
component along the $c$ axis. First, while the orientations of spins in
$C$-AF and $G$-AF phase are consistent with the expected anisotropy due
to spin-orbit coupling \cite{Hor03}, the observed magnetization reversal
with the weak FM component remains puzzling and found no
explanation in the theory so far. Second, it was also established by
neutron scattering experiments \cite{Ulr03} that the energy scale of
magnetic excitations is considerably reduced for the $C$--AF phase
(by a factor close to two) as compared with the magnon dispersion
measured in the $G$-AF phase. The magnetic order parameter in the $C$-AF
phase of LaVO$_3$ is also strongly reduced to $\simeq 1.3\mu_B$, which
cannot be explained by rather small quantum fluctuations in the $C$-AF
phase \cite{Rac02}. Finally, the $C$-AF phase of YVO$_3$ is dimerized.
Until now, only this last feature found a satisfactory explanation
in the theory \cite{Sir08}, see below.

The observed dimerization in the magnon spectra in YVO$_3$ motivated
the search for its mechanism within the spin-orbital superexchange
model. Dimerization of AF spin chains coupled to phonons is well
known and occurs in several systems \cite{Joh00}. In the
spin-orbital model for the $R$VO$_3$ perovskites a similar instability
might also occur without the coupling to the lattice when Hund's
exchange is sufficiently small. In particular, the GS at $\eta=0$ may
be approximated by the dimerized chain with strong FM
bonds alternating with the AF ones, if such chains are coupled by AF
interactions along the $a$ and $b$ axes \cite{She02} (the 1D chain
would give then the entangled disordered GS). For finite and
realistic $\eta\simeq 0.13$ the chain is FM (due to the weak coupling
to the neighbouring chains in $ab$ planes) \cite{Ole07} and at first
instance any dimerization appears surprising.

Before addressing the question of magnon excitations in the  $C$-AF phase
of YVO$_3$ stable at intermediate temperature, let us consider first the
1D spin-orbital superexchange model along the $c$ axis, as in YVO$_3$.
The Hamiltonian is given by
\cite{Sir08},
\begin{equation}
\label{SO1}
H_{S\tau}=J\sum_j\left(\vec{S}_j\!\cdot\!\vec{S}_{j+1}+1\right)
\left(\vec{\tau}_j\!\cdot\!\vec{\tau}_{j+1}+\frac{1}{4}-\gamma_H\right),
\end{equation}
where $\vec{S}_j$ represent $S=1$ spins and $\vec{\tau}_j$ are
$\tau=\frac12$ orbital pseudospins, respectively, and $\gamma_H$ is a
constant proportional to Hund's exchange which stabilizes FM spin
correlations. This expression is
somewhat simplified with respect to the full spin-orbital model for
YVO$_3$ \cite{Ole07}, but reflects its essential features and guarantees
that the GS is FM when $\gamma_H\simeq 0.1$. The FM GS state is
disentangled --- it is allowed to use the MF decoupling \cite{Ole06},
and to decompose the above Hamiltonian \eref{SO1} into the spin
($H_{S}$) and orbital ($H_{\tau}$) part, $H_{S\tau}\simeq H_S+H_\tau$.
This disentangled chain may be now studied either by density-matrix
renormalization group applied to transfer matrices (TMRG) \cite{Sir02},
or by an analytical approach, the so-called modified spin-wave theory
of Takahashi \cite{Tak86}.

It is easy to understand why the spin-orbital dimerization occurs at
finite temperature. The crucial concept is the interrelation between spin
and orbital correlations in the 1D spin-orbital chain: spin correlations
determine the exchange interactions in the orbital channel $H_{\tau}$,
and the orbital ones are responsible for the spin exchange in $H_{S}$.
In the GS the spin state is rigid, and spin correlations on the bonds
$\langle j,j+1\rangle$ along the $c$ axis are saturated, i.e.,
$\langle {\vec S}_j\cdot{\vec S}_{j+1}\rangle=1$, and do not allow for
any alternation in the orbital interactions which are determined by them.
But when temperature increases the thermal fluctuations soften the FM
order and the spin-orbital chain may dimerise \cite{Sir08}. Important
here is the rather dense spectrum of low energy excited states in the
spin-orbital chain \cite{Sir03}, which are entangled and all contribute
to the thermal averages used to calculate spin and orbital correlations.
We emphasize that the dimerization in the spin-orbital chain
may be seen as a signature of {\it entanglement in excited states\/} in
the $C$-AF phase which contribute at finite temperature. The exchange
constants alternate along the $c$ direction between a stronger
(${\cal J}_{c}^{(1)}$) and weaker (${\cal J}_{c}^{(1)}$) exchange with
$\delta_S>0$,
\begin{equation}
\label{jcdim}
{\cal J}_{c}^{(1)}\equiv{\cal J}_c(1+\delta_S)\,,\hskip .5cm
{\cal J}_{c}^{(2)}\equiv{\cal J}_c(1-\delta_S)\,.
\end{equation}
Similar expressions are also found for the orbital exchange interactions
which favour AO order and have alternating strength with $\delta_\tau>0$,
\begin{equation}
\label{jcorb}
{\cal J}_{\tau}^{(1)}\equiv{\cal J}_\tau(1+\delta_\tau)\,,\hskip .5cm
{\cal J}_{\tau}^{(2)}\equiv{\cal J}_\tau(1-\delta_\tau)\,.
\end{equation}

\begin{figure}[t!]
\begin{center}
\includegraphics[width=7.5cm]{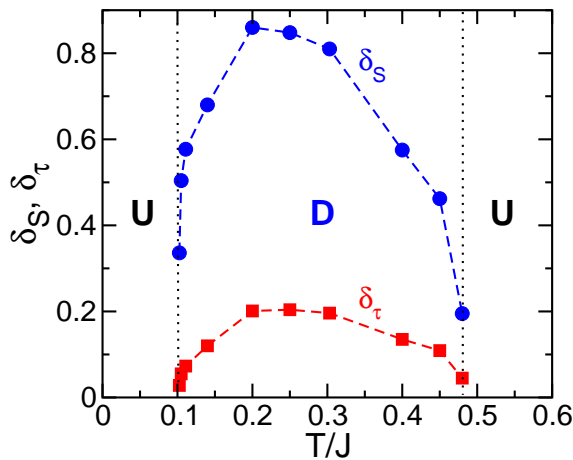}
\end{center}
\caption{(Color online)
Phase diagram with two undimerized (U) phases and a dimerized (D) phase
in between, and dimerization  parameters $\delta_S$ and $\delta_\tau$,
obtained for the spin-orbital model \eref{SO1} in the MF decoupling.
Dotted vertical lines indicate the onset of a dimerised phase
under increasing/decreasing temperature. For more details see
\cite{Sir08}, where these data were presented and discussed.
}
\label{fig:dim}
\end{figure}

While the spin and orbital operators are disentangled in the FM ground
state, one may consider a coupled FM spin chain to an orbital chain with
interactions which favour weak AO order accompanied by orbital
fluctuations, as realized in the $C$-AF phase.
The spin (orbital) exchange interaction along the chain is then
determined by the bond orbital (spin) correlations.
They are defined as follows:
\begin{eqnarray}
\label{bothjc}
{\cal J}_c&\equiv&\frac12\Big\langle\vec{\tau}_j\!\cdot\!\vec{\tau}_{j+1}
+\vec{\tau}_j\!\cdot\!\vec{\tau}_{j-1}\Big\rangle +\frac14-\gamma_H\,,\\
{\cal J}_\tau&\equiv&\frac12\Big\langle\vec{S}_j\!\cdot\!\vec{S}_{j+1}
+\vec{S}_j\!\cdot\!\vec{S}_{j-1}\Big\rangle +1\,,
\end{eqnarray}
and have to be determined self-consistently, together with spin and
orbital correlations along the chain. For the parameters selected in
equation \eref{SO1}, one finds that FM spin correlations with
${\cal J}_c<0$ are accompanied by ${\cal J}_\tau>0$ that favours AO
order along the chain. Such complementary spin and orbital correlations
are indeed found in the entire temperature regime, in agreement with
the Goodenough-Kanamori rules. But in the intermediate temperature
range, when the spins start to fluctuate and their correlations are not
rigid anymore, the dimerization sets in, see \fref{fig:dim}. Finite
temperature is here essential as dimerized spin correlations support
then the dimerized orbital correlations. Hence, the dimerization occurs
here simultaneously in both channels and has a dome-shaped form, with a
maximum at $T\simeq 0.2J$ \cite{Sir08}. The dimerization in the FM chain
is much stronger than the one in the AO chain, but they have to coexist
in the present self-consistent treatment of this phenomenon. When
temperature is high enough, in the present case for $T>0.49J$, the
dimerization vanishes again as the spins and the orbitals are
disordered by thermal fluctuations. The phase transition at finite
temperature between a uniform and a dimerised phase, shown in
\fref{fig:dim}, is a consequence of the MF decoupling.

\begin{figure}[t!]
\begin{center}
\includegraphics[width=8cm]{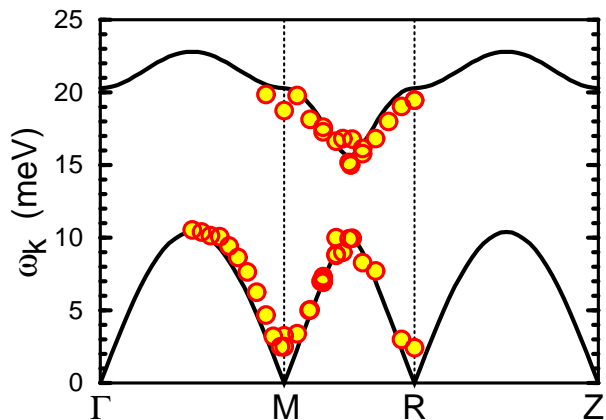}
\end{center}
\caption{(Color online)
Spin-wave dispersions $\omega_{\bf k}$ (solid lines) as obtained in the
LSW theory along the representative directions in the Brillouin zone
for the dimerized $C$-AF phase of YVO$_3$ with experimental
exchange constants \cite{Ulr03}:
$J_{ab}=2.6$ meV, $J_c=3.1(1\pm\delta_S)$ meV and $\delta_S=0.35$.
Parameters: $J=30$ meV, $\eta=0.13$, $\delta_S=0.35$, $K_z=0$.
The experimental points \cite{Ulr03} measured by neutron scattering at
$T=85$ K are reproduced by circles. The high symmetry points are:
$\Gamma=(0,0,0)$, $M=(\pi,\pi,0)$, $R=(\pi,\pi,\pi)$, $Z=(0,0,\pi)$.
This figure is reproduced from \cite{Ole07}.
}
\label{fig:sw}
\end{figure}

The microscopic model \eref{SO1}, which explains the anisotropy in the
exchange constants \eref{jcdim} as following from the joint dimerization
that occurs in the spin-orbital chain with FM spin order at finite
temperature \cite{Sir08}, helps to understand the magnon dispersion
found in YVO$_3$ by the neutron scattering \cite{Ulr03}. The observed
spin-wave dispersion may be explained by the following effective spin
Hamiltonian for the $C$-AF phase, derived assuming again that the spin
and orbital operators may be disentangled which is strictly valid only
at $T=0$:
\begin{eqnarray}
\label{hcafd}
{\cal H}_{S}&=&J_{c}\sum_{\langle i,i+1\rangle\parallel c}
   \left\{1+(-1)^i\delta_S\right\}{\vec S}_{i}\cdot{\vec S}_{i+1}
\nonumber\\
   &+&J_{ab}\sum_{\langle ij\rangle\parallel ab}{\vec S}_i\cdot{\vec S}_j
   +K_z\sum_i\left(S_i^z\right)^2\,.
\end{eqnarray}
Following the linear spin-wave (LSW) theory the magnon dispersion is given by
\begin{equation}
\label{spinw}
\omega_{\pm}({\bf k})= 2\sqrt{\left(2J_{ab}+|J_c|+\frac12 K_z
\pm J_c\;\eta_{\bf k}^{1/2}\right)^2-\big(2J_{ab}\gamma_{\bf k}\big)^2},
\end{equation}
with
\begin{eqnarray}
\label{gamma}
\gamma_{\bf k}&=&\frac12\left(\cos k_x+\cos k_y\right)\;,  \\
\label{etak}
\eta_{\bf k}&=&\cos^2k_z+\delta_S^2\sin^2k_z\,.
\end{eqnarray}
The single-ion anisotropy term $\propto K_z$ is responsible for the gap
which opens in spin excitations. Two modes measured by the neutron
scattering \cite{Ulr03} are well reproduced by $\omega_{\pm}({\bf k})$
given by equation \eref{spinw} when the experimental exchange constants
are inserted: $J_{ab}=2.6$ meV, $J_c=-3.1$ meV, $\delta_S=0.35$, see
\fref{fig:sw} (further improvement including a finite gap at the $\Gamma$
point are obtained taking finite $K_z>0$ \cite{Ulr03}. This shows that
while the essential features seen in the experiment are well reproduced
already by the present simplified spin exchange model ${\cal H}_S$
\eref{hcafd}, the spin interactions are more complex in reality
\cite{Ulr03}.

Summarizing, spin-orbital entanglement in the excited states is also
responsible for the exotic magnetic properties of the $C$-AF phase of
YVO$_3$. They arise from the coupling between the spin and orbital
operators which triggers the dimerization of the FM interactions
\eref{jcdim} as a manifestation of a universal instability which occurs
in FM spin chains at finite temperature, either by the coupling to the
lattice or to purely electronic degrees of freedom \cite{Sir08}.
This latter mechanism could play a role in many transition metal
oxides with (nearly) degenerate orbital states.

\section{Entanglement in the ground states of spin-orbital models}
\label{sec:T=0}

\subsection{Kugel-Khomskii model}
\label{sec:d9}

As shown in section 3, the GSs of certain spin-orbital models are
entangled and this will likely influence future experimental studies. As an
example we discuss here the Kugel-Khomskii $d^9$ model on a bilayer and
analyse the $d^1$ spin-orbital on a frustrated triangular lattice in
\sref{sec:rvb}. While the coexistence the $A$-type AF ($A$-AF) order and
the $C$-AO order below $T_{\rm N}\simeq 39$ K is well established in the
KCuF$_3$ perovskite \cite{Cuc02,Lee12} and this phase is well reproduced
by the spin-orbital superexchange $d^9$ model \cite{Ole00}, the model
poses an interesting theoretical question: which types of coexisting
spin-orbital order (or disorder) are possible when its microscopic
parameters are varied?
So far, it was only established that the long-range AF order is
destroyed by strong quantum fluctuations \cite{Fei98,Kha97,Ole00},
and it has been shown that certain spin disordered phases with VB
correlations may be stabilized by local orbital correlations
\cite{Fei97,Kha05}. However, the phase diagram of the Kugel-Khomskii
$d^9$ model was not studied systematically beyond the MF approximation
or certain simple variational wave functions and it remains an
outstanding problem in the theory \cite{Fei97}.

The simplest spin-orbital models are obtained when transition metal ions
are occupied by either one electron ($m=1$), or by nine electrons
($m=9$); in these cases the Coulomb interactions (\ref{Hint}) contribute
only in the excited states (in the $d^2$ or the $d^8$ configuration)
after a charge excitation between two neighboring ions,
$d_i^md_j^m\rightleftharpoons d_i^{m+1}d_j^{m-1}$.
A paradigmatic example of the spin-orbital physics is obtained in the
case of a single hole in the $d$ shell, as realized for the $d^9$ ($m=9$)
configuration of Cu$^{2+}$ ions in KCuF$_3$. Due to the
splitting of the $3d$ states in the octahedral field within the CuF$_6$
octahedra, the hole at each magnetic Cu$^{2+}$ ion occupies one of two
degenerate $e_g$ orbitals. The superexchange coupling \eref{som} is
usually analysed in terms of $e_g$ holes in this case \cite{Kug82}, and
this has become a textbook example of spin-orbital physics by now
\cite{vdB11,Ole09}.

The bilayer spin-orbital model is obtained following \cite{Ole00}; it
describes $S=\frac12$ spins with the Heisenberg SU(2) interaction coupled
to the $e_g$ orbital $\tau=\frac12$ pseudospins, with orbital operators
$\tau^{(\gamma)}_i$ \eref{tabc} obeying much lower cubic symmetry of the
orbital exchange:
\begin{eqnarray} \label{hamik}
{\cal H}&=&-\frac{1}{2}J\!\!\sum_{\langle ij\rangle||\gamma}
\left\{(r_1\,\Pi_t^{(ij)}+r_2\,\Pi_s^{(ij)})
\left(\frac{1}{4}-\tau^{(\gamma)}_i\tau^{(\gamma)}_j\right)\right. \nonumber \\
& &\hskip .7cm
+\left.\left(r_2+r_4\right)\Pi_s^{(ij)}\left(\frac{1}{2}-\tau^{(\gamma)}_i\right)
\left(\frac{1}{2}-\tau^{(\gamma)}_j\right)\right\}  \nonumber \\
&-&E_z\sum_{i}\tau_i^{(c)}\,.
\end{eqnarray}
The energy scale is given by the superexchange constant \eref{J}, with
$t$ standing here for the $(dd\sigma)$ effective hopping element.
The terms proportional to the coefficients $\{r_1,r_2,r_4\}$ originate
from the charge excitations to the upper Hubbard band \cite{Ole00} which
occur in $d^9_id^9_j\rightleftharpoons d^8_id^{10}_j$ processes and
depend on Hund's exchange \eref{eta} parameter, with $0<\eta<\frac{1}{3}$,
\begin{eqnarray}
r_1=\frac{1}{1-3\eta},\hskip .7cm r_2=\frac{1}{1-\eta},\hskip .7cm
r_4=\frac{1}{1+\eta}.
\end{eqnarray}
Note that $\tau^{(\gamma)}_i$ operators are not independent because they
satisfy the local constraint $\sum_{\gamma}\tau^{(\gamma)}_i\equiv 0$.
The bilayer model \eref{hamik} depends thus on two parameters:
($i$) Hund's exchange coupling $\eta$ \eref{eta}, and
($ii$) the crystal-field splitting of $e_g$ orbitals $E_z/J$.

The $\Pi^{s(t)}_{ij}$ operators stand for projections of spin states on
a bond $\langle ij\rangle$ on a singlet ($\Pi^s_{ij}$) and triplet
($\Pi^t_{ij}$) configuration for $S=\frac12$ spins, i.e.,
\begin{eqnarray}
\label{proje}
\Pi_s^{(ij)}=\left(\frac{1}{4}-{\bf S}_i\cdot{\bf S}_j\right),\hskip .5cm
\Pi_t^{(ij)}=\left(\frac{3}{4}+{\bf S}_i\cdot{\bf S}_j\right).
\end{eqnarray}
Their form suggests that singlet spin correlations will play an
important role in particular parameter regimes. The usual on-site MF
approximation captures only global symmetry breaking in the
bilayer, with essentially four different magnetic phases:
($i$) two $G$-AF phases with FO order characterized by either $x$ or $z$
orbitals occupied by the holes and stable at large values of $|E_z|$,
($ii$) the $A$-AF phase stabilized by finite Hund's exchange $\eta$
\eref{eta} near the orbital degeneracy $E_z=0$, and
($iii$) the FM phase which has the lowest energy at a sufficiently
large value of $\eta$. It is clear that a better approach than the MF
approximation with on-site order parameters $\langle S_i^z\rangle$ and
$\langle \tau_i^{(\gamma)}\rangle$ has to be employed to capture subtle
effects of spin fluctuations which may stabilize VB or resonating VB
(RVB) phases. Indeed, it has been shown that the phase diagram obtained
in the on-site MF approach changes drastically and is much richer when
the cluster MF approach is used instead \cite{Brz11}, see below.

A more sophisticated approach which goes beyond the single-site MF
approximation takes a cubic $2\times 2\times 2$ cluster as a reference,
with eight corner sites coupled to their neighbours along the bonds in
the $ab$ planes by the MF terms. This choice is motivated by the form
of the Hamiltonian containing different interactions in three different
directions, and the cube is the smallest cluster which couples the $ab$
planes and does not break the symmetry between the $a$ and $b$ axes as it
contains equal numbers of $a$ and $b$ bonds. In the considered case of a
bilayer there are no further neighbours of the cube along the $c$ axis.
After dividing the entire bilayer into identical cubes which cover the
bilayer lattice, the Hamiltonian (\ref{hamik}) can be written in a
cluster MF form as follows,
\begin{eqnarray}
{\cal H}_{\rm MF}=\sum_{m\in{\cal C}}\left({\cal H}^{\rm int}_m+{\cal
H}^{\rm ext}_m\right)\,, \label{mfham}
\end{eqnarray}
where the sum runs over the set of cubes ${\cal C}$, with each individual
cube labeled by $C_m\in{\cal C}$. Here ${\cal H}^{\rm int}_m$ contains
all bonds that belong to a given cube $C_m$ and the crystal-field terms
$\propto E_z$ at cube sites, i.e., it depends only on the operators on
the cube sites, while ${\cal H}^{\rm ext}_m$ contains all bonds outgoing
from a given cube $m$ and connecting it with neighbouring clusters.

The basic idea of
the cluster MF approach is to approximate ${\cal H}^{\rm ext}_m$ by
$\tilde{{\cal H}}^{\rm ext}_m$ containing only operators from the cube
$m$. This can be accomplished in many different ways depending on which
type of symmetry breaking is investigated. A natural choice is to take
$\tilde{{\cal H}}^{\rm ext}_m$ of the following general form \cite{Brz11},
\begin{equation}
\tilde{{\cal H}}^{\rm ext}_m=\frac{1}{2}\!\sum_{\gamma=a,b\atop i\in C_m}
\left\{S^z_ia^{\gamma}_i+S^z_i\tau^{\gamma}_ib^{\gamma}_i
+\tau^{\gamma}_ic^{\gamma}_i+d^{\gamma}_i \right\},
\label{htild}
\end{equation}
containing SU(2) symmetry-breaking spin field $S^z_i$, orbital field
$\tau^{\gamma}_i$ and spin--orbital field $S^z_i\tau^{\gamma}_i$.
Coefficients $\{a^{\gamma}_j,b^{\gamma}_j,c^{\gamma}_j,d^{\gamma}_j\}$
are the Weiss fields determined self--consistently for various values of
the parameters $\{E_z/J,\eta\}$. Note that we introduce here spin-orbital
field $S^z_j\tau^{\gamma}_j$ because, as seen {\it a posteriori\/} in
some phases, spins and orbitals alone do not suffice to describe the
symmetry breaking which may occur when these operators act together.

The standard way, as in any MF approach, is to derive self--consistency
equations for the Weiss fields. This can be done in a straightforward
fashion: we take the operator products from ${\cal H}^{\rm ext}_m$
and divide them into a part depending only on operators from the
cube $m$ itself, and from a neighboring cube $n$. This procedure can
be applied to all operator products in ${\cal H}^{\rm ext}_m$ and full
$\tilde{{\cal H}}^{\rm ext}_m$ can be recovered in the MF form
\eref{htild}. After repeating this procedure for all clusters,
one finds a set of commuting cubes interacting in a self--consistent
way. The Weiss fields at site $i$,
\begin{eqnarray}
a^{\gamma}_i&=&\frac{1}{2}(r_2+r_4)u^{\gamma}_i+\frac{1}{4}(r_2-r_1)s^{\gamma}_i,
  \\
b^{\gamma}_i&=&-(r_4+r_1)u^{\gamma}_i-\frac{1}{2}(r_2-r_1)s^{\gamma}_i,
  \\
c^{\gamma}_i&=&\frac{1}{4}(3r_1-r_4)t^{\gamma}_i+\frac{1}{8}(r_2+r_4),
  \\
d^{\gamma}_i&=&-\frac{1}{2}(r_1+r_4)u^{\gamma}_iu^{\gamma}_{m,i}
-\frac{1}{4}(r_2-r_1)(s_iu^{\gamma}_i+s^{\gamma}_iu^{\gamma}_{m,i}),
\nonumber \\
&-&\frac{1}{16}(r_2+r_4)(t^{\gamma}_{m,i}-t^{\gamma}_i)
+\frac{1}{8}(r_4-3r_1)t^{\gamma}_it^{\gamma}_{m,i} \nonumber \\
&-&\frac{1}{32}(3r_1+2r_2+r_4),
\end{eqnarray}
are determined together with the order parameters at site $i$,
\begin{eqnarray}
s_i&\equiv &\left\langle S^z_i\right\rangle,\\
\label{mdefs}
t^{\gamma}_{m,i}&\equiv &\left\langle\tau^{(\gamma)}_i\right\rangle,\\
\label{mdeft}
u^{\gamma}_{m,i}&\equiv &\left\langle
S^z_i\left(\frac{1}{2}-\tau^{(\gamma)}_i\right)\right\rangle.
\label{mdefu}
\end{eqnarray}

The next crucial step is to impose a condition that
$\{s^{\gamma}_i,t^{\gamma}_{i},u^{\gamma}_{i}\}$ are related to the
order parameters obtained on the internal sites of the considered
cluster. Thereby it is convenient to assume that two neighbouring cubes
can differ in orbital (and spin-orbital) configuration by the
interchange of $a$ and $b$ direction, i.e.,
\begin{eqnarray}
\label{szf}
s^{\gamma}_i=\pm s_i,\hskip .7cm \label{tzf}
t^{\gamma}_{i}=t^{\bar{\gamma}}_{m,i},\hskip .7cm \label{stf}
u^{\gamma}_{i}=\pm u^{\bar{\gamma}}_{m,i} \label{legs}
\end{eqnarray}
with $\bar{\gamma}$ being the complementary direction in the $ab$ plane
to $\gamma$, i.e., $(\gamma,\bar{\gamma})=(a,b),(b,a)$. This relation
gives the same results as the one when $a\leftrightarrow b$ symmetry in
the cube is not broken, but keeps the whole system $a$-$b$ symmetric in
the other case. It is also important that the spin-orbital field in not
factorized (but surprisingly it turns out that such a factorization does
not prevent spin-orbital entanglement to occur \cite{Brz11}).

\begin{figure}[t!]
\begin{center}
    \includegraphics[width=8.2cm]{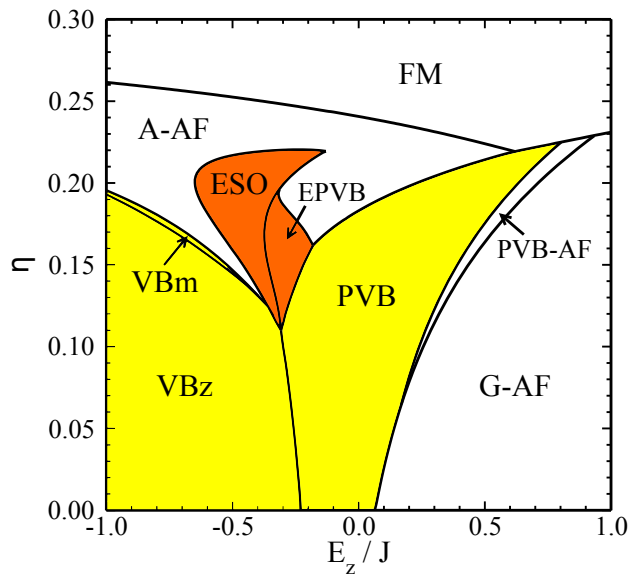}
\end{center}
\caption{(Color online) The phase diagram of the cluster MF Hamiltonian
(\ref{mfham}) of the $d^9$ spin-orbital model for a bilayer, with
independent spin \eref{mdefs}, orbital \eref{mdeft} and spin-orbital
\eref{mdefu} MFs. VB phases (mainly PVB and VB$z$) with spin disorder
are stable in the light shaded (yellow) area, and phases with
spin-orbital entanglement are indicated by dark grey (orange) shading.
This figure is reproduced from \cite{Brz11}.}
\label{fig:bilaphd}
\end{figure}

When the spin-orbital MF \eref{mdefu} is not factorized but calculated
self-consistently, one finds the phase diagram of \fref{fig:bilaphd}.
Here two $G$-AF phases for $E_z>0$ with occupied $x$ orbitals, $A$-AF
with accompanied $C$-AO order and FM are familiar from the simplest on-site
MF approach, and appear also in the 3D KK model \cite{Ole00}. The part
of the diagram for $\eta<0.2$ and $E_z/J<0.5$ is dominated by two
new phases compared with a one-site MF approach (shown in \cite{Brz11}):
VB$z$ phase with $z$ orbitals occupied stable for negative $E_z$, and
plaquette VB (PVB) phase in a range of $E_z\simeq 0$, both with
vanishing magnetization. For small $E_z/J<0.25$ one finds that quantum
fluctuations included within the present approach select the former
spin-disordered VB$z$ phase. The latter PVB phase has singlets formed on
the bonds either along the $a$ or $b$ direction of the cluster,
depending on the cube. This phase breaks the $a\leftrightarrow b$
symmetry locally but the global symmetry is preserved thanks to the
$\pi/2$ rotation of neighbouring clusters mentioned above. The orbitals
take shape of cigars pointing in the direction of the singlets.

The most important result presented in \fref{fig:bilaphd} are the
regions of stability of three new entangled phases: ESO, EPVB and PVB-AF,
obtained only when the spin-orbital order parameter $u^{\gamma}_{i}$
\eref{mdefu} is not factorized into the spin and orbital part.
ESO stands for entangled
spin-orbital phase and is characterized by relatively high values of
spin-orbital order parameters, especially for high values of $\eta$ when
other order parameters are close to zero. This phase contains singlets
along the bonds parallel to the $c$ axis, its magnetization vanishes and
orbital configuration is nonuniform. EPVB stands for entangled PVB phase
and resembles it, but has in addition a finite spin-orbital field, and
weak global AF order. Finally, a different type of phase with
spin-orbital entanglement is the PVB-AF phase connecting PVB and $G$-AF
in a smooth way, stable only if $\eta$ is large enough.

Note that with the exception of this last phase, the other two entangled
phases arise near the quantum critical point (QCP) which is found at
$\eta=0$ in the MF approach and moves to a finite value of
$\eta\simeq 0.12$ when quantum fluctuations on singlet bonds are
explicitly included. Therefore, the phase diagram of \fref{fig:bilaphd}
implies that singlet formation suppresses frustration caused by Hund's
exchange coupling and moves the region of the most frustrated
interactions to finite $\eta$. This
shows once again that the simple single-site MF approach is insufficient
to describe faithfully the phase diagram of the present spin-orbital model
\eref{hamik}. We suggest that similar entangled phases are expected in
the 3D model which should be investigated within the cluster MF approach
in the near future.

\subsection{Spin-orbital resonating valence-bond liquid}
\label{sec:rvb}

In this section we consider another example of spin-orbital entangled
states which are found in the $d^1$ model on the frustrated triangular
lattice \cite{Nor08,Cha11}, as realized in the (111) planes of NaTiO$_2$.
In the limit of large intraorbital Coulomb interaction $U$ intersite
charge excitations are again transformed away and one finds the
following effective Hamiltonian \cite{Nor08},
\begin{equation}
\label{d1} {\cal H} = J \left\{ (1 - \alpha) \; {\cal H}_s
                 + \sqrt{(1 - \alpha) \alpha} \; {\cal H}_m
                 + \alpha \; {\cal H}_d \right\}\,,
\end{equation}
where $J$ is the exchange energy for $S=\frac12$ spins and three $t_{2g}$
orbital flavours active on a bond in different processes. Here the
interaction may arise either from superexchange via oxygen orbitals due
to transitions via the effective hopping $t$ that follows from the $d-p$
hybridization, or via direct (kinetic) exchange between $t_{2g}$ orbitals
with flavour $\gamma$ active on a bond along the direction $\gamma$ in a
triangular lattice via the hopping $t'$. The parameter $\alpha$
interpolates between the superexchange ($\alpha=0$) and direct exchange
($\alpha=1$) limit. It is the first parameter of the present
model \eref{d1} and is given by the ratio of these two hopping elements:
\begin{equation}
\label{alpha}
\alpha=\frac{t'^2}{t^2+t'^2}.
\end{equation}
The superexchange involves two $t_{2g}$ orbital flavors \eref{t2go}
different from $\gamma$ on each bond that are not active in direct
exchange. Consequently, the
superexchange is more quantum, similar to the superexchange in titanates
or vanadates, while the direct exchange is more classical in the orbital
channel, bearing some similarity to the superexchange for $e_g$ orbitals
analysed above in the bilayer KK model. The second parameter of
the spin-orbital model \eref{d1} is Hund's exchange $\eta$ \eref{eta},
as in the KK model.
More details on the model and its derivation can be found in \cite{Nor08}.

In the subsequent sections we focus first on the frustrated
interactions in the $d^1$ model \eref{som} at $\eta=0$ limit,
and we give here its explicit form in this case:
\begin{eqnarray}
\label{som0}
{\cal H}_0\!\! &=&\! \!
J \sum_{\langle ij \rangle \parallel \gamma}
\Big\{ (1 - \alpha) \left[ 2 \left( \vec{S}_i \cdot \vec{S}_j
 + \frac{1}{4} \right)\right. \nonumber \\
& &\!\!\left.\times\left[
\left( \vec{T}_{i} \cdot \vec{T}_{j}\right)^{(\gamma)}
 + \frac{1}{4} n_i^{(\gamma)} n_j^{(\gamma)} \right]\!
 + \frac{1}{2} (n_{i\gamma} + n_{j\gamma}) - 1 \right] \nonumber \\
& +&\!\!  \alpha \left[ \left( \vec{S}_i \! \cdot \! \vec{S}_j
 - \frac{1}{4} \right) n_{i\gamma} n_{j\gamma}
-\frac{1}{4} \Big(
n_{i\gamma}  n_j^{(\gamma)} +  n_i^{(\gamma)} n_{j\gamma} \Big)
\right] \nonumber \\
&- &\! \!\frac{1}{4} \sqrt{\alpha (1 - \alpha)} \; \Big(
T_{i\bar{\gamma}}^+ T_{j{\tilde \gamma}}^+ + T_{i{\tilde \gamma}}^-
T_{j\bar{\gamma}}^- + T_{i{\tilde \gamma}}^+ T_{j\bar{\gamma}}^+
+ T_{i\bar{\gamma}}^- T_{j{\tilde \gamma}}^- \Big)\Big\}.
\end{eqnarray}
The summations include the bonds $\langle ij \rangle \parallel \gamma$
of a triangular lattice, with $\gamma=a,b,c$ labeling three directions.
This case is rather special as the multiplet structure collapses to a
single excitation with energy $U$ (spin singlet and triplet excitations
are then degenerate), and the Hamiltonian simplifies. The operators
$n_{i\gamma}$ are electron number operators for the orbital flavour
$\gamma$ at site $i$, and $n_i^{(\gamma)}$ is the density in the
remaining two orbitals, which is related to $n_{i\gamma}$ by the
local constraint,
\begin{equation}
n_{i\gamma}+n_i^{(\gamma)}=1.
\end{equation}
The scalar products of the orbital operators in \eref{som0},
\begin{eqnarray}
\label{ttscalar}
\left(\vec{T}_{i} \cdot \vec{T}_{j}\right)^{(\gamma)}\! &\equiv&
\frac12\Big( T_{i\gamma}^+T_{j\gamma}^- + T_{i\gamma}^-T_{j\gamma}^+\Big)
+ T_{i\gamma}^z T_{j\gamma}^z\,,
\end{eqnarray}
involve two active orbital flavours on superexchange bonds.
For a bond along the axis $\gamma$ orbital operators at site $i$ are
defined by the electron creation
$\{a_i^\dagger,b_i^\dagger,c_i^\dagger\}$ and annihilation
$\{a_i,b_i,c_i\}$ operators for electrons with a given flavour.
For instance, for the bonds along the $a$ or $b$ axis they are:
\begin{eqnarray}
\label{TaTb}
T_{ia}^+ \!& =&\! b_i^{\dagger} c_i^{}\,, \hskip 2cm
T_{ib}^+   =      c_i^{\dagger} a_i^{}\,,
\\
T_{ia}^- \!& =&\! c_i^{\dagger} b_i^{}\,, \hskip 2cm
T_{ib}^-   =      a_i^{\dagger} c_i^{}\,,
\\
\label{TzTz}
T_{ia}^z\! & =&\!\frac12 ( n_{ib} - n_{ic} )\,, \hskip .5cm
T_{ib}^z     =   \frac12 ( n_{ic} - n_{ia} )\,.
\end{eqnarray}
The labels $\bar\gamma\neq\tilde\gamma$ in the quantum fluctuating part
$\propto\sqrt{\alpha(1-\alpha)}$ refer to the two orbital
operators on each bond along the direction $\gamma$ involved in the
fluctuating operators defined in
\eref{TaTb}. Orbital fluctuations are the only processes contributing
to the mixed exchange terms in this limit ($\eta=0$).

A remarkable feature of the Hamiltonian in the limit of $\eta=0$
\eref{som0} is the lack of higher symmetry in any of the points when
$\alpha$ is varied. Even at $\alpha=0.5$, where all electron transitions
have the same amplitude, no higher symmetry occurs as the superexchange
($\alpha=0$) and direct exchange ($\alpha=1$) result from quite distinct
processes and involve different subsets of orbital flavours which cannot
be transformed one into the other. The only analytical solution was found
in the $\alpha=1$ case, where at $\eta=0$ the extremely degenerate GS is a
liquid of hard-core dimers \cite{Jac07}. This degeneracy is removed at
finite $\eta>0$, and a VB crystal with a large unit cell of 20 sites is
formed.

\begin{figure}[t!]
\begin{center}
\includegraphics[width=7.5cm]{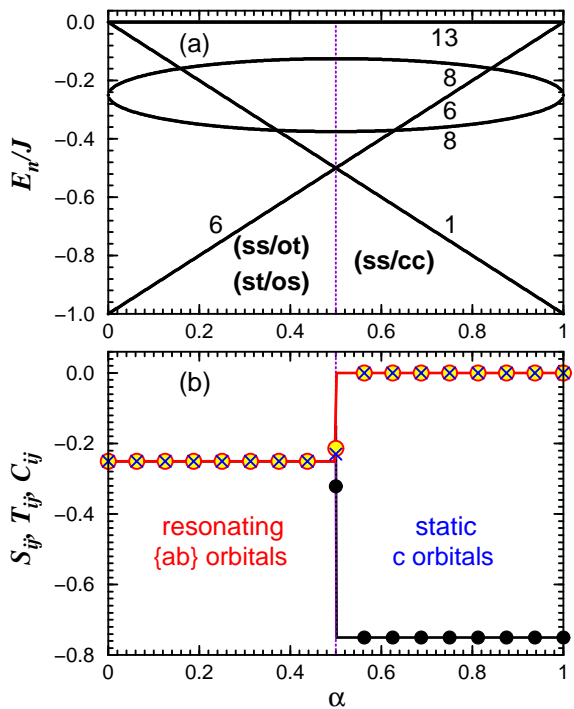}
\end{center}
\caption{(Color online)
Evolution of the properties of a single bond $\gamma \equiv c$ for the
spin-orbital model \eref{som0} on a triangular lattice as a
function of $\alpha$ at $\eta = 0$:
(a) energy spectrum $E_n$ (solid lines) with degeneracies given by numbers;
(b) spin $S_{ij}$ \eref{sij} (filled circles), orbital $T_{ij}$ \eref{tij}
(empty circles),
and spin--orbital $C_{ij}$ \eref{cij} ($\times$) correlations.
The transition between the two distinct regimes occurs by a level
crossing at $\alpha = 0.5$. For $\alpha < 0.5$, the two types of dimer
wave function [(ss/ot) and (os/st)] are degenerate ($d = 6$) for
resonating orbital configurations $\{ab\}$, while at $\alpha > 0.5$,
the nondegenerate spin singlet is supported by $c$ orbitals occupied
at both sites [(ss/cc)].
This figure is reproduced from \cite{Nor08}. }
\label{fig:bond}
\end{figure}

To understand a subtle interplay between the quantum superexchange at
$\eta=0$ and more classical direct exchange in the limit of $\eta=1$ we
consider first a single bond oriented along the $c$ axis. In the
superexchange limit the active orbitals are $a$ and $b$, while only $c$
orbitals contribute to the direct exchange \eref{som0}. A single bond
gives the GS energy $E_0=-J$, both in the superexchange ($\alpha = 0$)
and in the direct exchange ($\alpha = 1$) limit, see \fref{fig:bond}(a).
These two limits differ in a fundamental way --- the GS at $\alpha = 0$
has degeneracy $d = 6$ due to two complementary triply degenerate wave
functions, with spin singlet and orbital triplet (ss/ot), and spin
triplet accompanied by orbital
singlet (st/os), while in the opposite $\alpha = 1$ case limit spin
singlet is accompanied by frozen FO order of active $c$ orbital (ss/cc).
Energy increase when the $\alpha=0.5$ point is approached from either
side indicates increasing frustration.

The different character of wave functions prevents any energy gain that
might result from orbital fluctuations in the mixed exchange term, and
the GS energy $E_0$ increases linearly toward $E_0=-\frac12 J$
when the point $\alpha=0.5$ is approached from either side. At this
point the interactions are maximally frustrated, degeneracy is $d=7$,
and a QPT between the two described GSs takes place. Also the remaining
part of the spectrum (excited states) is symmetric with respect to the
$\alpha=0.5$ point. Altogether, the evolution of the spectrum with
$\alpha$ demonstrates not only that superexchange and direct exchange
are physically distinct, excluding each other and unable to contribute
at the same time, but also that the two wave functions
optimal in either limit are extremely robust.

The above interpretation is consistent with the spin, orbital, and
composite spin-orbital correlation functions for the considered bond,
defined in equations \eref{sij}-\eref{cij}. In the entire regime of
$0\leq\alpha<0.5$, averaging over two degenerate (ss/ot) and (st/os)
wave functions leads to equal spin and orbital correlation functions
$S_{ij} = T_{ij} = - \frac14$, see \fref{fig:bond}(b). As a singlet
for one quantity is matched by a triplet for the other one, the two
sectors are strongly correlated, and one also finds $C_{ij}=-\frac14$.
Although individual quantum states may be written here as products of
the respective spin and orbital states, this result suggests that strong
spin-orbital entanglement in expected in the GS of a larger system.
It may be shown that entanglement arises there mathematically because
the GS is a resonating superposition of a number of configurations
\cite{Cha11} which do not couple here with one another and form
degenerate states for a single bond. Note that the obtained
value, $C_{ij}= - \frac14$, is the same as in the SU(4) model,
see \sref{sec:su2}, and reflects maximal possible entanglement.

By contrast, for $\alpha > 0.5$ the above degenerate states favoured by
superexchange become excited states, and the spin-singlet GS
($S_{ij} = -\frac{3}{4}$) is the lowest, see \fref{fig:bond}(b). The
orbital configuration is characterized here by a rigid order of $c$
orbitals, $\langle n_{ic} n_{jc} \rangle = 1$, which quenches all
orbital fluctuations. Thus the spin and orbital parts are trivially
decoupled, giving $C_{ij} = 0$. Finally, at the transition point
$\alpha=0.5$ one finds degeneracy $d=7$ and averaging over all
degenerate states yields $S_{ij}=-0.321$, $T_{ij}=-0.214$, and somewhat
reduced composite function $C_{ij}\simeq -0.23$, as one of the degenerate
states gives here $C_{ij} = 0$. Summarizing, the regime of entangled
spin-orbital states $\alpha < 0.5$ is
complemented by a factorized (disentangled) (ss/cc) wave function.

\begin{figure}[t!]
\begin{center}
\includegraphics[width=8.2cm]{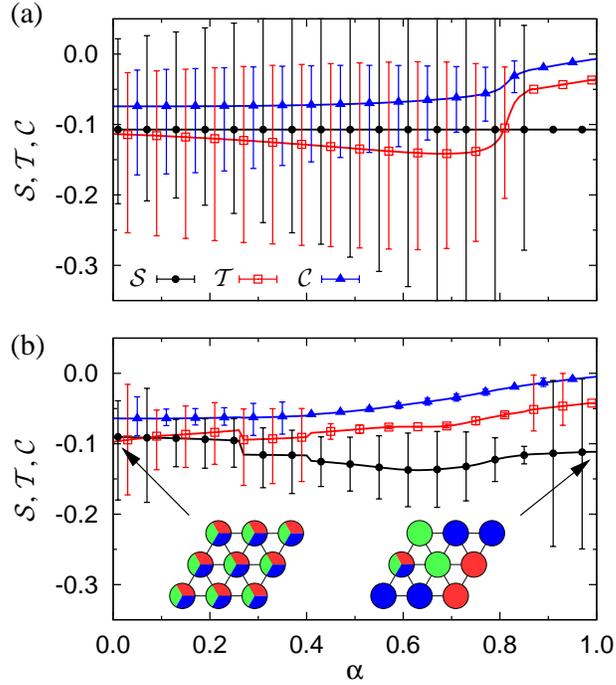}
\end{center}
\caption{Bond correlations for clusters with PBC at $\eta=0$ ---
spin ${\cal S}$ \eref{sij}, orbital ${\cal T}$ \eref{tij}, and spin-orbital
${\cal C}$ \eref{cij}:
(a) hexagonal cluster of $N=7$ sites, and
(b) rhombic cluster of $N=9$ sites.
Vertical lines indicate the exactly determined range of possible values
of each quantity at a given $\alpha$ that follows from the GS degeneracy.
The insets in (b) indicate typical orbital patterns in the superexchange
($\alpha=0$) and direct exchange ($\alpha=1$) limit for the rhombic N9
cluster; a more symmetric hexagonal cluster N7 is obtained by removing
top-right and bottom-left sites.
These data were presented in \cite{Cha11}.}
\label{fig:stc}
\end{figure}

Below we analyse how the above two distinct regimes of the $d^1$ model
are modified in case of large number of neighbours and frustration in
the triangular lattice. We consider two clusters with PBC:
a hexagonal cluster of $N=7$ sites (N7), and a rhombic cluster of $N=9$
sites (N9). Unfortunately, the cluster with even $N=12$ number of sites
is too large and would require very extensive calculations, but the
presented clusters are sufficient to demonstrate essential differences
between the nature of interactions in both limits. Due to the PBC all
sites and bonds are equivalent, so bond correlations are all the same
and independent of the bond direction $\gamma$. Each $t_{2g}$ orbital is
occupied on average by 1/3 electron, but similar to a single bond the
GSs are manifestly different in the limits of $\alpha=0$ and $\alpha=1$,
see the insets in \fref{fig:stc}(b).

In case of N7 cluster all three directions are equivalent and the spin
correlations are AF and constant, $S_{ij}\simeq -0.11$ independently of
$\alpha$. Note that this
value can be deduced as follows: In the low-spin phase one has
${\cal S}_t=\frac12$ total spin, and one can determine the intersite spin
correlations using the following identity \cite{Cha11}:
\begin{equation}
\label{ident}
\vec{\cal S}^2=7\vec{S}_i^2+42\langle\vec{S}_i\cdot\vec{S}_j\rangle\,.
\end{equation}
Due to PBC every pair of sites stands for a nearest neighbour bond and
this implies the above relation. The value
$\langle\vec{S}_i\cdot\vec{S}_j\rangle=-3/28=-0.107$ obtained from it
reflects AF spin correlations but is much reduced from the classical limit
of $-0.25$ found in the N\'eel state on a square lattice, in agreement with
high frustration of the triangular lattice. Note that the present situation
is radically different from the 1D SU(4) model, where frustration is absent
but spin correlations are reduced by their coupling to the orbital correlations.
In the N9 cluster spin correlations are in a similar range but vary
somewhat with $\alpha$ as the cluster shape breaks the symmetry between
bonds in three nonequivalent  directions $\{a,b,c\}$.
These correlations are weaker (${\cal S}\simeq -0.090$) at $\alpha=0$
than in the N7 cluster and become more pronounced (${\cal S}\simeq
-0.144$) when $\alpha\simeq 0.6$, while orbital fluctuations
gradually weaken to ${\cal C}\simeq -0.050$ at $\alpha=1$.

In both clusters orbital correlations are negative, ${\cal T}<0$, and the
Goodenough-Kanamori rule stating that these correlations should be
complementary to spin ones is violated. The orbital correlations
weaken in both clusters when $\alpha$ increases towards $\alpha=1$ and
the superexchange interactions are less important, particularly in the N9
cluster. Joint spin-orbital correlations are also similar in both clusters
(e.g. ${\cal C}\simeq -0.070$ at $\alpha=0$) and $|{\cal C}|$ gradually
decreases when spin and orbitals disentangle approaching $\alpha=1$.

An important question is whether spin order and excitations could
be described by an effective spin model derived from the spin-orbital
model (\ref{som}). Below we show that this is not the case in the $d^1$
model on the triangular lattice. MF procedure used frequently leads
here to \cite{Cha11}:
\begin{eqnarray}
\label{sommf}
{\cal H}_{\rm MF} &=& \sum_{\langle ij\rangle \parallel\gamma}\left\{
 \left\langle{\hat {\cal J}}_{ij}^{(\gamma)}\right\rangle
             {\vec S}_i\cdot {\vec S}_j
-\left\langle{\hat {\cal J}}_{ij}^{(\gamma)}\right\rangle
\left\langle {\vec S}_i\cdot {\vec S}_j\right\rangle
\right\}\nonumber \\
&+&                  \sum_{\langle ij\rangle \parallel\gamma} \left\{
{\hat {\cal J}}_{ij}^{(\gamma)} \left\langle {\vec S}_i\cdot {\vec S}_j
\right\rangle + {\hat {\cal K}}_{ij}^{(\gamma)} \right\}\,.
\end{eqnarray}
In this way spin and orbital degrees of freedom are disentangled and the
model reduces to a superposition of the spin model with self-consistently
determined orbital correlations, and the orbital model, with self-consistently
derived spin correlations, similar to the decoupling of spin and orbital
degrees of freedom introduced for the 1D spin-orbital chain in
\sref{sec:dim}. Following the spin model, one obtains the MF spin
interactions for N7 and N9 clusters by averaging the orbital operator
${\hat {\cal J}}_{ij}^{(\gamma)}$ (its explicit form is given in
\cite{Cha11}) over the MF GS $|\Phi_0\rangle$ \eref{Jij}.
Note that the orbital fluctuations in the term
$\propto\sqrt{\alpha(1-\alpha)}$ in \eref{som0} contribute here as
well as they couple different components of $|\Phi_0\rangle$. In
contrast, the exact exchange constant $J_{\rm exact}$ \eref{Jijex}
is found when the {\it exact\/} GS $|\Phi\rangle$ obtained after Lanczos
diagonalization is used for a given cluster.

\begin{figure}[t!]
\begin{center}
\includegraphics[width=8cm]{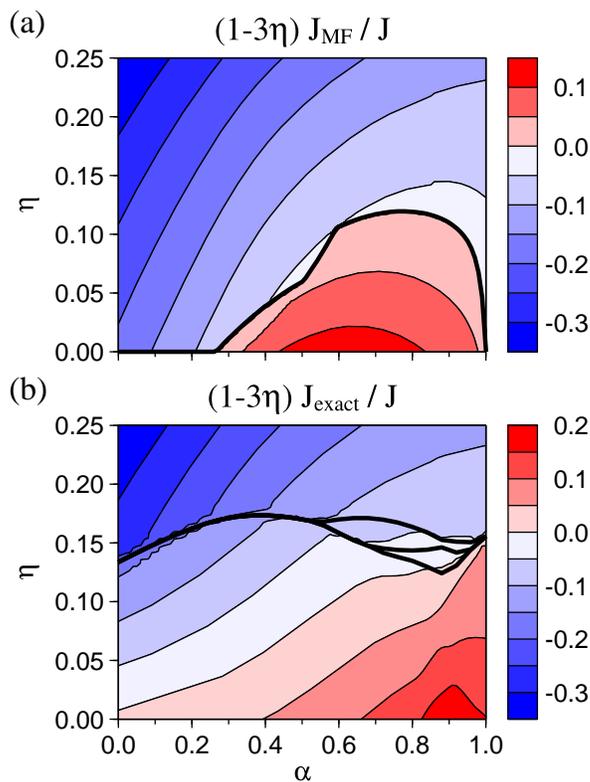}
\end{center}
\caption{(Color online)
Contour plots of the effective exchange constant $J_{\rm MF}$ as
obtained for the hexagonal cluster of $N=7$ sites with PBC:
(a) within the MF calculation which includes orbital fluctuations \eref{Jij},
and
(b) using the exact GS found in exact diagonalization \eref{Jijex}.
In case (a) the transition from low-spin to high-spin phase occurs when
the exchange constant $J_{\rm MF}$ changes sign and becomes negative.
Thick lines in (b) indicate the phase boundaries between phases with
increasing intermediate total spin value ${\cal S}_t=3/2$, 5/2 between
${\cal S}_t=1/2$ and 7/2 for increasing $\eta$.
This figure is reproduced from \cite{Cha11}.}
\label{fig:jeff}
\end{figure}

In \fref{fig:jeff} we compare the phase diagrams obtained from the above
MF procedure and from Lanczos diagonalization for the N7 cluster.
Consider first a QPT from the low-spin (${\cal S}_t=1/2$)
disordered phase to the high-spin (${\cal S}_t=7/2$) FM
phase which occurs for sufficiently large $\eta$. When spin and orbital
operators are disentangled \eref{sommf}, see \fref{fig:jeff}(a), and
${\cal C}\equiv 0$, it coincides with the sign change of
the MF exchange constant $J_{\rm MF}$ \eref{Jij} and no intermediate
phase (with $1/2<{\cal S}_t<7/2$) is found, as in a spin system.

Comparing the values of $J_{\rm MF}$ \eref{Jij} and $J_{\rm exact}$
\eref{Jijex} found from the MF and from exact diagonalization of the
N7 cluster, see \fref{fig:jeff}, one finds that
$J_{\rm exact}\ge J_{\rm MF}$ in a broad range of $\alpha$.
Therefore, the MF approximation turns out to be rather unrealistic and
overestimates (underestimates) the stability of states with FM (AF) spin
correlations. The value of $J_{\rm MF}$ decreases with increasing $\eta$,
but positive values $J_{\rm MF}>0$ are found at $\eta=0$ only if
$0.27<\alpha<1$. This demonstrates that FM states:
($i$) are favoured when joint spin-orbital fluctuations are suppressed,
and
($ii$) are stabilized by orbital fluctuations close to $\alpha=0$ even
in absence of Hund's exchange.
The transition from the low-spin (${\cal S}_t=1/2$) to the high-spin
(${\cal S}_t=7/2$) state occurs in Lanczos diagonalization at a much
higher finite value of $\eta\approx 0.14$, with only weak dependence
on $\alpha$, see \fref{fig:jeff}(b). In addition,
one finds two phases with intermediate spin values ${\cal S}_t=3/2,5/2$
in a range of $\eta$ values near $\alpha=0.8$. Note that the exchange
constant $J_{\rm exact}$ changes discontinuously at the onset of the FM
phase.

We have found that the qualitative trends presented here for the N7
cluster are similar to those observed for the N7 cluster \cite{PM11}
and thus they may be considered generic for the present $d^1$ model on
the triangular lattice. In both cases one finds that:
($i$) the FM phase is stable in the MF approximation close to $\alpha=0$
and becomes degenerate with the low-spin phase at $\alpha=1$,
($ii$) the MF procedure is exact in the regime of FM phase, and
($iii$) the transition to the FM phase occurs gradually through
intermediate values of total spin ${\cal S}_t$ (except at $\alpha=1$).
This suggests that partially polarized FM phase should occur in the
thermodynamic limit. It arises due to spin-orbital entanglement which
is gradually suppressed when $\eta$ increases.

We argue that the  results presented in \cite{Cha11} and \cite{PM11}
provide evidence that the present $d^1$ spin-orbital model realizes a
paradigm of a {\it spin-orbital liquid phase\/} in the superexchange
regime, and the order-out-of-disorder mechanism does not occur when the
Hilbert space contains coupled spin and orbital sectors. Previous search
for this quantum state of matter in other systems, particularly in
LiNiO$_2$ where $e_g$ orbitals are active on the $d^7$ configurations of
Ni$^{3+}$ ions on the triangular lattice
\cite{Ver04}, were unsuccessful \cite{Mos02,Rei05}. After considering
the present model in more detail we suggest that the triple degeneracy
of $t_{2g}$ orbitals plays a crucial role in the onset of a spin-orbital
liquid, as the number of orbital flavours fits to the geometry of the
triangular lattice. In contrast, the direct exchange regime is dominated
by VB states with spin singlets accompanied by static configurations of
directional orbitals providing the energy gain for the direct exchange.
Note that the frustrated triangular lattice plays here an important role
and removes any kind of orbital order in the entire range of $\alpha$.

\section{Hole propagation in a Mott insulator with coupled spin-orbital order}
\label{sec:hole}

Finally, we give an example of a single hole doped into the half-filled
Mott insulator with a pseudo-entangled AF/AO order \cite{Woh09}. This
type of order is realized in the $ab$ planes of LaVO$_3$ with $S=1$
spins \cite{Miy06} and in Sr$_2$VO$_4$ with $S=\frac12$ spins
\cite{Ere11}. It might appear as entangled as it contradicts the
Goodenough-Kanamori rules and has certain similarity to the entangled
spin-orbital states in a 1D chain, see \sref{sec:t2g}. It is well
known that a hole doped into a quantum antiferromagnet couples to the
collective (delocalized) spin excitations  (magnons), and propagates
through the lattice surrounded by a "cloud" of magnons which is the
essence of a QP behaviour \cite{Kan89}. Thereby the energy scale of the
"coherent" hole propagation is strongly renormalized from the hopping
$t$ and is given by the AF superexchange constant $J$. This QP is
frequently called a {\it spin polaron} \cite{Mar91} and was observed
in the photoemission spectra of the parent compounds of high-$T_c$
cuprates, as e.g. in Sr$_2$CuO$_2$Cl$_2$ \cite{Dam03}.

A hole doped in a Mott insulator with orbital order could behave in a
similar way when orbital fluctuations or interorbital hopping
\cite{vdB00} would also repair the defects created by a hole. Orbital
excitations are decoupled from spin ones in
disentangled states, such as FM spin and AO order in the $ab$ planes of
LaMnO$_3$ \cite{Fei99}. It has been shown \cite{vdB00} that a hole
introduced into such a state indeed does not disturb the FM spin order and
couples to the collective excitations of the AO state (orbitons). Here
again a QP is formed which is called this time an {\it orbital polaron}
\cite{Kil99}. Usually, however, the bandwidth of an orbital polaron is
much smaller than that of the spin one. In fact, the orbitons are in
general considerably less mobile than the magnons (or even immobile) due
to almost directional Ising-like superexchange
\cite{Fei99,vdB99}. Actually, one can understand the hole motion in this
case in terms of the string picture \cite{Wro10,Wro08}.

The orbital order created by $t_{2g}$ orbitals in a 2D square lattice is
more robust than that of $e_g$ ones while only 1D hopping for each
orbital flavour is allowed in the former case
\cite{Dag08}. The case of a hole doped into the plane with FM spin order
accompanied by the $t_{2g}$ AO order, which could correspond not only to
the hole introduced into the ordered ground state of Sr$_2$VO$_4$ with
$t_{2g}$ orbitals but also, surprisingly, to K$_2$CuF$_4$ or
Cs$_2$AgO$_4$ with active $e_g$ orbitals. As the GS has AO order, also
here a QP might formed due to the dressing of a hole by the collective
orbital excitations. However, due to the specific $t_{2g}$ orbital
symmetries the orbitons are not mobile at all, and the QP acquires a
finite bandwidth only due to the frequently neglected three-site terms
\cite{Dag08}. Thus, the string picture determines the nature of the
$t_{2g}$ orbital polarons even more than in
systems with $e_g$ orbital degrees of freedom.

\begin{figure}[t!]
\begin{center}
\includegraphics[width=7cm]{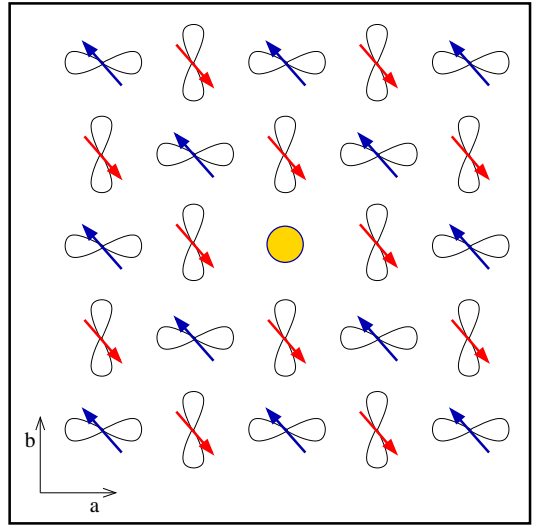}
\end{center}
\caption{Artist's view of a single hole introduced into the spin
and orbitally ordered $ab$ plane of LaVO$_3$ with coexisting AF/AO
order. One electron at each V$^{3+}$ ion occupies the $xy$ orbital (not
shown) and the other one either $yz$ or $zx$ degenerate orbital, forming
the classical AO state (the projections of these orbitals onto the $ab$
plane are shown) whereas their spins $S=\frac12$ alternate on two
sublattices, forming the AF N\'eel state.
This figure is reproduced from \cite{Woh09}. }
\label{fig:hole}
\end{figure}

A more complex situation arises, however, when a doped hole may couple
in a $t_{2g}$ system both to magnon and to orbiton excitations
\cite{Zaa93}. This situation is found in the $ab$ plane of LaVO$_3$,
shown schematically in \fref{fig:hole} and studied in \cite{Woh09}. The
coexisting AF/AO order consists of the AF
order induced by the superexchange interactions between the occupied $c$
orbitals, and the AO order of active $\{a,b\}$ orbitals. The GS in an
undoped system is therefore a perfect two-sublattice order, with quantum
fluctuations in the spin channel but a classical AO order in the orbital
channel as the superexchange interations in this part are Ising-like.
The coexistence of the AO and AF order is extremely rare as it formally
violates \cite{Ole06} the Goodenough-Kanamori rules predicting
complementary spin and orbital order in the GS. This GS has long-range
order, with an up (down) spin component accompanied by a $zx$ ($yz$)
orbital, but this composite nature of the GS has no dramatic
consequences as long as the system is undoped.

The problem of a single hole doped into the $ab$ plane of LaVO$_3$, see
\fref{fig:hole}, is challenging as the hole doping occurs in the orbital
$\{a,b\}$ doublet and disturbs locally both AF and AO order \cite{Hor11}.
Unlike in the FM/AO case \cite{Woh08}, for the present coexisting AF/AO
order neither the spin nor the orbital background is transparent for a
propagating hole, and the hole has to couple simultaneously to
{\it both\/} magnons and orbitons when it moves by one lattice spacing.
It is then unclear whether a QP may form and which role is played
individually by spins and orbitals in possible formation of a
spin-orbital polaron.

The 2D spin-orbital $t$-$J$ model may be seen as a generalization of the
spin $t$-$J$ model \cite{Cha77} and the orbital $t_{2g}$ $t$-$J$ model
\cite{Dag08} to the spin-orbital superexchange --- it consists of
three terms \cite{Woh09},
\begin{equation}
\label{eq:str-coupl}
H=H_t+H_J+H_{3s},
\end{equation}
where the last one, $H_{3s}$, stands for the three-site effective hopping.
The second term $H_J$ is the spin-orbital superexchange model for the
$R$VO$_3$ perovskites \eref{som} introduced above in \sref{sec:optic}.
The first term in (\ref{eq:str-coupl}) describes the hopping of
$\{a,b\}$ electrons in the constrained Hilbert space, i.e., in the space
with singly occupied (at hole position) or doubly occupied sites. This
means that electrons in the $ab$ plane, which is under consideration
here, can hop only along the $b$ ($a$) direction when they carry $a$
($b$) orbital flavour.
The $c$ orbitals do not participate in hopping processes as they are
always occupied by one electron \eref{nabc}, see \cite{And07,Fuj06}.
Hence, we arrive at the kinetic $\propto t$ part of Hamiltonian
(\ref{eq:str-coupl}),
\begin{equation}\label{eq:ht}
H_t = -t \sum_{{\bf i},\sigma} {\cal P}\left( \tilde{b}^\dag_{{\bf i}\sigma}
\tilde{b}_{{\bf i}+\hat{\bf a}\sigma}^{}
+\tilde{a}^\dag_{{\bf i}\sigma}\tilde{a}_{{\bf i}+\hat{\bf b}\sigma}^{}
+ {\rm H.c.}\right) {\cal P}.
\end{equation}
Here the constrained operators
$\tilde{a}^\dag_{{\bf i}\sigma},\{\tilde{b}^\dag_{{\bf i}\sigma}\}$ mean
that the hopping is allowed only in the restricted Hilbert space with not
more than one $\{a,b\}$ electron at each site ${\bf i}$ ($\bar{\sigma}$
stands for the spin component opposite to $\sigma$). Besides, since Hund's
exchange coupling is large ($J_H\gg t$) \cite{Miz96}, we project the final
states resulting from the electron hopping onto the high-spin states,
which occurs due to the ${\cal P}$ operators in \eref{eq:ht}.
More details can be found in \cite{Woh09}.

Low energy excitations are magnons and orbitons, with their energies
determined by the respective exchange constants derived from the
spin-orbital superexchange \cite{Woh09},
\begin{eqnarray}\label{eq:js}
J_S &=& \frac{1-3 \eta-5 \eta^2}{4 (1-3\eta)(1+2\eta)}\,J\,, \\
\label{eq:jo}
J_O &=& \frac{\eta(2-\eta)}{(1-3\eta)(1+2\eta)}\,J\,.
\end{eqnarray}
One finds that $J_O>0$ and $J_S>0$ in the expected range of $\eta<0.2$,
which means that the classical GS has indeed coexisting AF and
AO order. By rotating first spins and orbitals at sublattice $A$ to the
FM/FO order and introducing Schwinger bosons $f^\dagger_{{\bf i}\sigma}$
for spin and $t^\dagger_{{\bf i}\alpha}$ for orbital excitations, one
finds the energies of magnons and orbitons using the LSW theory. As
usually in a quantum antiferromagnet, magnons are dispersive and have
a Goldstone mode,
\begin{equation}
\omega_{\bf k}=J_SzS\sqrt{1-\gamma^2_{\bf k}}\,,
\end{equation}
where $\gamma_{\bf k}$ is given by \eref{gamma}, $S=1$ and $z=4$ is the
coordination number.
In contrast, the orbital excitations at energy $J_O$ are local and have
no dispersion as the orbital interactions are Ising-like.

A crucial step in deriving the hole spectral function $A({\bf k},\omega)$
is the following representation of the electron operators in terms of the
$\{f^\dag_{{\bf i}\sigma},t^\dag_{{\bf i}\alpha}\}$ ($\alpha=a,b$)
Schwinger bosons:
\begin{eqnarray}
\tilde{a}^\dag_{{\bf i} \sigma} &=&\frac{1}{\sqrt{2}}
f^\dag_{{\bf i} \sigma} t^\dag_{{\bf i}a} h^{}_{\bf i}\,, \\
\tilde{b}^\dag_{{\bf i} \sigma} &=& \frac{1}{\sqrt{2}} f^\dag_{{\bf i}\sigma}
t^\dag_{{\bf i}b} h^{}_{\bf i}\,.
\end{eqnarray}
The above equations demonstrate that a spin excitation is always
generated together with an orbital excitation, which implies that the
diagrams contributing to the self-energy contain only vertices with two
outgoing or incoming excitation lines.
The factor $\frac{1}{\sqrt{2}}$ follows from spin algebra for spins
$S=1$ in the spin-orbital model \cite{Woh09}. Here the projection onto
the high-spin states has already been done, so the projection operators
${\cal P}$ in $H_t$ are no longer needed.

The spectral function can be obtained from the Green's functions which
are defined separately for ${\cal A}$ and ${\cal B}$ sublattice and depend on the
corresponding self-energy $\Sigma_{\alpha}({\bf k},\omega)$ ($\alpha=a,b$):
\begin{equation}
\label{eq:dyson1}
G_{\alpha}({\bf k}, \omega)=\frac{1}{\omega+ \varepsilon_{\alpha}({\bf k})
-\Sigma_{\alpha}({\bf k}, \omega)}\,.
\end{equation}
Here $\varepsilon_a({\bf k})=2\tau\cos (2 k_y)$ and
$\varepsilon_b({\bf k})=2\tau\cos (2 k_x)$, and $\tau=\frac14 J$
describes the three-site hopping responsible for the 1D dispersion
within each sublattice. In what follows we show only the result for
$\tau=0$ for more clarity. Green's functions are solved self-consistently
together with the self-energies derived in the self-consistent Born
approximation (SCBA). After solving the Green's functions, the spectral
functions for a hole created in $\alpha=\{a,b\}$ orbital are:
\begin{equation}
\label{aa}
A_\alpha({\bf k},\omega)=-\frac{1}{\pi}\lim_{\delta\to 0}\,\mbox{Im}\,
G_{\alpha}({\bf k},\omega+i\delta)\,.
\end{equation}
At $\tau=0$ the spectral function does not depend on the orbital flavour,
and we define $A({\bf k},\omega)\equiv A_\gamma({\bf k},\omega)$.

\begin{figure}[t!]
\begin{center}
\includegraphics[width=8cm]{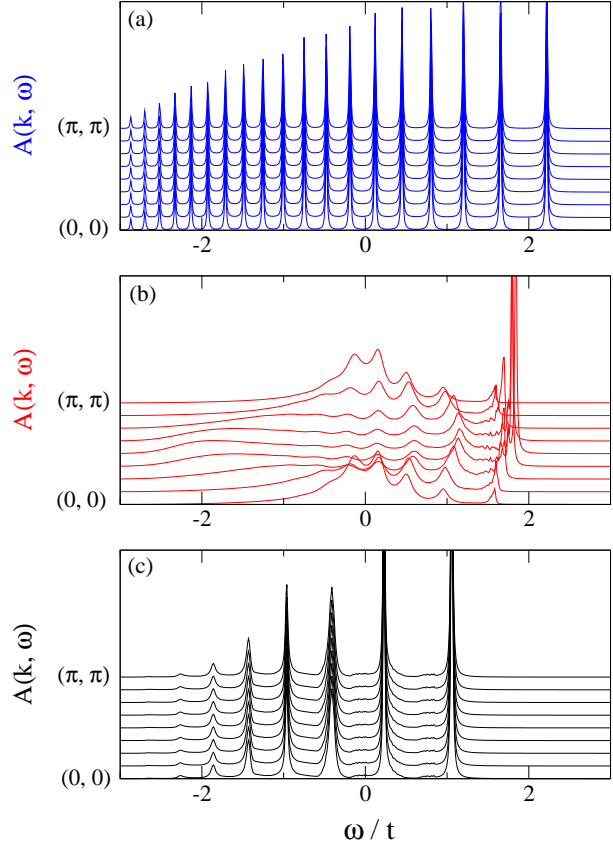}
\end{center}
\caption{
Spectral functions for a single hole in the AF/AO phase along the
$\Gamma-M$ direction of the Brillouin zone for:
(a) toy orbital model \eref{eq:orbitalmodel} (top);
(b) toy spin model \eref{eq:spinmodel} (middle);
(c) the full spin-orbital model \eref{som} (bottom).
Parameters: $J=0.4t$ and $\eta=0.15$ (i.e., $J_S = 0.06t$ and $J_O=0.16t$).
Broadening $\delta=0.01t$ in the definition \eref{aa} of the spectral
function is assumed. Image courtesy of Krzysztof Wohlfeld.
}
\label{fig:ak}
\end{figure}

The spectral functions $A({\bf k},\omega)$ \eref{aa} shown in
\fref{fig:ak} are representative for the situation in the strongly
correlated transition metal oxides. They are obtained by solving the SCBA
equations for $J=0.4t$ on a mesh of $16\times16$ ${\bf k}$-points.
Similar data for other values of $J=0.2t$ and $0.6t$ are presented in
\cite{Woh09}. A realistic value of Hund's exchange in LaVO$_3$ is
$\eta=0.15$ \cite{Ole07}. The peak in the low-energy
part of the spectrum, see \fref{fig:ak}(c), has no dispersion which
indicates hole confinement
(a rather weak 1D dispersion along the $k_x$ ($k_y$) direction for holes
doped into the $b$ ($a$) orbitals is obtained at finite $\tau=\frac14 J$).
The spin-orbital spectral functions form ladder-like spectra and
have many similarities with the spectra obtained for the $t_{2g}$ orbital
$t$-$J$ model \cite{Dag08}. However, there are few subtle differences
with the orbital model which demonstrate that the spin-orbital case is
more complex and the spin part also contributes.

To understand better the nature of the obtained spectra we present also
the spectral functions obtained for the related spin and orbital problem
in \fref{fig:ak}(a-b). These two models read as follows \cite{Woh09}:
\begin{eqnarray}
\label{eq:spinmodel}
H_{S}&=&-t \sum_{\langle {\bf i}{\bf j}\rangle, \sigma}
{\cal P}( \tilde{c}^\dag_{{\bf i}\sigma} \tilde{c}_{{\bf j}\sigma} + {\rm
H.c.}){\cal P}+ J_S \sum_{\langle {\bf i}{\bf j}\rangle} {\bf S}_{\bf i}\cdot{\bf
S}_{\bf j}\,, \\
\label{eq:orbitalmodel}
H_{O}&=&-t \sum_{\bf i}\left(
\tilde{b}^\dag_{\bf i} \tilde{b}_{{\bf i}+\hat{\bf a}}
+\tilde{a}^\dag_{{\bf i}}\tilde{a}_{{\bf i}+\hat{\bf b}} + {\rm H.c.}\right)
+ J_O \sum_{\langle {\bf i}{\bf j} \rangle}T^z_{\bf i} T^z_{\bf j}.
\end{eqnarray}
Here spin operators $\{{\bf S}_{\bf i}\}$ stand for $S=1$ spins, $T^z_i$
are $z$th components of pseudospin $T=1/2$, and the operators ${\cal P}$
project onto the high-spin states. The constrained electron operators,
$\tilde{c}^\dag_{{\bf i}\sigma}=c^\dag_{{\bf i}\sigma}(1-n_{{\bf i}\bar{\sigma}})$
in \eref{eq:spinmodel} and
$\tilde{b}^\dag_{\bf i}=b^\dag_{\bf i}(1-n_{{\bf i}a})$ and
$\tilde{a}^\dag_{\bf i}=a^\dag_{\bf i}(1-n_{{\bf i}b})$  in
\eref{eq:orbitalmodel} exclude double occupancies from the Hilbert space
in each case, similar to \eref{eq:ht}. The superexchange energy scale is
$J_S$ \eref{eq:js} for the spin model \eref{eq:spinmodel} and $J_O$
\eref{eq:jo} for the orbital one \eref{eq:orbitalmodel}, which mimics
the influence of the orbital part on the spins and of the spin part
(with the AF order) on the orbitals. With the present parameters one
finds the spin-only exchange constant $J_S=0.06t$ (somewhat higher than
that deduced from the observed value of the N\'eel temperature
$T_N\simeq 143$ K in LaVO$_3$ \cite{Miy06}), and $J_O=0.16t$.

The spectral functions for the above models were obtained using the SCBA
on a mesh of $16\times16$ ${\bf k}$-points,
following the derivations presented in \cite{Mar91} and \cite{Woh08}
in the case of the spin and orbital model. One only has to make the
following substitutions in the respective SCBA equations: $J\rightarrow -J_S$,
$S\rightarrow 1$ and (due to the quantum double exchange factor) also
$t\rightarrow t/\sqrt{2}$ in the spin case \cite{Mar91}, and
$J\rightarrow -J_O$, $E_0\rightarrow 0$, and $\tau\rightarrow 0$
in the orbital case \cite{Woh08}. The spectral functions obtained for
the above toy models are quite different from those obtained for the
spin-orbital $t$-$J$ model. The spectral
function for the spin model, see \fref{fig:ak}(b), consists of a QP and
the incoherent part at higher energies, both having considerable
${\bf k}$-dependence. It is evident that this ${\bf k}$-dependence is
removed by the orbital interactions. The orbital model at $\tau=0$ has
a strictly localized spectral function with a ladder spectrum \cite{Mar91},
see \fref{fig:ak}(a). The spin-orbital spectral function, see
\fref{fig:ak}(c), resembles qualitatively the ladder spectrum found for
the orbital model in \fref{fig:ak}(a). The momentum dependence of both
the QP state and the incoherent part of the spectrum obtained for the
spin model, see \fref{fig:ak}(b), is entirely suppressed. These results
demonstrate that the quantum spin fluctuations are to a large extent
quenched in the spin-orbital model by the simultaneous coupling of the
hole to {\it both\/} spin and orbital excitations. It is remarkable that
the hole still couples to the spin degrees of freedom by generating
string potential due to defects created by hole motion. Thus, {\it the
string} which acts on the hole moving in the plane with AF/AO
order {\it is of the composite spin-orbital character}. This
not only explains the peculiar correspondence between the orbital and
spin-orbital model, but also shows that the spins play
an active role in the lightly doped spin-orbital system.

We remark that the orbitally induced string formation considered here
could be understood as a topological effect. It happens even if the
orbital excitation energy is turned to zero, i.e., when the hole moves
in the orbital sector incoherently. Hence, the mere presence of
orbitals is sufficient to obtain the (almost) classical behaviour of a
hole doped into the GS with AF/AO order. This result suggests that
further investigation of the hole propagation in spin-orbital systems is
a fascinating subject for future studies.

\section{Discussion and summary}
\label{sec:summa}

Spin-orbital entanglement discussed in this topical review concerns
entanglement on the exchange bonds, similar to entanglement in e.g. spin
singlets which build VB states \cite{Nor09,Bal10}. We have shown that
this concept is important is several spin-orbital models and in general
either the ground state or excited states are entangled. Two recent
examples of entangled ground states were presented: the Kugel-Khomskii
model on a bilayer, and the spin-orbital $d^1$ model on the triangular
lattice, in addition to the well known 1D SU(2)$\otimes$SU(2) model.
It may be expected that more spin-orbital model systems with entangled
ground states will be found in the near future. Whether or not such
states are indeed realized in nature depends on the coupling to the
lattice. For instance, the $e_g$ orbital order is robust in LaMnO$_3$
\cite{Fei99} and KCuF$_3$ \cite{Lee12}, and for this reason the ground
states of these compounds are disentangled.

We have also demonstrated that the Goodenough-Kanamori rules are
violated in the regime of weak Hund's exchange in several situations. It
is in this regime that the mean field decoupling procedure of spin and
orbital operators fails and the magnetic properties can be determined
only by solving the full entangled spin-orbital many-body problem. Also
for weak (or vanishing) Hund's exchange {\it spin-orbital liquid\/}
phase is stabilized by spin-orbital entanglement in the $d^1$
spin-orbital model on the triangular lattice.

However, a frustrated lattice does not guarantee that a disordered
spin-orbital liquid-like state arises. A good counterexample is the
frustrated lattice of alkali $R$O$_2$ hyperoxides (with $R$=K, Rb, Cs),
where the interactions induced by the lattices compete with the
superexchange and stabilize the orbital order, lifting the geometric
frustration of the lattice \cite{Woh11}. Then the spin-orbital
interactions may be consideed as disentangled, and the
Goodenough-Kanamori rules require certain reinterpretation. In fact,
they have been generalized by including large interorbital hopping
terms \cite{Woh11}.

It has been shown that ground states which contradict the celebrated
Goodenough-Kanamori rules may appear in two different situations:
($i$) either when quantum fluctuations in the spin-orbital system are
strong and stabilize the entangled ground state with coexisting AF and
AO order, or
($ii$) when the AF superexchange interactions follow from charge
excitations in other orbitals not involved the actual orbital order, or
($iii$) when the coupling to the lattice stabilizes the AO order in the
regime where the AF spin correlations are expected due to superexchange.
In these latter situation
the ground state is more classical and not entangled as it follows from
interactions induced by the lattice and not from the spin-orbital
superexchange. Also in the regime of large Hund's exchange $J_H$, ground
states with coexisting FM and FO order may appear \cite{Bog10}, similar
to the FM/FO states in the SU(2)$\times$SU(2) model. They contradict
again the Goodenough-Kanamori rules, but are in fact disentangled ---
here quantum fluctuations are absent and play no role for their stability.

On the example of the $R$VO$_3$ perovskites we have shown that the
experimental data in this unique family of correlated oxides indicate
that low-energy excited states are entangled and the energies of spin
and orbital excitations are similar. This happens because lattice
distortions and interactions of $t_{2g}$ orbitals with Jahn-Teller modes
are weak and only the electronic interactions such as superexchange
decide about the system behaviour. In
the $R$VO$_3$ systems the temperature dependence of the optical spectral
weight, the phase diagram as a function of the ionic radius of $R$ ions,
and the dimerized magnon excitations, are all determined by the
presence of entangled states with low excitation energies.

The interplay between spin, orbital and spin-orbital excitations poses
a very interesting problem for future theoretical studies as well as
an experimental challenge. We also point out that in case of FO order
accompanied by AF interactions, spin fluctuations couple to orbital
excitations and thus they cannot be considered separately \cite{WoDa},
in spite of formal separation of spin and orbital degrees of freedom in
the ground state. Another important finding is recent observation that
composite spin-orbital excitations fractionalize in the quasi-1D Mott
insulator Sr$_2$CuO$_3$ \cite{Sch12}. The nature of spin-orbital
excitations and the circumstances of their possible decay are very
challenging and unresolved questions. The case considered here of a hole
moving in the AF/AO order \cite{Woh09} seems to suggest that, at least
in doped systems, spin-orbital excitations may play a very important
role in transport as they may not decay and impose some topological
constraints on carrier propagation.

The present topical review focused on spin-orbital entanglement on the
superexchange bonds, but a different kind of entanglement arises in
presence of on-site relativistic spin-orbit interaction. In this case
a Kramers doublet gives the lowest energy states of a single ion and
these states are next considered to derive interactions between
neighbouring ions \cite{Jac09}. Such local on-site entangled states play
an important role and decide about the magnetic properties of
Sr$_2$IrO$_4$ with weak ferromagnetism and Sr$_2$VO$_4$ with hidden
spin-orbital order. In a 1D superexchange model with Ising-like orbital
superexchange locally entangled states introduce orbital dynamics and lead
to a phase diagram with a novel phase having long-range N\'eel order of
spin and orbital angular momenta \cite{Ner11}. Excitations is such models are
a challenge and they are under investigation at present. If localized states
with an effective angular momentum ${\vec J}_i=2{\vec S}_i-{\vec L}_i$ at
site $i$ (with ${\vec S}$ and ${\vec L}$ being spin and orbital operators)
are considered, one finds a gapped spectrum at finite spin-orbit coupling
\cite{Che09}.

Finally, we would like to emphasize that a better understanding of the
concept of spin-orbital entanglement is important and could help to make 
progress in other fields as this subject is interdisciplinary. 
Recently a scheme to generate spin-orbit-path hybrid
Greenberg-Horne-Zeilinger entanglement \cite{Gre90} was proposed
\cite{She10} for photons which are entangled with different degrees of
freedom. In quantum chemistry entanglement is considered in chemical 
bonds which are classified using measures of electron correlation 
\cite{Ole86} and entanglement \cite{Alc10}. Spin-orbital-like
entanglement is also applicable to nuclear systems, where nucleons
possess as well two degrees of freedom --- spin and isospin.
It has been established that the entanglement length of the nucleons is
significantly larger than that one expects \cite{Chen}. Entanglement may
also play an important role in quantum computations if the spin state
used for information storage would be measured by investigating orbital
qubits in entangled states \cite{Ion05}.

\ack

It is a great pleasure to thank particularly
Lou-Fe' Feiner,
Peter Horsch,
Giniyat Khaliullin and
Jan Zaanen
for a very friendly collaboration over many years which significantly
contributed to my present understanding of the subject. I thank also
all other collaborators on specific projects for insightful discussions:
A Avella,
W Brzezicki,
J Chaloupka,
L Cincio,
M Daghofer,
J Dziarmaga,
R Fr\'esard,
A Herzog,
B Keimer,
B Normand,
K Parlinski,
K Ro\'sciszewski,
G A Sawatzky,
J Sirker,
F Trousselet,
K Wohlfeld and
W-L You.
Kind help of Krzysztof Wohlfeld on preparing \fref{fig:ak} is warmly
acknowledged.
We acknowledge financial support by the Foundation for Polish Science
(FNP) and by the Polish National Science Center (NCN) under Project
No. N202 069639.


\section*{References}

\begin{thebibliography}{199}

\bibitem{Ole06} Ole\'s A M, Horsch P, Feiner L F and Khaliullin G 2006
                   {\it Phys. Rev. Lett.\/} \textbf{96} 147205

\bibitem{Nie00} Nielsen M A and Huang I L 2000
                   {\it Quantum Computation and Information\/}
                   (Cambridge University Press, Cambridge, England)

\bibitem{Ben06} Bengtsson I and \.Zyczkowski K 2006
                   {\it Geometry of Quantum States --- An Introduction
                        to Quantum Entanglement\/}
                   (Cambridge University Press, Cambridge, England)

\bibitem{Hor09} Horodecki R, Horodecki P, Horodecki M and Horodecki K 2009
                   {\it Rev. Mod. Phys.\/} \textbf{81} 865

\bibitem{Ami08} Amico L, Fazio R, Osterloh A and Vedral V 2008
                   {\it Rev. Mod. Phys.\/} \textbf{80} 517

\bibitem{Ami09} Amico L and Fazio R 2009
                   {\it J. Phys. A: Math. Theor.\/} \textbf{42} 504001 \\
                Latorre J I and Riera A 2009
                   {\it J. Phys. A: Math. Theor.\/} \textbf{42} 504002

\bibitem{Pes09} Peschel I and Eisler V 2009
                   {\it J. Phys. A: Math. Theor.\/} \textbf{42} 504003

\bibitem{Blo08} Bloch I 2008
                   {\it Nature\/} \textbf{453} 1016

\bibitem{Aff09} Affleck I, Laflorencie N and Sorensen E 2009
                   {\it J. Phys. A: Math. Theor.\/} \textbf{42} 504009

\bibitem{Kug82} Kugel K I and Khomskii D I 1982
                   {\it Sov. Phys. Usp.\/} \textbf{25} 231

\bibitem{Ole09} Ole\'s A M 2009
                   {\it Acta Phys. Polon.} A \textbf{115} 36

\bibitem{Fei99} Feiner L F and Ole\'s A M 1999
                   {\it Phys. Rev.} B \textbf{59} 3295

\bibitem{Ole05} Ole\'s A M, Khaliullin G, Horsch P and Feiner L F 2005
                   {\it Phys. Rev.\/} B \textbf{72} 214431

\bibitem{Goode} Goodenough J B 1963
                   {\it Magnetism and the Chemical Bond\/}
                   (Interscience, New York) \\
                Kanamori J 1959
                   {\it J. Phys. Chem. Solids\/} \textbf{10} 87

\bibitem{Dag01} Dagotto E, Hotta T and Moreo A 2001
                   {\it Phys. Rep.\/} \textbf{344} 1 \\
                Dagotto E 2005 {\it New J. Phys.\/} \textbf{7} 67

\bibitem{Kov10} Kovaleva N N, Ole\'s A M, Balbashov A M, Maljuk A,
                   Argyriou D N, Khaliullin G and Keimer B 2010
                   {\it Phys. Rev.} B \textbf{81} 235130

\bibitem{Feh04} Wei\ss{}e A and Fehske H 2004
                  {\it New J. Phys.\/} \textbf{6} 158

\bibitem{Dag06} Daghofer M, Ole\'s A M, Neuber D M and von der Linden W 2006
                  {\it Phys. Rev.\/} B \textbf{73} 104451 \\
                Daghofer M and Ole\'s A M 2007
                  {\it Acta Phys. Polon.\/} A \textbf{111} 497

\bibitem{Ros07} Ro\'sciszewski K and Ole\'s A M 2007
                   {\it J. Phys.: Condensed Matter} \textbf{19} 186223 \\
                Ro\'sciszewski K and Ole\'s A M 2008
                   {\it J. Phys.: Condensed Matter} \textbf{20} 365212 \\
                Ro\'sciszewski K and Ole\'s A M 2010
                   {\it J. Phys.: Condensed Matter} \textbf{22} 425601

\bibitem{Tok06} Tokura Y 2006
                   {\it Rep. Prog. Phys.\/} \textbf{69} 797

\bibitem{Miy06} Miyasaka S, Okimoto Y, Iwama M and Tokura Y 2003
                   {\it Phys. Rev.\/} B \textbf{68} 100406 \\
                Miyasaka S, Fujioka J, Iwama M, Okimoto Y and Tokura Y 2006
                   {\it Phys. Rev.\/} B \textbf{73} 224436 \\
                Fujioka J, Yasue T, Miyasaka S, Yamasaki Y, Arima T,
                   Sagayama H, Inami T, Ishii K and Tokura Y 2010
                   {\it Phys. Rev.} B \textbf{82} 144425

\bibitem{Fei97} Feiner L F, Ole\'s A M and Zaanen J 1997
                   {\it Phys. Rev. Lett.\/} \textbf{78} 2799

\bibitem{Fei98} Feiner L F, Ole\'s A M and Zaanen J 1998
                   {\it J. Phys.: Condens. Matter\/} \textbf{10} L555

\bibitem{Kha97} Khaliullin G and Oudovenko V 1997
                   {\it Phys. Rev.\/} B \textbf{56} R14243

\bibitem{Miy02} Miyasaka S, Okimoto Y and Tokura Y 2002
                   {\it J. Phys. Soc. Jpn.\/} \textbf{71} 2086

\bibitem{Ulr03} Ulrich C, Khaliullin G, Sirker J, Reehuis M, Ohl M,
                   Miyasaka S, Tokura Y and Keimer B 2003
                   {\it Phys. Rev. Lett.\/} \textbf{91} 257202

\bibitem{Ole10} Ole\'s A M 2010
                   {\it Acta Phys. Polon.} A \textbf{118} 212

\bibitem{Mar91} Mart\'inez G and Horsch P 1991
                   {\it Phys. Rev.\/} B \textbf{44} 317

\bibitem{Zaa92} Zaanen J, Ole\'s A M and Horsch P 1992
                   {\it Phys. Rev.\/} B \textbf{46} 5798

\bibitem{vdB00} van den Brink J, Horsch P and Ole\'s A M 2000
                   {\it Phys. Rev. Lett.\/} \textbf{85} 5174

\bibitem{Dag08} Daghofer M, Wohlfeld K, Ole\'s A M, Arrigoni E
                   and Horsch P 2008
                   {\it Phys. Rev. Lett.\/} \textbf{100} 066403

\bibitem{Woh08} Wohlfeld K, Daghofer M, Ole\'s A M and Horsch P 2008
                   {\it Phys. Rev.\/} B \textbf{78} 214423

\bibitem{Bal01} Ba\l{}a J, Sawatzky G A, Ole\'s A M and Macridin A 1991
                   {\it Phys. Rev. Lett.\/} \textbf{87} 067204

\bibitem{Woh09} Wohlfeld K, Ole\'s A M and Horsch P 2009
                   {\it Phys. Rev.\/} B \textbf{79} 224433

\bibitem{Kha04} Khaliullin G, Horsch P and Ole\'s A M 2004
                   {\it Phys. Rev.\/} B \textbf{70} 195103

\bibitem{Brz11} Brzezicki W and Ole\'s A M 2011
                   {\it Phys. Rev.\/} B \textbf{83} 214408 \\
                Brzezicki W and Ole\'s A M 2012
                   {\it Acta Phys. Polon.\/} A \textbf{121} 1045

\bibitem{Nor08} Normand B and Ole\'s A M 2008
                   {\it Phys. Rev.\/} B \textbf{78} 094427

\bibitem{Cha11} Chaloupka J and Ole\'s A M 2011
                   {\it Phys. Rev.\/} B \textbf{83} 094406

\bibitem{Nor09} Normand B 2009
                   {\it Cont. Phys.\/} \textbf{50} 533

\bibitem{Bal10} Balents L 2010
                   {\it Nature\/} {\bf 464} 199

\bibitem{Lon80} Longa L and Ole\'s A M 1980
                   {\it J. Phys. A: Math. Theor.\/} \textbf{13} 1031

\bibitem{vdB04} van den Brink J 2004
                   {\it New J. Phys.\/} \textbf{6} 201

\bibitem{vdB99} van~den Brink J, Mack F, Horsch P and Ole\'s A M 1999
                   {\it Phys. Rev.\/} B \textbf{59} 6795

\bibitem{Dag04} Daghofer M, von der Linden W and Ole\'s A M 2004
                   {\it Phys. Rev.\/} B \textbf{70} 184430

\bibitem{Nus04} Khomskii D I and Mostovoy M V 2003
                   {\it J. Phys. A: Math. Theor.\/} \textbf{36} 9197 \\
                Nussinov Z, Biskup M, Chayes L and van den Brink J
                   {\it Europhys. Lett.\/} \textbf{67} 990

\bibitem{Cin10} Cincio L, Dziarmaga J and Ole\'s A M 2010
                   {\it Phys. Rev.\/} B \textbf{82} 104416

\bibitem{Brz09} Brzezicki W and Ole\'s A M 2009
                   {\it Phys. Rev.\/} B \textbf{80} 014405

\bibitem{Mil05} Dorier J, Becca F and Mila F 2005
                   {\it Phys. Rev.\/} B \textbf{72} 024448

\bibitem{Brz07} Brzezicki W, Dziarmaga J and Ole\'s A M 2007
                   {\it Phys. Rev.\/} B \textbf{75} 134415 \\
                Brzezicki W and Ole\'s A M 2008
                   {\it Acta Phys. Polon.\/} A \textbf{115} 162

\bibitem{Wen08} Wenzel S and Janke W 2008
                   {\it Phys. Rev.\/} B \textbf{78} 064402

\bibitem{Brz10} Brzezicki W and Ole\'s A M 2010
                   {\it Phys. Rev.\/} B \textbf{82} 060401

\bibitem{Tro10} Trousselet F, Ole\'s A M and Horsch P 2010
                   {\it Europhys. Lett.\/} \textbf{91} 40005

\bibitem{Dou05} Dou\c{c}ot B, Feigel'man M V, Ioffe L B
                   and Ioselevich A S 2005
                   {\it Phys. Rev.\/} B \textbf{71} 024505

\bibitem{Gla09} Gladchenko S, Olaya D, Dupont-Ferrier E, Dou\c{c}ot B,
                   Ioffe L B and Gershenson M E 2009
                   {\it Nature Physics\/} \textbf{5} 48

\bibitem{Ryn10} van Rynbach A, Todo S and Trebst S 2010
                   {\it Phys. Rev. Lett.\/} \textbf{105} 146402

\bibitem{Wen11} Wenzel S and L\"auchli A M 2011
                   {\it Phys. Rev. Lett.\/} \textbf{106} 197201

\bibitem{Fei05} Feiner L F and Ole\'s A M 2005
                   {\it Phys. Rev.\/} B \textbf{71} 144422

\bibitem{Ole02} Ole\'s A M and Feiner L F 2002
                   {\it Phys. Rev.\/} B \textbf{65} 052414

\bibitem{Oit11} Oitmaa J and Hamer C J 2011
                   {\it Phys. Rev.\/} B \textbf{83} 094437

\bibitem{Kha00} Khaliullin G and Maekawa S 2000
                   {\it Phys. Rev. Lett.\/} \textbf{85} 3950

\bibitem{Khali} Khaliullin G 2001
                   {\it Phys. Rev.\/} B \textbf{64} 212405

\bibitem{Kha02} Khaliullin G and Okamoto S 2002
                   {\it Phys. Rev. Lett.\/} \textbf{89} 167201 \\
                Khaliullin G and Okamoto S 2003
                   {\it Phys. Rev.\/} B \textbf{68} 205109

\bibitem{Ima98} Imada M, Fujimori A and Tokura Y 1998
                   {\it Rev. Mod. Phys.\/} \textbf{70} 1039

\bibitem{Cha77} Chao K A, Spa\l{}ek J and Ole\'s A M 1977
                   {\em J. Phys.\/} C \textbf{10} L271 \\
                Chao K A, Spa\l{}ek J and Ole\'s A M 1978
                   {\it Phys. Rev.\/} B \textbf{18} 3453

\bibitem{Gri71} Griffith J S 1971
                   {\it The Theory of Transition Metal Ions\/}
                   (Cambridge University Press, Cambridge, England)

\bibitem{Zaa93} Zaanen J and Ole\'s A M 1993
                   {\it Phys. Rev.\/} B \textbf{48} 7197

\bibitem{Kha01} Khaliullin G, Horsch P and Ole\'s A M 2001
                   {\it Phys. Rev. Lett.\/} \textbf{86} 3879

\bibitem{Ole83} Ole\'s A M 1983
                   {\it Phys. Rev.} B \textbf{28} 327

\bibitem{Bae86} Baeriswyl D, Carmelo J and Luther A 1986
                   {\it Phys. Rev.\/} B \textbf{33} 7247 \\
                Aichhorn M, Horsch P, von der Linden W and Cuoco M 2002
                   {\it Phys. Rev.\/} B \textbf{65} 201101

\bibitem{Ole00} Ole\'s A M, Feiner L F and Zaanen J 2000
                   {\it Phys. Rev.\/} \textbf{61} 6257

\bibitem{Ole07} Ole\'s A M, Horsch P and Khaliullin G 2007
                   {\it Phys. Rev.\/} B \textbf{75} 184434

\bibitem{Aff00} Itoi C, Qin S and Affleck I 2000
                   {\it Phys. Rev.\/} B \textbf{61} 6747

\bibitem{Bal99} van den Brink J, Stekelenburg W, Khomskii D I,
                    Sawatzky G A and Kugel K I 1988
                   {\it Phys. Rev.\/} B \textbf{58} 10276 \\
                 Ba\l{}a J, Ole\'s A M and Sawatzky G A 2001
                   {\it Phys. Rev.\/} B \textbf{63} 134410

\bibitem{Her11} Herzog A, Horsch P, Ole\'s A M and Sirker J 2011
                   {\it Phys. Rev.\/} B \textbf{83} 245130

\bibitem{Fri99} Frischmuth B, Mila F and Troyer M 1999
                   {\it Phys. Rev. Lett.\/} \textbf{82} 835

\bibitem{Pss07} Ole\'s A M, Horsch P and Khaliullin G 2007
                   {\it Phys. Stat. Solidi\/} (b) \textbf{244} 2378

\bibitem{Maj69} Majumdar C K and Ghosh D K 1969
                   {\it J. Math. Phys.\/} \textbf{10} 1388

\bibitem{Chen}  Chen D, Wang W and Zou L-J 2010
                   {\it Phys. Lett.\/} A \textbf{374} 1393

\bibitem{Che07} Chen Y, Wang Z D, Li Y D and Zhang F C 2007
                   {\it Phys. Rev.\/} B \textbf{75} 195113

\bibitem{And07} De Raychaudhury, Pavarini E and Andersen O K 2007
                  {\it Phys. Rev. Lett.\/} \textbf{99} 126402

\bibitem{Hor08} Horsch P, Ole\'s A M, Khaliullin G and Feiner L F 2008
                   {\it Phys. Rev. Lett.\/} \textbf{100} 147205

\bibitem{Vid07} Vidal G 2007
                  {\it Phys. Rev. Lett.\/} \textbf{99} 220405 \\
                Vidal G 2008
                  {\it Phys. Rev. Lett.\/} \textbf{101} 110501 \\
                Cincio L, Dziarmaga J and Rams M M 2008
                   {\it Phys. Rev. Lett.\/} \textbf{100} 240603

\bibitem{Hor03} Horsch P, Khaliullin G and Ole\'s A M 2003
                   {\it Phys. Rev. Lett.\/} \textbf{91} 257203

\bibitem{Goo06} Zhou J-S and Goodenough J B 2006
                   {\it Phys. Rev. Lett.\/} \textbf{96} 247202

\bibitem{Pav05} Pavarini E, Yamasaki A, Nuss J and Andersen O K 2005
                   {\it New J. Phys.\/} \textbf{7} 188

\bibitem{Ree06} Reehuis M, Ulrich C, Pattison P, Ouladdiaf B, Rheinst\"adter M C,
                   Ohl M, Regnault L P, Miyasaka M, Tokura Y and Keimer B 2006
                   {\it Phys. Rev.\/} B \textbf{73} 094440

\bibitem{Sag06} Sage M H, Blake G R and Palstra T T M 2006
                   {\it Phys. Rev. Lett.\/} \textbf{96} 036401 \\
                  Sage M H, Blake G R and Palstra T T M 2008
                   {\it Phys. Rev.\/} B \textbf{77} 155121

\bibitem{Sil03} Da Silva T N, Joshi A, Ma M and Zhang F C 2003
                   {\it Phys. Rev.\/} B \textbf{68} 184402

\bibitem{God07} Yan J-Q, Zhou J-S, Goodenough J B, Ren Y, Cheng J G,
                  Chang S, Zarestky J, Garlea O, Llobet A, Zhou H D,
                  Sui Y, Su W H and McQueeney R J 2007
                  {\it Phys. Rev. Lett.\/} \textbf{99} 197201

\bibitem{Ren00} Ren Y, Palstra T T M, Khomskii D I, Nugroho A A,
                   Menovsky A A and Sawatzky G A 2000
                   {\it Phys. Rev.\/} B \textbf{62} 6577

\bibitem{Rac02} Raczkowski M and Ole\'s A M 2002
                   {\it Phys. Rev.\/} B \textbf{66} 094431

\bibitem{Sir08} Sirker J, Herzog A, Ole\'s A M and Horsch P 2008
                   {\it Phys. Rev. Lett.\/} \textbf{101} 157204

\bibitem{Joh00} Johnston D C, Kremer R K, Troyer M, Wang X, Kl\"umper A,
                   Bud'ko S L, Panchula A F and Canfield P C 2000
                   {\it Phys. Rev.\/} B \textbf{61} 9558

\bibitem{She02} Shen S Q, Xie X C and Zhang F C 2002
                   {\it Phys. Rev. Lett.\/} \textbf{88} 027201

\bibitem{Sir02} Sirker J and Kl\"umper A 2002
                   {\it Europhys. Lett.\/} \textbf{60} 262

\bibitem{Tak86} Takahashi M 1986
                   {\it Prog. Theor. Phys. Suppl.\/} \textbf{87} 233

\bibitem{Sir03} Sirker J and Khaliullin G 2003
                   {\it Phys. Rev.\/} B \textbf{67} 100408

\bibitem{Cuc02} Caciuffo R, Paolasini L, Sollier A, Ghigna P, Pavarini E,
                   van den Brink J and Altarelli M 2002
                   {\it Phys. Rev.\/} B \textbf{65} 174425 \\
                Binggeli N and  Altarelli M 2005
                   {\it Phys. Rev.\/} B \textbf{70} 085117 \\
                Deisenhofer J, Leonov I, Eremin M V, Kant Ch, Ghigna P,
                   Mayr F, Iglamov V V, Anisimov V I and van der Marel D 2008
                   {\it Phys. Rev. Lett.\/} \textbf{101} 157406

\bibitem{Lee12} Lee J C T, Yuan S, Lal S, Joe Y II, Gan Y, Smadici S,
                   Finkelstein K, Feng Y, Rusydi A, Goldbart P M,
                   Cooper S L and Abbamonte P M 2012
                   {\it Nature Phys.\/} \textbf{8} 63

\bibitem{Kha05} Khaliullin G 2005
                   {\it Prog. Thepr. Phys. Suppl.\/} \textbf{160} 155

\bibitem{vdB11} van den Brink J, Nussinov Z and Ole\'s A M 2011
                   in: {\it Introduction to Frustrated Magnetism: Materials,
                   Experiments, Theory\/} edited by Lacroix C, Mendels P
                   and Mila F, Springer Series in Solid-State Sciences
                   Vol. \textbf{164} (Springer, New York) pp 629-670

\bibitem{Jac07} Jackeli G and Ivanov D A 2007
                   {\it Phys. Rev.\/} B \textbf{76},132407

\bibitem{PM11}  Ole\'s A M and Chaloupka J 2012
                   {\it Acta Phys. Polon.\/} A \textbf{121} 1026

\bibitem{Ver04} Vernay F, Penc K, Fazekas P and Mila F 2004
                   {\it Phys. Rev.\/} B \textbf{70} 014428

\bibitem{Mos02} Mostovoy M V and Khomskii D I 2002
                   {\it Phys. Rev. Lett.\/} \textbf{89} 227203

\bibitem{Rei05} Reitsma A J W, Feiner L F and Ole\'s A M 2005
                   {\it New J. Phys.\/} \textbf{7} 121

\bibitem{Ere11} Eremin M V, Deisenhofer J, Eremina R M, Teyssier J,
                   van der Marel D and Loidl A 2011
                   {\it Phys. Rev.\/} B \textbf{84} 212407

\bibitem{Kan89} Schmitt-Rink S, Varma C M and Ruckenstein A E 1988
                   {\it Phys. Rev. Lett.} \textbf{60} 2793 \\
                Kane C L, Lee P A and Read N 1989
                   {\it Phys. Rev.\/} B \textbf{39} 6880 \\
                Brunner B, Assaad F F and Muramatsu A 2000
                   {\it Phys. Rev.\/} B \textbf{62} 15480  \\
                Bejas M, Greco A and Foussats A 2006
                   {\it Phys. Rev.\/} B \textbf{73} 245104

\bibitem{Dam03} Nazarenko A, Vos K J E, Haas S, Dagotto E and Gooding R J 1995
                   {\it Phys. Rev.\/} B \textbf{51} 8676 \\
                Bala J, Ole\'s A M and Zaanen J 1995
                   {\it Phys. Rev.\/} B \textbf{52} 4597  \\
                Damascelli A, Hussain Z and Shen Z-X 2003
                   {\it Rev. Mod. Phys.\/} \textbf{75} 473

\bibitem{Kil99} Kilian R and Khaliullin G 1999
                   {\it Phys. Rev.\/} B \textbf{60} 13458

\bibitem{Wro08} Wr\'obel P, Suleja W and Eder R 2008
                   {\it Phys. Rev.\/} B \textbf{78} 064501

\bibitem{Wro10} Wr\'obel P and Ole\'s A M 2010
                   {\it Phys. Rev. Lett.} \textbf{104} 206401

\bibitem{Hor11} Horsch P and Ole\'s A M 2011
                   {\it Phys. Rev.\/} B \textbf{84} 064429

\bibitem{Fuj06} Fujioka J, Miyasaka S and Tokura Y 2006
                   {\it Phys. Rev. Lett.}  \textbf{97} 196401

\bibitem{Miz96} Mizokawa T and Fujimori A 1996
                   {\it Phys. Rev.\/} B \textbf{54} 5368

\bibitem{Woh11} Wohlfeld K, Daghofer M and Ole\'s A M 2011
                   {\it Europhys. Lett.\/} {\bf 96} 27001

\bibitem{Bog10} Bogdanski P, Halaoui M, Ole\'s A M and Fr\'esard R 2010
                   {\it Phys. Rev.\/} B \textbf{82} 195125

\bibitem{WoDa}  Wohlfeld K, Daghofer M, Nishimoto S, Khaliullin G
                   and van den Brink J 2011
                   {\it Phys. Rev. Lett.\/} \textbf{107} 147201

\bibitem{Sch12} Schlappa J, Wohlfeld K, Zhou K J, Mourigal M, Haverkort M W,
                   Strocov V N, Hozoi L, Monney C, Nishimoto S, Singh S,
                   Revcolevschi A, Caux J-S, Patthey L, R\o{}nnow H M,
                   van den Brink J and Schmitt T 2012
                   {\it Nature\/} \textbf{485} 82

\bibitem{Jac09} Jackeli G and Khaliullin G 2009
                   {\it Phys. Rev. Lett.} \textbf{102} 017205 \\
                Jackeli G and Khaliullin G 2009
                   {\it Phys. Rev. Lett.} \textbf{103} 067205 \\
                Chaloupka J, Jackeli G and Khaliullin G 2010
                   {\it Phys. Rev. Lett.} \textbf{105} 027204 \\
                Ament L J P, Khaliullin G and van den Brink J 2011
                   {\it Phys. Rev.\/} B \textbf{84} 020403

\bibitem{Ner11} Chern G-W, Perkins N B and Japaridze G I 2010
                   {\it Phys. Rev.\/} B \textbf{82} 085106 \\
                Nersesyan A, Chern G-W and Perkins N B 2011
                   {\it Phys. Rev.\/} B \textbf{83} 205132

\bibitem{Che09} Chern G-W and Perkins N B 2009
                   {\it Phys. Rev.\/} B \textbf{80} 180409

\bibitem{Gre90} Greenberger D M, Horne M A, Shimony A and Zeilinger A 1990
                   {\it Am. J. Phys.\/} \textbf{58} 1131

\bibitem{She10} Chen L and She W 2010
                   {\it Phys. Rev.\/} A \textbf{83} 032305

\bibitem{Ole86} Ole\'s A M, Pfirsch F, Fulde P and B\"ohm M C 1986
                   {\it J. Chem. Phys.\/} \textbf{85} 5183 \\
                Ole\'s A M, Pfirsch F, Fulde P and B\"ohm M C 1987
                   {\it Z. Phys. B\/} \textbf{66} 359

\bibitem{Alc10} Alcoba D R, Bochicchio R C, Lain L and Torre A 2010
                   {\it J. Chem. Phys.\/} \textbf{133} 144104

\bibitem{Ion05} Ionicioiu R and Popescu A E 2005
                   {\it New J. Phys.\/} \textbf{7} 120



\end{thebibliography}
\end{document}